\documentclass{aa}  

\usepackage{graphicx}
\usepackage{txfonts}
\usepackage[colorlinks=true,linkcolor=bluea,allcolors=blue]{hyperref}
\usepackage{booktabs}
\usepackage{lscape}
\usepackage{tabularx,colortbl}
	
\usepackage{lineno}
\usepackage[T1]{fontenc}
\usepackage[utf8]{inputenc}
\begin{document}

   \title{The properties of the obscuring material of an AGNs sample from mid-IR and X-ray simultaneous fitting}
   
   \author{D. Esparza-Arredondo\inst{1,2,3}
          \and
          O. Gonz\'alez-Mart\'in \inst{3}
          \and
          D. Dultzin\inst{4}
          \and
          C. Ramos Almeida \inst{1,2}
        \and
          B. Garc\'ia-Lorenzo \inst{1,2}
          \and
          A. Alonso-Herrero\inst{5}
          \and
          I. Garc\'ia-Bernete \inst{6}
          \and
          J. Masegosa\inst{7}
          }

   \institute{Instituto de Astrof\'isica de Canarias (IAC), C/Vía Láctea, s/n, E-38205 LaLaguna, Spain 
         \email{donaji@iac.es}
         \and 
            Departamento de Astrof\'isica, Universidad de La Laguna (ULL), E-38205 La Laguna, Spain
         \and
            Instituto de Radioastronom\'ia y Astrof\'isica (IRyA-UNAM), 3-72 (Xangari), 8701, Morelia, Mexico
         \and
             Instituto de Astronom\'ia (IA-UNAM), Mexico city, Mexico
        \and
            Centro de Astrobiolog\'{\i}a (CAB), CSIC-INTA, Camino Bajo del Castillo s/n, E-28692 Villanueva de la Ca\~nada, Madrid, Spain
        \and
            Department of Physics, University of Oxford, Keble Road, Oxford OX1 3RH, UK
        \and
            Instituto de Astrofísica d Andalucía, CSIC. C/Glorieta de la Astronomía s/n, E-18005 Granda, Spain
             }

   \date{Received 31/08/24 ; Accepted 23/03/2025}
\titlerunning{Dusty-gas torus properties}
 
  \abstract
   {Over ten mid-infrared (mid-IR) and X-ray models are currently attempting to describe the nuclear obscuring material of active galactic nuclei (AGNs), but many questions remain unresolved. }
   {This study aims to determine the physical parameters of the obscuring material in nearby AGNs and explore their relationship with nuclear activity.}
   {We selected 24 nearby Seyfert AGNs with X-ray luminosities ranging from $10^{41}$ to $\rm{10^{44}}$ erg/$\rm{s^{-1}}$, using \emph{NuSTAR} and \emph{Spitzer} spectra. Our team fitted the spectra using a simultaneous fitting technique. Then, we compared the resulting parameters with AGN properties, such as bolometric luminosity, accretion rate, and black hole mass.}
   {Our analysis shows that dust and gas share a similar structure in most AGNs. Approximately 70\% of the sample favour a combination of the X-ray \emph{uxclumpy} model with the \emph{clumpy} and \emph{two-phases} models at IR wavelengths. We found that linking the half-opening angle and torus angular width parameters from X-ray and mid-IR models helps constrain other parameters and break degeneracies. The study reveals that Sy1 galaxies are characterized by low covering factors, half-opening angles, and column densities but high Eddington rates. In contrast, Sy2 galaxies display higher covering factors and column densities, with a broader range of half-opening angles. We also observed that the distribution of obscuring material is closer to the nucleus in intermediate-luminosity sources, while it is more extended in more luminous AGNs.}
   {Our findings reinforce the connection between the properties of gas--dust material within 10 pc and AGN activity. Applying this methodology to a larger sample and incorporating data from facilities such as \emph{JWST} and \emph{XRISM} will be crucial in further refining these results.}


   \maketitle
%

\section{Introduction}
\label{sec:intro}
Active Galactic Nuclei (AGNs) are highly energetic and compact regions found at the centers of galaxies, harboring supermassive black holes with masses ($\rm{M_{BH}}$) ranging from $\rm{\sim 10^{6}}$-$\rm{10^{9} \, M_{\odot}}$. The primary source of the intense radiation emitted by AGNs is the accretion of surrounding matter onto the supermassive black hole, forming a hot and rotating accretion disk. This disk is surrounded by an obscuring region composed of gas and dust, called torus classically. This torus component is key to understanding the nature of the AGNs and its relationship with its host galaxy, being the region that connects the AGNs with the host galaxy \citep{RamosAlmeida17}.

One of the first ideas to explain the optical differences between AGNs spectra suggested that the line-of-sight (LOS) between the observer and central engine is intercepted by this torus \citep[Unification model by][]{Antonucci85, Urry95}. However, the intrinsic properties of the torus might be linked to changes in the accretion state and/or SMBH mass \citep{Fabian06, Khim17, Ricci17}. Several works have found that several parameters, such as the covering factor, defined as the fraction of sky that is obscured, correlates with the AGN luminosity or Eddington ratio due to radiative feedback on dusty gas \citep{Maiolino07, Treister08, Assef13, Ricci17, Ricci23, Buchner17, Ezhikode17, Ananna22}. Furthermore, this possible dependence also changes with redshift \citep{Aird15, Buchner15}.

Resolving the torus component is a challenging task due to its small angular size. Over the past two decades, significant efforts have been made to develop instruments and techniques to determine its morphology and properties. For instance, the Atacama Large Millimeter / Submillimeter Array (ALMA) has detected molecular tori with diameters of around $\rm{\sim 50\,pc}$ \citep[e.g.,][]{Garcia-Burillo24, Imanishi18, Imanishi20, Alonso-Herrero23, Combes19}. The submillimeter images are associated with dust but also synchrotron emission. Several molecular lines can be isolated and related to the molecular torus component \citep[see][]{Pasetto19, Garcia-Burillo21}. Other examples are observations obtained with the VLT telescopes (through the MIDI and MATISSE instruments) that have obtained resolved dust emission morphology at near- and mid-IR wavelengths to some AGNs. These observations have revealed a dust structure with a size smaller than 10\,pc \citep[e.g.,][]{Hoenig13, Tristram14, Lopez-Gonzaga14, Lopez-Gonzaga16, Leftley19, Lopez-Rodriguez20, Isbell22, Gamez-Rosas22}. The near- and mid-IR emission originates from hot dust outside the sublimation radius and is heated by a large portion of the optical/UV photons generated by the AGN accretion disk. These interferometric studies are currently limited to nearby and bright AGNs. With the advent of the extremely large telescope at the end of this decade, we can unambiguously resolve the near and mid-infrared morphology of this dust emission and compare it with model images \citep{Nikutta21a, Nikutta21}.

At the same time, different torus models have been developing in the last decades. These models try to explain the properties of the torus through radiative transfer simulations. Fitting infrared data with these models allows us to obtain physical parameters (hereafter, we call them model parameters) that describe the geometry and distribution of the material (dust or neutral gas). In this work, we use the static models, which assume different geometries and compositions of the dust in a particular lifetime of the sources. In mid-IR, the first static models assumed a toroidal geometry where the dust distribution is smooth with different radial and vertical density profiles \citep[``\emph{smooth} models'', ][]{Pier92, Efstathiou95, Fritz06}. The following models assumed the same geometry but explored a dust distribution in clumps \citep[``\emph{clumpy} models'', ][]{Nenkova02, Nenkova08b, Hoenig10}. In the last decade, more complex models have emerged. The ``\emph{wind-disk} models'' assume that the dust is located in two components: a geometrically thin disk of optically thick dust clumps and an outflowing wind described by a hollow cone composed of dusty clouds \citep[e.g.,][]{Hoenig17}. Several works have also explored the ``\emph{two-phase}'' models in the last few years. These models assume that the distribution of the dust is a combination of smooth and clumpy inside a torus-geometry component \citep[e.g.,][]{Siebenmorgen15, Gonzalez-Martin23}.

The static models produce infrared spectral energy distributions (SEDs), which can be compared with observations from near-IR, mid-IR, and sub-mm instruments. In the last two decades, several works applied this approach to samples with different types of AGNs and found that model parameters depend on classification \citep[e.g.,][]{RamosAlmeida09, RamosAlmeida11, Hoenig10, Alonso-Herrero11, Lira13, Garcia-Bernete24, Efstathiou22} and even a dependency with the AGN luminosity \citep{Gonzalez-Martin19b}. \cite{Gonzalez-Martin19a} also investigated which is the best mid-IR model to reproduce the infrared emission of several Seyferts galaxies \citep[see also][]{Garcia-Bernete22}. They found large residuals irrespective of the model used, indicating either that the AGN dust continuum emission is more complex than predicted by the models or that the parameter space is not well sampled \citep[see also][]{Martinez-Paredes20, Garcia-Bernete19, Victoria-Ceballos22, Garcia-Bernete22}. \cite{Gonzalez-Martin23} presented new SED models that significantly improve the reproduction of AGN-dominated mid-IR spectra, addressing many of the discrepancies found in previous works.

Alternatively, the torus of AGNs can also be studied using X-ray wavelengths, which trace obscuration due to gas. One part of the accretion disk emission is also reprocessed by an optically thin corona of hot electrons plasma above the accretion disc that scatters the energy in the X-ray bands due to inverse Compton \citep[][and references therein]{Netzer15, RamosAlmeida17}. This comptonization produces one of the three main components seen in AGN at X-rays: the intrinsic continuum. The second and third components are the reflection of the intrinsic continuum and the iron emission line at $\rm{6.4\, keV}$ (FeK$\rm{\alpha}$), respectively. The scattering of X-ray emission, reflected by the inner walls of the torus and/or the BLR, produces these two components. While the FeK$\rm{{\alpha}}$ line can be produced by material with column densities ($\rm{N_H}$) as low as $\rm{N_{H} = 10^{21-23} \, cm^{-2}}$, the Compton hump can only be seen by the reprocessing of X-ray photons in a Compton thick material\footnote{Note that the FeK$\rm{\alpha}$ line and the Compton hump components might also be associated with reprocessing of the intrinsic emission at the accretion disc \citep{Fabian98, Laor91}.} ($\rm{N_{H} > 10^{24} \, cm^{-2}}$).

As in the case of mid-IR, some works have developed several models to model the reprocessed X-ray emission, assuming a \emph{smooth} or \emph{clumpy} distribution of neutral gas in a torus with different geometries \citep[e.g.,][]{Balokovic18, Buchner19}. Although there are slight differences in morphology, these models have made it possible to constrain several material properties that originate from the reprocessed emission. Only a few studies have compared these models to constrain the model parameters \citep[e.g.,][]{Liu14, Furui16, Balokovic18}. Recently, \cite{Saha22} used Bayesian methods to investigate the degeneracies between these X-ray model parameters, distinguish between models, and determine the dependence of the parameter constraints on the instruments used. They found that the model parameters are highly unconstrained using only X-ray data.

There is a growing consensus that the gas-producing X-ray reflection and the dust-producing IR emission giving rise to the mid-IR continuum have a common origin and share the same geometry in the central parsecs \citep[][for a review]{RamosAlmeida17}. However, the research community remains unclear on how they establish this connection. Only a few works combine mid-IR and X-ray observations and models for a large collection of objects \citep[e.g.,][]{Ogawa20, Esparza-Arredondo21}. In this work, we apply the simultaneous fitting X-ray-Mid-IR technique (SFT) proposed and tested for the case of IC\,5063 by \citet{Esparza-Arredondo19} to the sample of local AGNs studied by \citet{Esparza-Arredondo21}. The SFT combines the static X-ray and mid-IR models previously developed at each wavelength and their respective data sets. The advantage of the SFT is its ability to constrain better all the model parameters \citep{Esparza-Arredondo19} by simultaneously fitting mid-IR and X-ray data. Since models based on a single wavelength range often suffer from parameter degeneracies, the SFT mitigates these uncertainties by incorporating independent constraints from both bands.

Our goal is to obtain the physical parameters of the obscuring material through our SFT and to shed light on their dependency on the AGN intrinsic properties, looking for signatures of torus evolution. Section \ref{sec:IndvModels_And_SFT} provides a brief overview of the individual models and their incorporation into our SFT. Section  \ref{sec:sample} explains the sample selection process. In Section \ref{sec:specfit}, we detail the method used for spectral fitting. Section\,\ref{sec:Results} presents the outcomes of fitting our sample. Furthermore, in Section\,\ref{sec:Disc}, we analyze our results in the context of our objectives. Section\,\ref{sec:conclusion} gives a summary and conclusions. Throughout this work, we assume a cosmology with $\rm{H_{0} = 70\, kms^{-1}\,Mpc^{-1}}$, $\rm{\Omega_{M} = 0.27}$, and $\rm{\Omega_{\lambda} = 0.73}$.

\section{Models and its implementation in the Simultaneous fitting technique}
\label{sec:IndvModels_And_SFT}
\subsection{Individual model description}
\label{subsec:IndividualModels}
The mid-IR and X-ray models aim to understand the structure and composition of material that obstructs the accretion disk when viewed from certain angles. These models are developed using radiative transfer codes and encapsulate the required physics to interpret the main continuum features at both wavelengths. Mid-IR models account for reemission due to dust, which can reproduce the spectra continuum at this range. Meanwhile, the X-ray models replicate the reflection component and $\rm{FeK_{\alpha}}$ emission line, assuming reflection mainly in neutral gas.

\begin{figure*}
    \centering
    \includegraphics[width=2\columnwidth, trim={0 0 0 0}, clip]{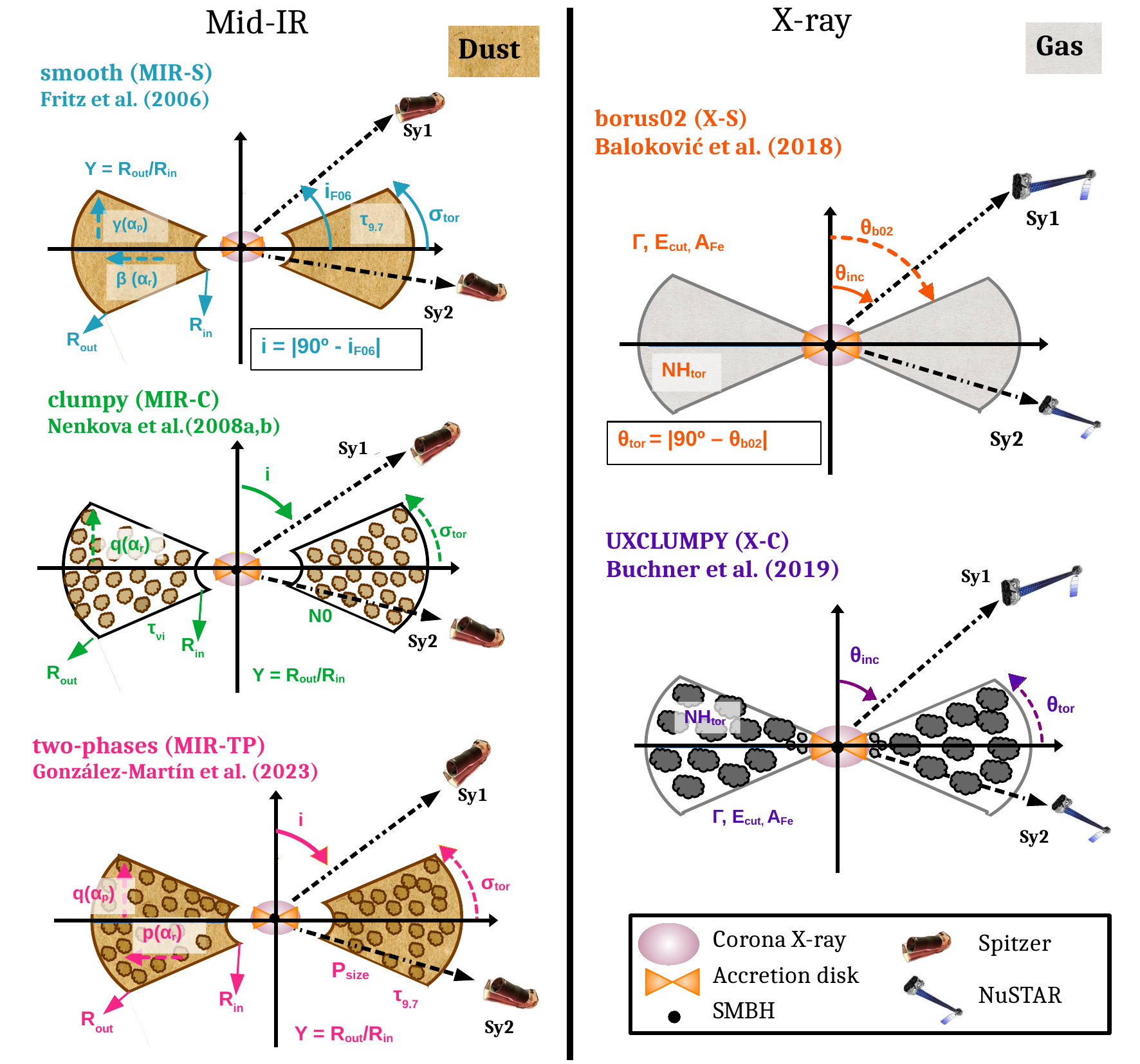}
    \caption{Illustrations of the different geometries of the dust (left) and gas (right) models used in this work. The models are described in Section \ref{subsec:IndividualModels}. In tables \ref{tab:IRparameters} and \ref{tab:XrayParameters}, we describe the parameters of these models. The individual names used to refer to these models, from subsection \ref{sec:sample} onwards, are shown in parentheses.}
    \label{fig:ModelsSchema}
\end{figure*}

In this work, we used two X-ray models: the \emph{borus02} by \citet{Balokovic18} and the \emph{uxclumpy} by \citet{Buchner19} and three mid-IR models: \emph{smooth} by \citet{Fritz06}, the \emph{clumpy} by \citet{Nenkova08a}, and the \emph{two-phases} (GoMar23) by \citet{Gonzalez-Martin23}. The \emph{smooth} and \emph{clumpy} mid-IR models are chosen to match those available at X-rays in geometry. We also included the \emph{two-phases} torus model because it is the one that reproduces the SEDs of a sample of 100 nearby AGNs observed with \emph{Spitzer} \citep{Gonzalez-Martin23}, and its geometry also coincides with that proposed by the X-ray models. In Sect. \ref{sec:sample}, we also used the \emph{clumpy disk$+$wind} model by  \citet{Hoenig17} to fit only the mid-IR data. This approach helped to refine the sample selection because previous works have shown that a significant number of AGNs prefer this model, especially those with the highest luminosities, low column density, or type 1 \citep[e.g.,][]{Gonzalez-Martin19b, Martinez-Paredes21, Garcia-Bernete22}.

In Figure \ref{fig:ModelsSchema}, we show an illustration of the geometry of the three mid-IR (\emph{smooth}, \emph{clumpy}, and \emph{two-phases}) and two X-ray (\emph{borus02} and \emph{uxclumpy}) models used in all this work. Tables \ref{tab:XrayParameters} and \ref{tab:IRparameters} briefly describe the free parameters of the X-ray and mid-IR models, respectively. The models at both wavelengths allow us to calculate the inclination angle (called $\theta_{inc}$ at X-ray and $i$ at IR) and torus angular size ($\sigma_{tor}$ at IR and $\theta_{tor}$ at X-ray). Additionally, the X-ray models also provide the average column density ($N_{H_{tor}}$), the relative abundance of iron ($A Fe K_{\alpha}$, fixed to the solar value), and the power law index ($\Gamma$) multiplied by an exponential cutoff ($e^{-E/E_{cut}}$) describing the torus incident emission. On the other hand, the standard parameters for mid-IR models include the inner and outer radius ratio ($Y$), the radial and polar density distribution slopes ($\alpha_r$ and $\alpha_p$), and the optical depth ($\tau_{\nu}$, derived from $\tau_{97}$ and $\tau_{\nu_0}$). From the \emph{clumpy} and \emph{two-phases} models, we can also obtain the number of clouds in the equatorial plane ($N0$) and the maximum grain size ($P_{size}$), respectively. The tables also provide the ranges for each parameter evaluated. We also suggest referring to the primary papers for a complete description of the models and their parameters \citep[see also our previous works][]{Esparza-Arredondo19, Esparza-Arredondo21}.

\begin{table*}[t]
    \centering
    \setlength{\tabcolsep}{1pt}
    \renewcommand{\arraystretch}{1.2}
    \begin{small}
    \begin{tabular}{lcccc}
    \hline \hline
    \multicolumn{2}{c}{Name and notation} & \multicolumn{3}{c}{Ranges}  \\
     & Symbol & \emph{smooth} & \emph{clumpy} & \emph{two-phases}  \\
    \hline
    Inclination angle & $\rm{i}$ & [$0.01^{\circ}$, $90^{\circ}$] &[$0.01^{\circ}$, $90^{\circ}$] & [$0^{\circ}$, $90^{\circ}$]  \\
    Torus angular width & $\rm{\sigma_{tor}}$ & [$20^{\circ}$, $60^{\circ}$] & [$15^{\circ}$, $70^{\circ}$] & [$10^{\circ}$, $80^{\circ}$] \\
    {Ratio between the inner and outer radius} & $\rm{Y}$ & [10, 150] &   [5, 100] & [2, 40]\\
    Equatorial optical depth at $\rm{9.7\mu m}$ & $\rm{\tau_{9.7 \, \mu m}}$ &  [0.1, 10] & & [3, 13]\\
    {Number of clouds in the equatorial plane} & $\rm{N0}$ & - &  [1, 15] & - \\
    {Slope of the radial density distribution} & $\rm{\alpha_{r}}$ & [1, 0.01] & [0.01, 2.5] &  [0,1.5] \\ 
    {Slope of the polar dust density distribution} & $\rm{\alpha_{p}}$ & [0.01, 6] &  &  [0,1.5] \\
    Optical depth of each cloud & $\rm{\tau_{\nu i}}$ & - & [10, 300] & - \\
    Maximum dust grain size & $\rm{P_{size}}$ & - & - & [0.01, 10]\\[2pt]
    \hline 
    Distribution of the dust & & \emph{smooth} & \emph{clumpy} & \emph{smooth} and \emph{clumpy} \\[1pt]
    \hline
    \end{tabular}
    \end{small}
    \caption{Description and range of the parameters from the IR models used in this work. The inclination angle from \emph{smooth} is measured with respect to the equatorial plane; in the case of the \emph{clumpy} and \emph{two-phases} models, these angles are measured with respect to the polar plane. The torus angular width is measured with respect to the equatorial plane in all models. The $\alpha_{r}$ parameter is abbreviated as $\rm{\beta}$, $q$, and $p$ in the original works: \citet{Fritz06, Nenkova08b, Gonzalez-Martin23}, respectively. The $\alpha_{p}$ parameter is abbreviated as $\rm{\gamma}$ and $\rm{q}$ by \citet{Fritz06} and \citet{Gonzalez-Martin23}, respectively. The dust is distributed in these models following a variable radial power law and polar exponential density distribution of dust: $\rho (r,\theta) \propto r^{-\alpha_r} e^{- \alpha_p |cos (\sigma_{tor})|}$, where $r$ and $\theta$ are polar coordinates.}
    \label{tab:IRparameters}
\end{table*}

\begin{table*}[t]
    \centering
    \setlength{\tabcolsep}{1pt}
    \renewcommand{\arraystretch}{1.2}
    \begin{small}
    \begin{tabular}{lccc}
    \hline \hline
    \multicolumn{2}{c}{Name and notation} & \multicolumn{2}{c}{Ranges}  \\
     & Symbol  & \emph{borus02} & \emph{uxclumpy}  \\
    \hline
    \hline
    Inclination angle & $\rm{\theta_{inc}}$ & [$19^{\circ}$, $87^{\circ}$] &[$0^{\circ}$,$90^{\circ}$] \\
    Half opening angle of the polar cutouts & $\rm{\theta_{tor}}$ & [$0^{\circ}$, $84^{\circ}$] & [$6^{\circ}$, $90^{\circ}$] \\
    Average column density ($cm^{-2}$) & $\rm{N_{H_{tor}}}$ & [22.0, 25.5] & [20.0, 26.0] \\
    Photon index & $\rm{\Gamma}$ & [1.4, 2.6] & [1.0, 3.0] \\
    High-energy cut-off & $\rm{E_{cut}}$ & [20,2000] & [60,400] \\
    Relative abundance of Iron & $\rm{A_{Fe}/A_{Fe,\odot}}$ & [0.01,10.0] & 1 (fix) \\
    Covering fraction of the inner ring & $\rm{CTkcover}$ & - & [0, 0.6] \\[2pt]
    \hline 
    Distribution of the neutral gas & & \emph{smooth} & \emph{clumpy} \\[1pt]
    \hline \hline 
    \end{tabular}
    \end{small}
    \caption{Description and range of the parameters from the X-ray models used in this work. The $\rm{\theta_{inc}}$ angle is measured with respect to the polar plane (where the source is observed face-on when the $\rm{\theta_{inc} = 19^{\circ}}$, and edge-on if $\rm{\theta_{inc} = 84^{\circ}}$). The half-opening angles are measured with respect to the equatorial plane in all models. A description of physical parameters is given in Table 1 of \cite{Esparza-Arredondo21}, and a cartoon of the models is shown in Figure 2 of \cite{Esparza-Arredondo19}.}
    \label{tab:XrayParameters}
\end{table*}

\subsection{Baseline model definition in the SFT: Mixing the X-ray and mid-IR models.}
\label{subsec:ModelsDesc}
In this work, we follow the simultaneous fitting technique (SFT) proposed by \citet{Esparza-Arredondo19}. This technique combines X-ray and mid-IR models previously developed at each wavelength and their respective data sets. The SFT is performed through the XSPEC package, a command-driven interactive, spectral-fitting tool within the HEASOFT software\footnote{https://heasarc.gsfc.nasa.gov}. XSPEC allows fitting data from different satellites such as \emph{ROSAT}, \emph{Chandra}, \emph{XMM-Newton}, and \emph{NuSTAR} with models constructed from single emission components coming from different mechanisms and physical regions. Most X-ray models are included in XSPEC, and it is possible to include new ones using the ATABLE task. We converted the mid-IR SED libraries of models to multi-parametric models within the spectral-fitting tool XSPEC as an additive table (see Section 4 of \cite{Esparza-Arredondo19} and \cite{Gonzalez-Martin19a}).

In this work, we define our baseline models as the following command sequence in XSPEC:
\begin{equation}
\begin{split}
    & phabs * (atable \lbrace Xmodel \rbrace + zdust*zphabs * cabs \\ & * cutoffpl) + zdust*atable \lbrace IRmodel \rbrace
\end{split}
\label{eq1}
\end{equation}
\noindent where ${phabs}$ is the foreground galactic absorption (see Col.\,3 in Table \ref{tab:Sample}), and ${ zdust * zphabs * cabs}$ represents the line-of-sight absorption at the redshift of the source. The $cutoffpl$ component is a power law with high energy exponential rolloff, which models the intrinsic component of X-ray. These X-ray absorbers are not evaluated at energies below $\rm{10^{-4}}$ keV. However, the mid-IR and X-ray simultaneous fit requires the X-ray intrinsic emission to be properly absorbed below those energies. Therefore, we introduced a $zdust$ component to neglect any artificial contribution of this component to mid-IR wavelengths and incorporated a foreground extinction at this range. The mid-IR and X-ray models are introduced through the command $atable$. We have the option to use either \emph{borus02} or \emph{uxclumpy} model for the reflection component and reproduce the dust continuum with the \emph{smooth}, \emph{clumpy}, or \emph{two-phases} model (see Subsection \ref{subsec:IndividualModels} for an individual description of models).

In our work \cite{Esparza-Arredondo19}, we have shown that combining the mid-IR and X-ray models in a single baseline model can bear significant advantages. One of them is the ability to link similar parameters. For instance, the inclination angle for the mid-IR models can be set to the same value as the X-ray model ($i = \theta_{inc}$), thereby reducing the number of free parameters and simplifying the model.

\section{Sample}
\label{sec:sample}
The sample analyzed here contains 24 nearby AGNs ($\rm{z < 0.4}$) drawn from our previous work by \citet{Esparza-Arredondo21}. In this preliminary work, we selected 36 sources from an initial sample of 169 AGNs with available archival \emph{Spitzer} and \emph{NuSTAR} data. These 36 AGNs have a non-negligible contribution to the reflection component at X-rays and are dominated by the AGN-heated dust at mid-IR. In \citet{Esparza-Arredondo21}, we individually fitted the \emph{Spitzer} and \emph{NuSTAR} spectra of these sources with the \emph{smooth} or \emph{borus02} and \emph{clumpy} or \emph{uxclumpy} models at mid-IR and X-ray.

Out of the 36 AGNs that we selected, we could not fitted three of them (Mrk\,1018, PG\,1535$+$547, and ESO\,428-G014) at X-rays, and we could not fit four sources (Mrk\,3, NGC\,4388, ESO-097-G013, and MCG+07-41-03) at mid-IR with any of the four models that we tested. This means that the $\rm{\chi^{2}/d.o.f.}$ value was greater than 1.2. As a result, we excluded these seven objects from the analysis. We found that the \emph{smooth} (\emph{borus02}) and \emph{clumpy} (\emph{uxclumpy}) models cannot explain their mid-IR (X-ray) properties. Therefore, we reduced the sample size to 29 sources.

We now improve the analysis by fitting the 29 \emph{Spitzer} spectra to the \emph{disk$+$wind} and \emph{two-phases} torus models. This analysis shows four sources (RBS\,0770, PG\,1211+143, RBS\,1125, and Mrk\,1393) that require the disk$+$wind model at mid-IR wavelengths. We also exclude them from this analysis because no simultaneous fit can be performed due to the lack of a self-consistent X-ray model for the disk$+$wind distribution. We also discard MCG+01-57-016 because its X-ray spectrum is complex and could be associated with the accretion disk \citep{Akylas22}. We will study these 11 discarded AGNs in a forthcoming investigation using new and more complex models.

Therefore, the sample for this study comprises the remaining 24 nearby Seyferts, consisting of eight type 1 (Sy1) and sixteen type 2 (Sy2). Our sample covers three orders of magnitude in X-ray intrinsic luminosity ($\rm{log(L_{2-10}\, keV) \sim 41.7 - 44.6}$), and three orders of magnitude in black hole mass ($\rm{log(M_{BH}) \sim 6.2 - 8.9}$). Table\,\ref{tab:Sample} (Cols. 1-7) shows the main observational details of the final sample.

\begin{table*}[ht]
\caption{General information of the sample}
    \label{tab:Sample}
    \centering
    \setlength{\tabcolsep}{1pt}
    \begin{small}
    \begin{tabular}{lcccccccc}
    \hline \hline
     Objname & z & $N_{H_{Gal}}$ & $\rm{L_{X_{(2-10KeV)}}}$ & $\rm{L_{bol}}$ & $\rm{log(M_{BH})}$ & $\rm{L/L_{edd}}$ &  mid-IR \\
           &    & $\rm{cm^{-2}}$ & erg s$\rm{^{-1}}$ & erg s$\rm{^{-1}}$ & $\rm{M_{\odot}}$ & & model\\
    (1) & (2) & (3) & (4) & (5) & (6) & (7) & (8) \\
    \hline \hline
\hline
    &  & Seyfert 1 &  & \\
\hline 
Mrk590  & 0.0213 & 0.040 & 42.57$\pm_{0.02}^{0.02}$ & 43.36 & 7.57$^{1}$ & -2.32    & \emph{clumpy}  \\
PG0804+761  &  0.1000 & 0.038 & 42.55$\pm_{0.01}^{0.01}$ & 45.30 & 8.73$^{1}$  & -1.54 & \emph{two-phases}  \\ 
I11119+3257 & 0.1890 & 0.024 & 44.12$\pm_{0.08}^{0.06}$ & 45.17 & 7.30$^{2}$ & -0.24 & \emph{two-phases} \\  
Mrk231 & 0.0422 & 0.010 & 42.78$\pm_{0.04}^{0.04}$ & 43.6 & 8.40$^{3}$ & -2.91  &     \emph{two-phases} \\
Mrk1383  & 0.0866 & 0.035 &44.30$\pm_{0.01}^{0.01}$ & 45.23 & 9.01$^{1}$ &  -1.89 &  \emph{two-phases}  \\
Mrk1392  & 0.0361 & 0.050 & 43.18$\pm_{0.02}^{0.02}$ & 44.09 &  7.87$^{1}$ & -1.89  & \emph{clumpy}  \\
ESO141-G055  & 0.0371 & 0.118  & 43.94$\pm_{0.01}^{0.01}$ & 44.91 & 7.99$^{1}$ & -1.19  & \emph{two-phases} \\
NGC7213  & 0.0051 & 0.017 & 41.98$\pm_{0.01}^{0.01}$  & 42.80 &  7.13$^{1}$ &   -2.44 & \emph{clumpy}   \\
\hline
  &  & Seyfert 2 &   & \\
\hline 
UM146  & 0.0144 & 0.079 & 41.84$\pm_{0.07}^{0.06}$  & 42.82 & 6.83$^{4}$ & -2.12 &  \emph{clumpy}    \\
NGC788  & 0.0136 & 0.029 & 42.43$\pm_{0.07}^{0.06}$ & 43.41 & 8.18$^{1}$ &  -2.88 &  \emph{clumpy}   \\
NGC1052  & 0.0048 & 0.029 & 41.74$\pm_{0.01}^{0.01}$ & 42.55 & 8.82$^{1}$ & -4.38 &  \emph{clumpy} \\ 
NGC1358  &  0.0134 & 0.066 & 42.38$\pm_{0.03}^{0.02}$ & 43.83 & 8.10$^{3}$ &  -2.38 & \emph{clumpy} \\
J05081967+1721483  & 0.0175 & 0.383 & 42.97$\pm_{0.02}^{0.02}$ & 43.83 & 8.35$^{1}$ & -2.63 & \emph{two-phases} \\ 
Mrk78   & 0.0371 & 0.038 & 43.11$\pm_{0.05}^{0.05}$ & 44.05 &  8.53$^{1}$ &  -2.59 &  \emph{clumpy}   \\ 
Mrk1210 & 0.0135 & 0.033 & 44.63$\pm_{0.02}^{0.02}$ & 43.78 & 6.86$^{1}$  & -1.19 &  \emph{two-phases} \\ 
J10594361+6504063  & 0.0836 & 0.028 & 43.66$\pm_{0.02}^{0.02}$ & 44.63  & 8.25$^{1}$  & -1.73 &   \emph{smooth}  \\
NGC4507  & 0.0118 & 0.105 & 43.17$\pm_{0.01}^{0.01}$& 44.21 & 7.81$^{1}$ &  -1.71 & \emph{clumpy}   \\
NGC4939  & 0.0085 & 0.044 & 42.0$\pm_{0.04}^{0.04}$ & 43.00 & 7.75$^{1}$ &  -2.86 & \emph{two-phases}   \\ 
IC4518W   & 0.0162 & 0.170 & 42.68$\pm_{0.06}^{0.05}$ & 43.61 & 7.64$^{1}$ & -2.14 &    \emph{two-phases} \\
ESO138-G1  & 0.0091 & 0.217 & 42.43$\pm_{0.04}^{0.04}$ & 42.57 & 6.57$^{1}$ & -2.11 &   \emph{clumpy}  \\
NGC6300 & 0.0029 & 0.105 & 41.86$\pm_{0.01}^{0.01}$ & 42.69  &  6.77$^{1}$  & -2.19 &  \emph{two-phases} \\ 
ESO103-G35 & 0.0133 & 0.082 & 43.23$\pm_{0.01}^{0.01}$ & 44.15 & 7.37$^{1}$ &  -1.33 &   \emph{clumpy}     \\
IC5063  & 0.0088 & 0.067 & 42.55$\pm_{0.02}^{0.01}$ & 43.47  &  8.24$^{1}$ &  -2.88 &  \emph{clumpy}   \\
PKS2356-61  & 0.0963 & 0.015 & 43.96$\pm_{0.04}^{0.04}$ & 45.15 & 8.55$^{1}$ &  -1.51 &   \emph{clumpy}     \\
\hline \hline
\end{tabular}
\end{small}
\tablefoot{Columns from 1 to 8 correspond to the object name, redshift, galactic column density (obtained by the $\rm{N_H}$ tool within HEASOFT), intrinsic (not absorbed) X-ray luminosity (obtained through the clumin tool from Xspec), black hole mass, Eddington ratio, and the best mid-IR model obtained from individual fits, respectively. The bolometric luminosity is computed from the X-ray luminosities reported in Col.\,5 of this table and using the relationship $\rm{L_{bol} = k L_X(2-10keV)}$, where the bolometric correction (k) is determined by a fourth-order polynomial dependent on $\rm{L(2-10\,keV)}$ \citep{Marconi04}. $\rm{M_{BH}}$ references: (1) \cite{Koss22}, (2) \cite{Yang20}, (3) \cite{Osorio-Clavijo22}, and (4) \cite{vandenBosch16}.}
\end{table*}

\begin{figure*}
    \centering
    \includegraphics[width=0.9\columnwidth, trim={5 10 20 20}, clip]{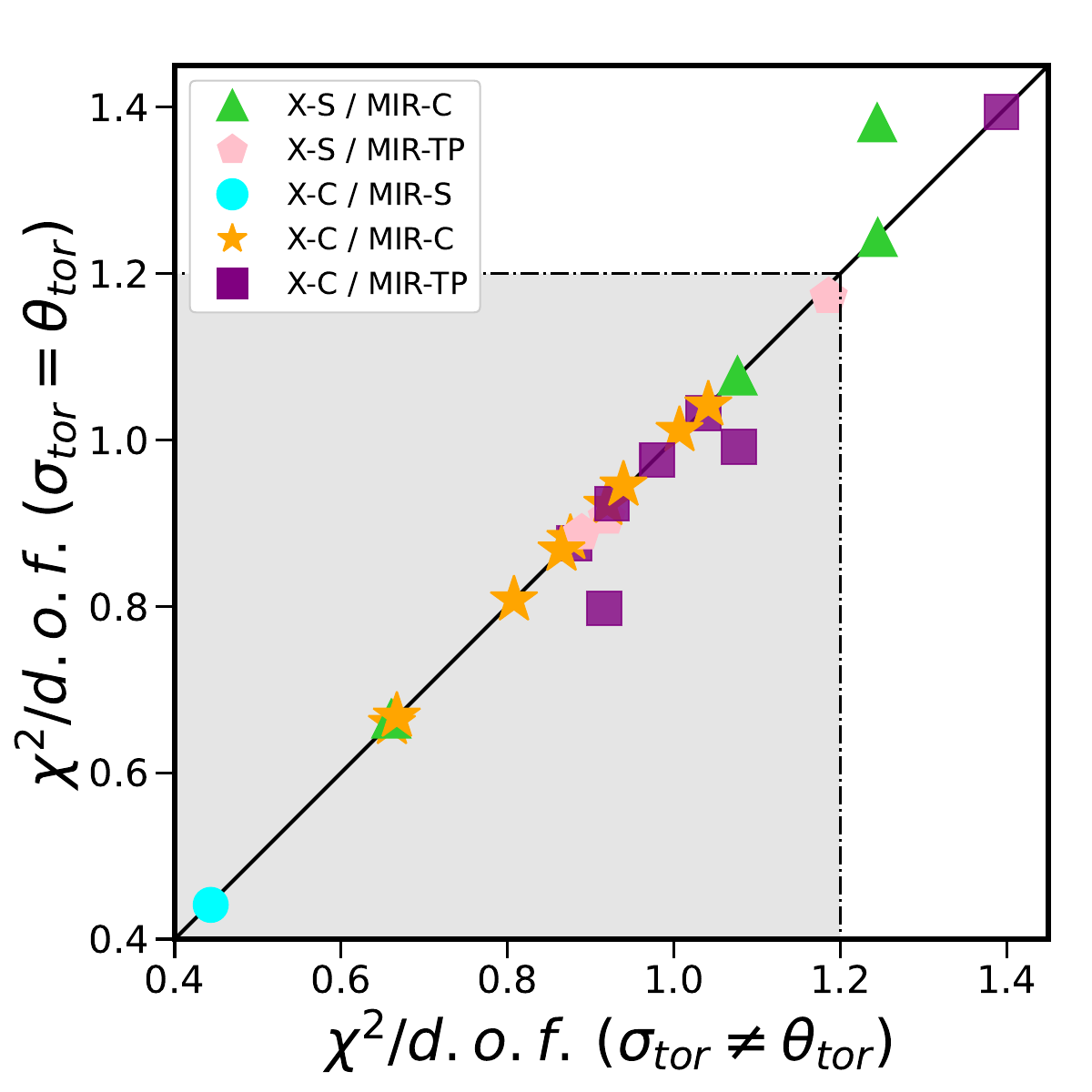}
    \includegraphics[width=0.87\columnwidth, trim={12 0 24 24}, clip]{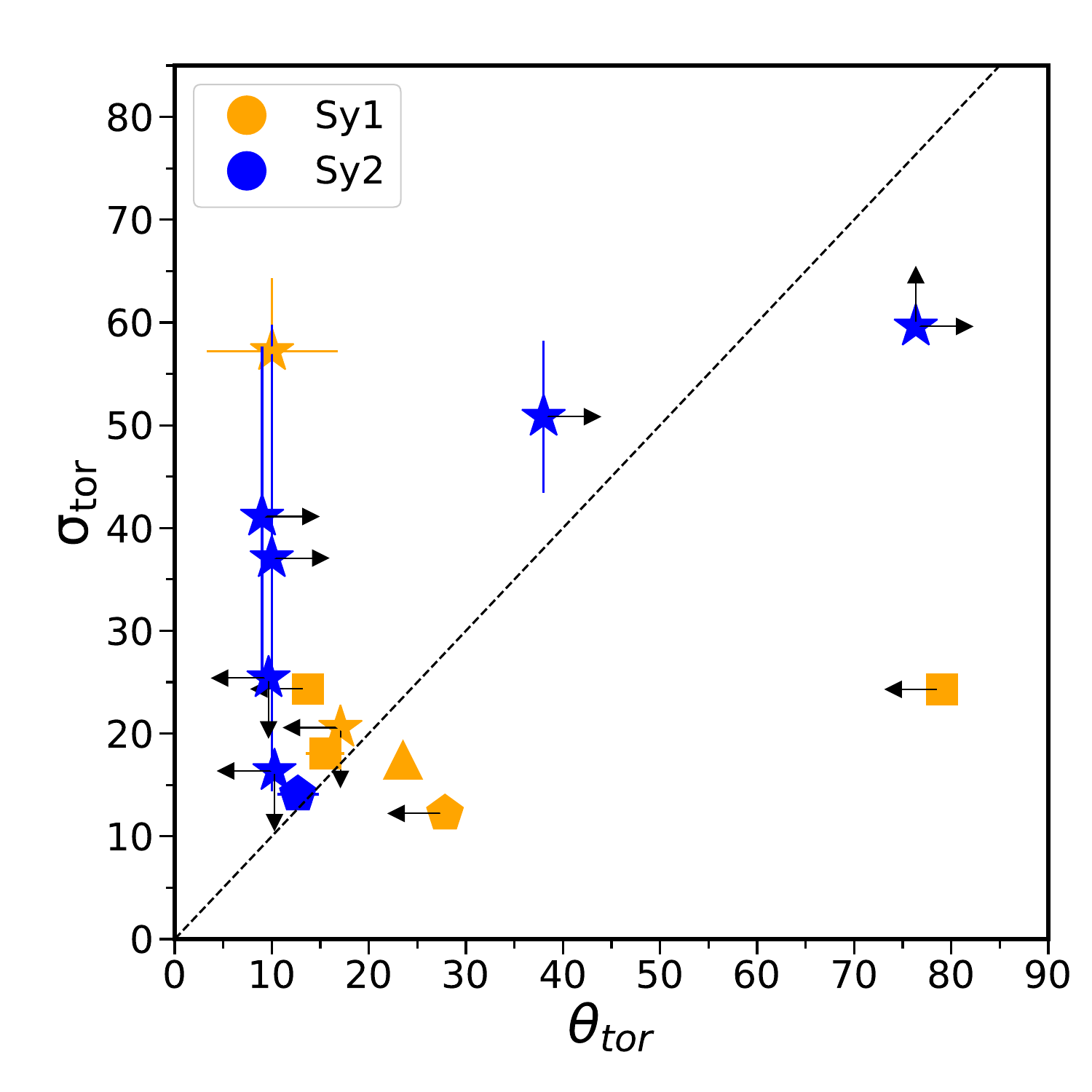}
    \caption{Left: Comparison between $\chi^{2}/d.o.f.$ values before and after linking the half-opening angle from X-ray models ($\rm{\theta_{tor}}$) and torus angular width from IR models ($\rm{\sigma_{tor}}$), in all sources. The shaded area shows the $\chi^{2}/d.o.f.$ values considered as a plausible solution. The baseline models are shown with different symbols and colors: X-S/MIR-C (green triangles), X-S/MIR-TP (pink pentagons), X-C/MIR-S (cyan circles), X-C/MIR-C (orange stars), and X-C/MIR-TP (purple squares). Right: Comparison between the half-opening angle from X-ray models and torus angular width ($\rm{\theta_{tor}}$ vs $\rm{\sigma_{tor}}$), using baseline models where the inclination angles are linked ($\rm{\theta_{inc} = i}$). Sy1 and Sy2 are shown as orange and blue symbols. The arrows show the upper and lower limits in $\theta_{tor}$ and $\rm{\sigma_{tor}}$ along the axes x and y, respectively.}
    \label{fig:Comp_ChiReduced}
\end{figure*}

\begin{figure*}
    \centering
    \includegraphics[width=0.48\columnwidth]{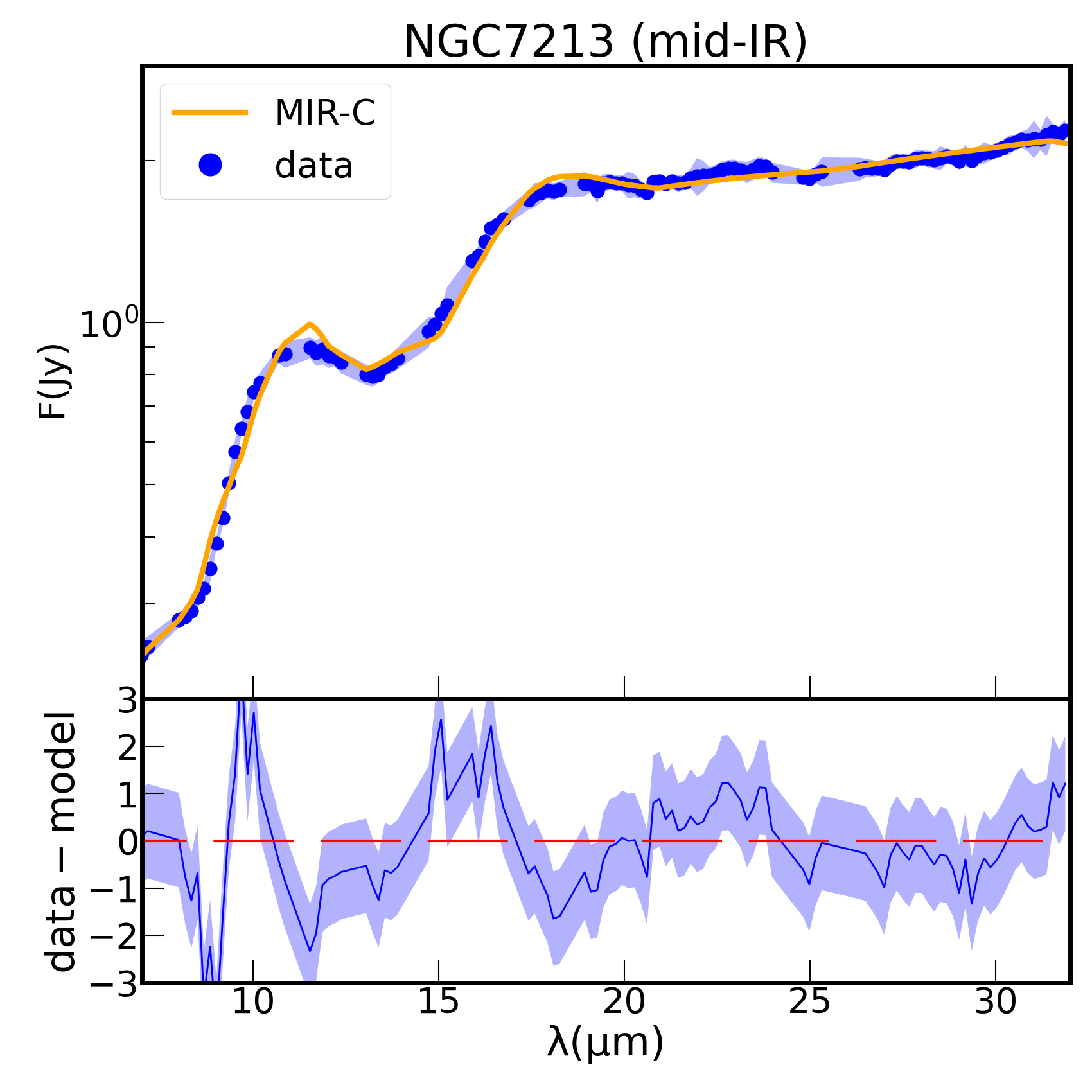}
    \includegraphics[width=0.48\columnwidth]{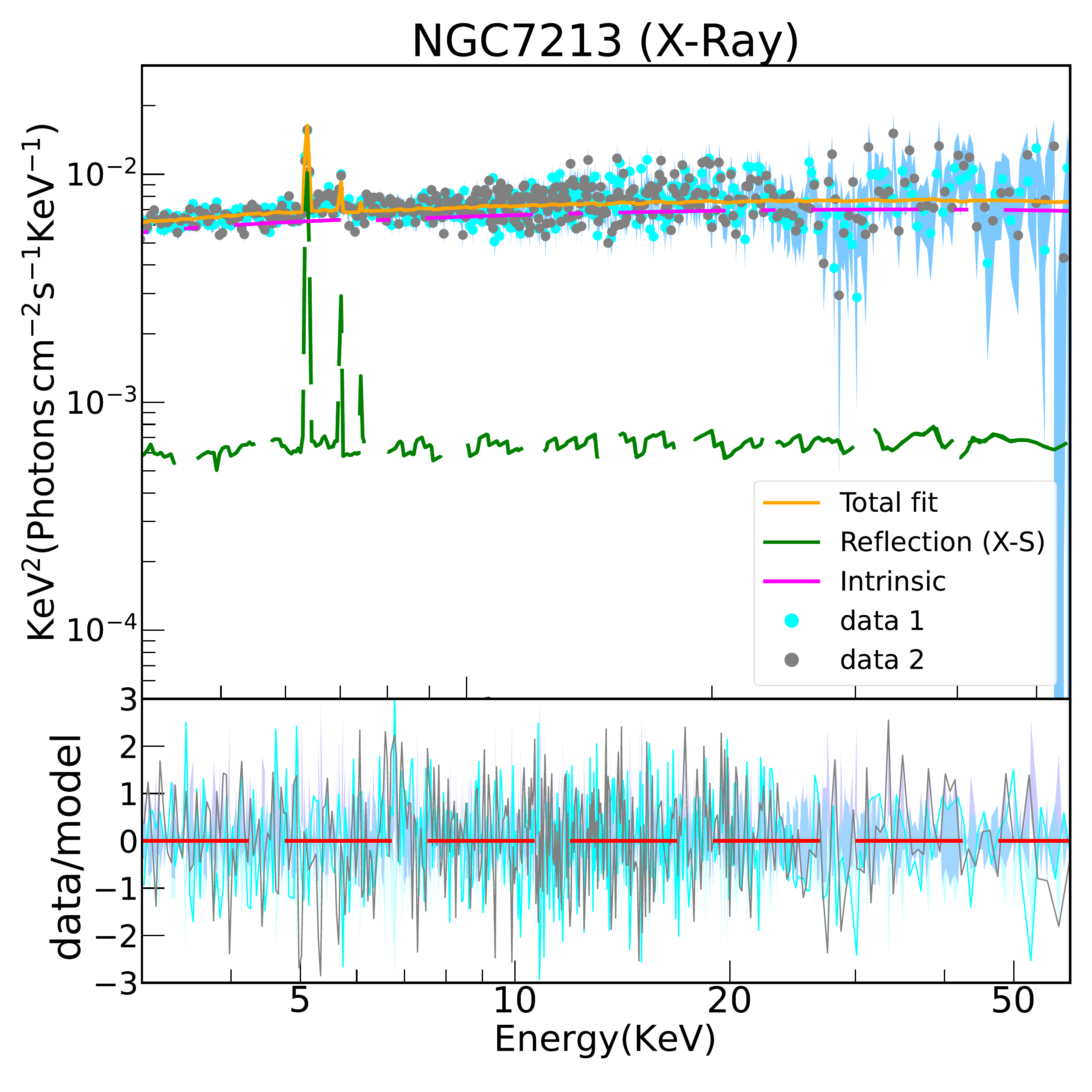}
    \includegraphics[width=0.48\columnwidth]{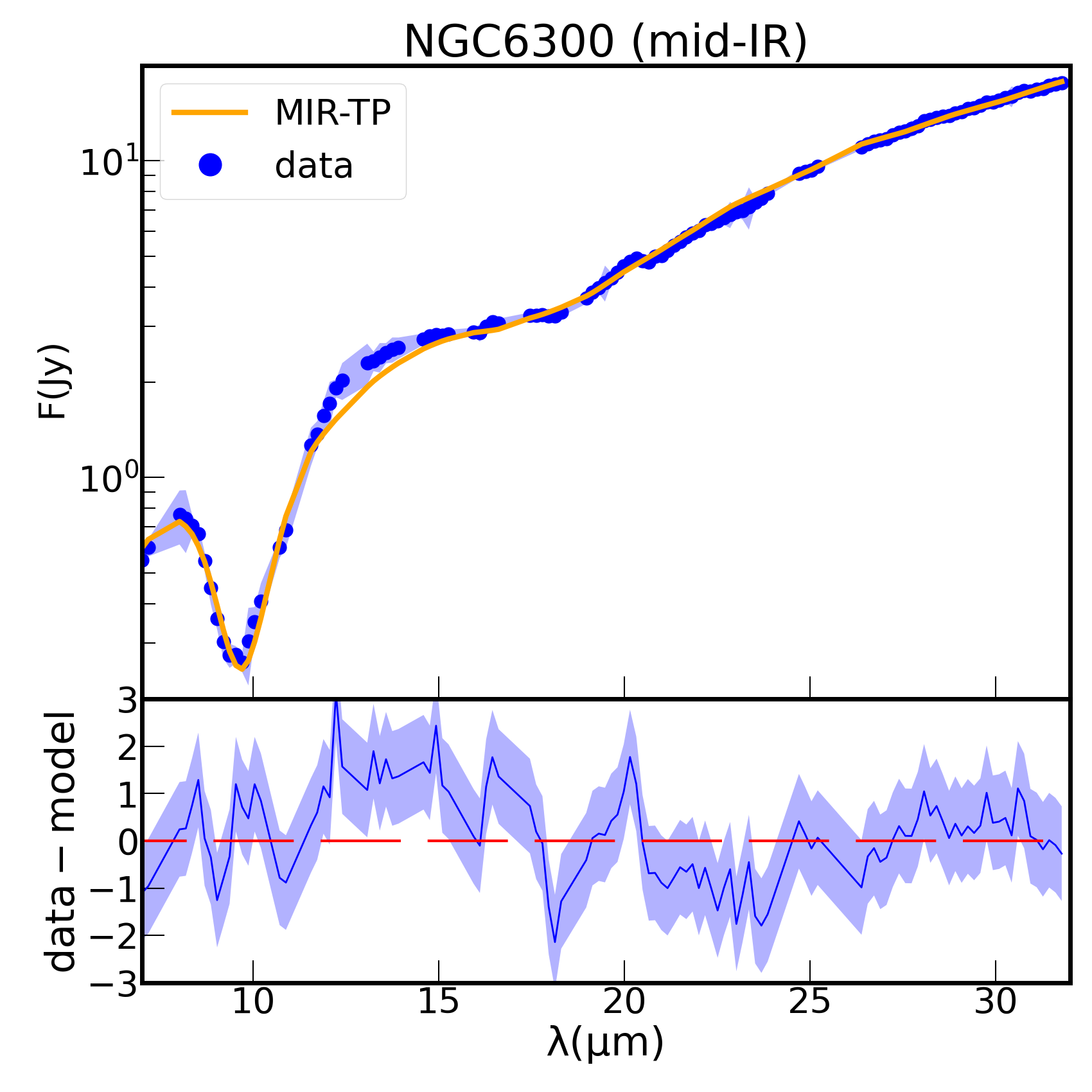}
    \includegraphics[width=0.48\columnwidth]{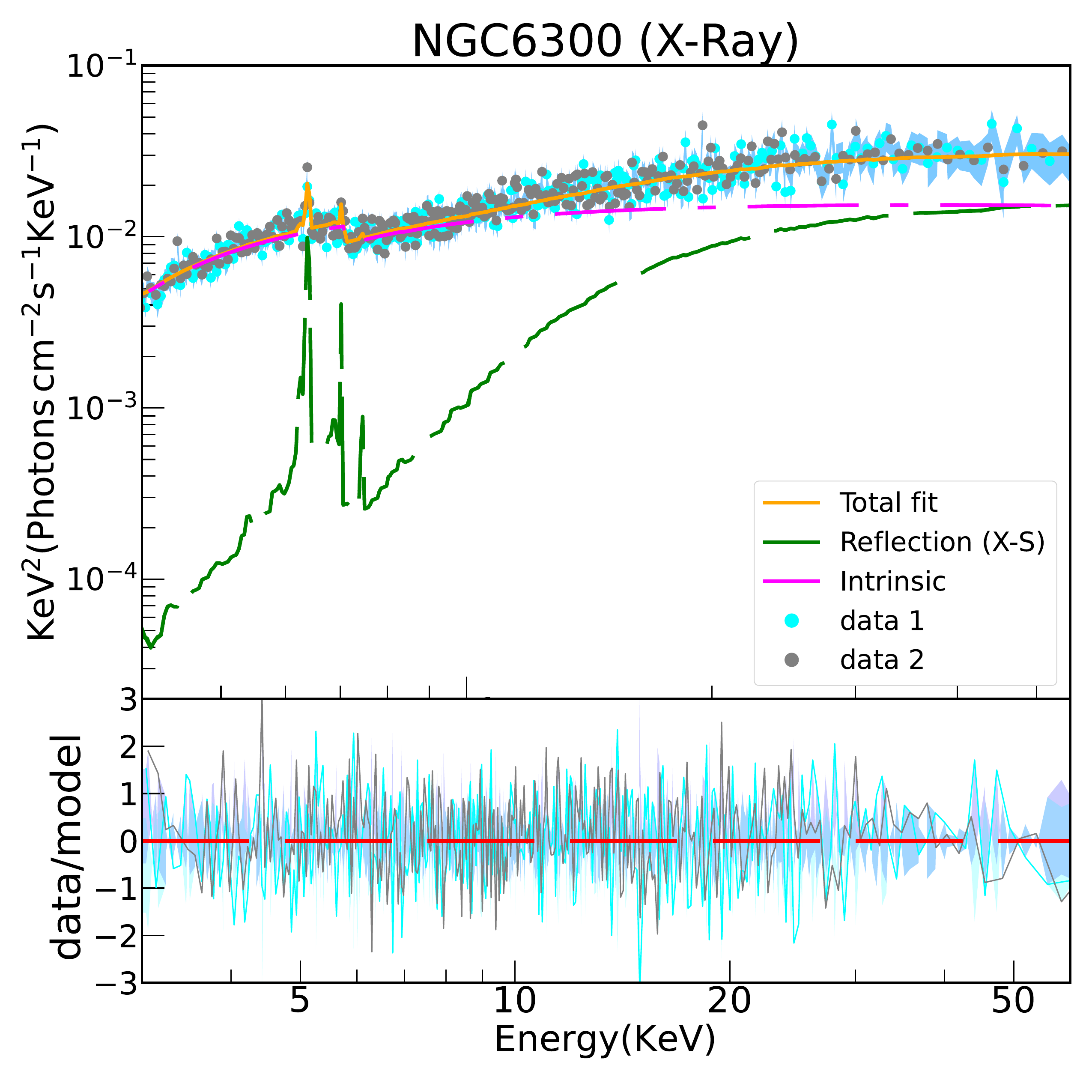}
    \includegraphics[width=0.48\columnwidth]{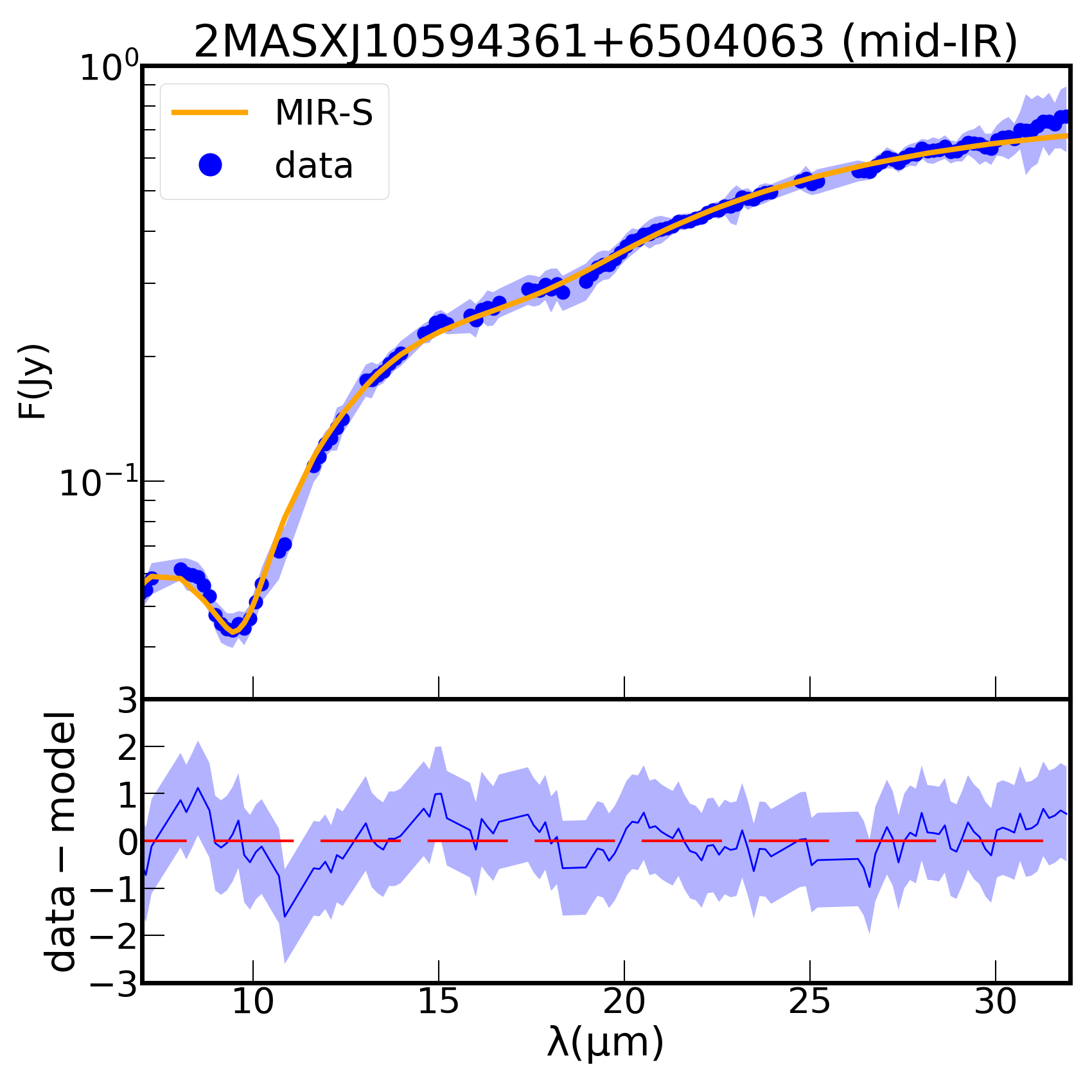}
    \includegraphics[width=0.48\columnwidth]{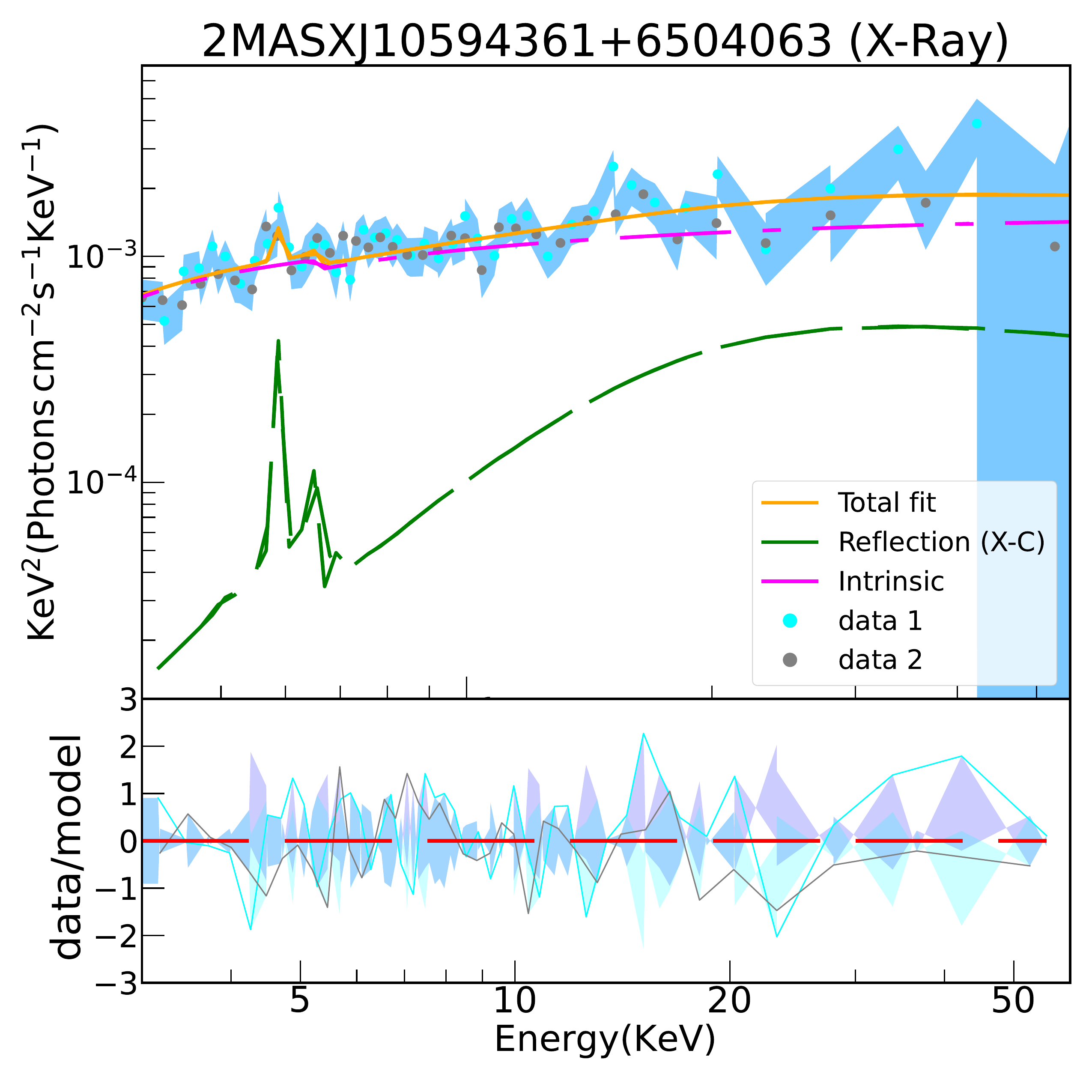}
    \includegraphics[width=0.48\columnwidth]{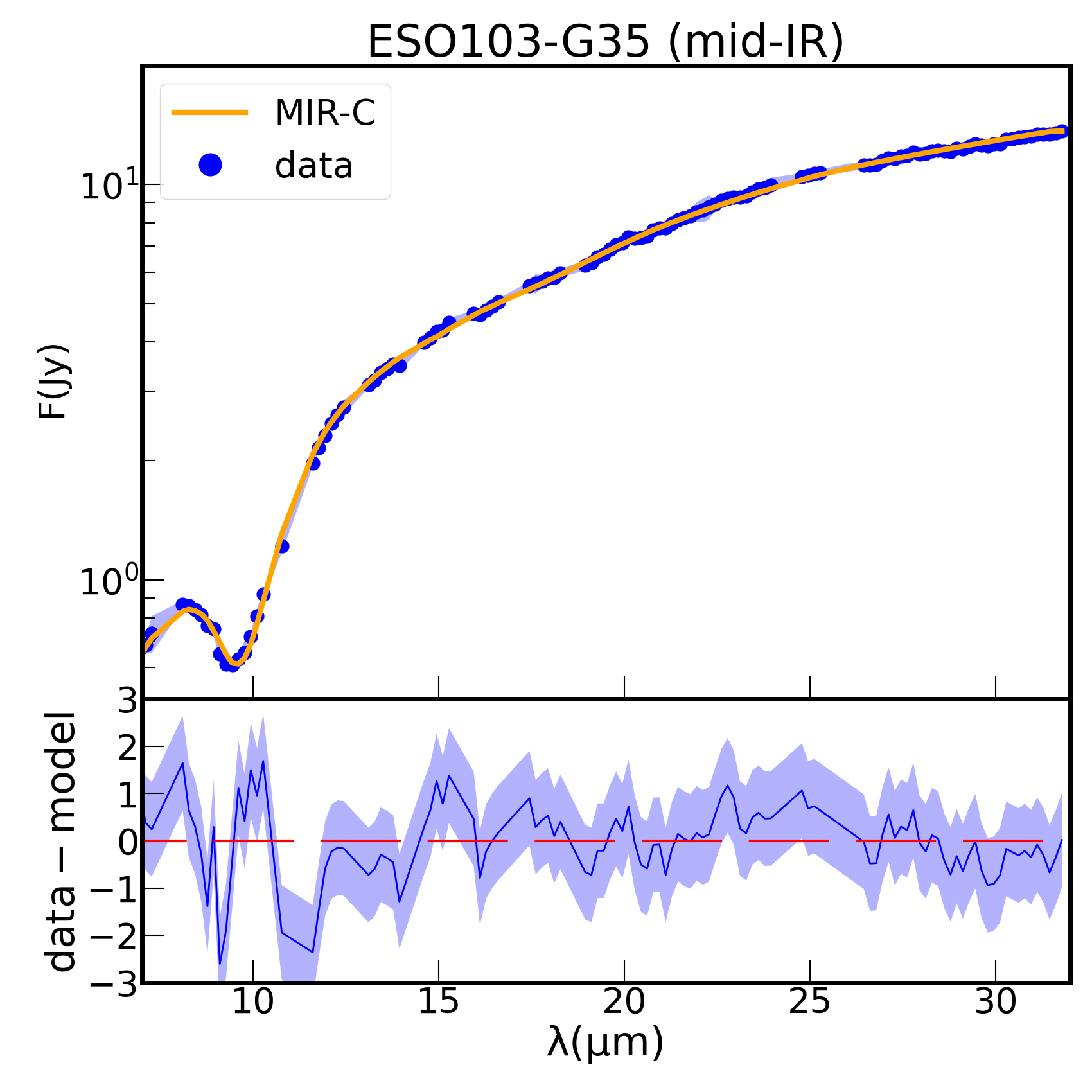}
    \includegraphics[width=0.48\columnwidth]{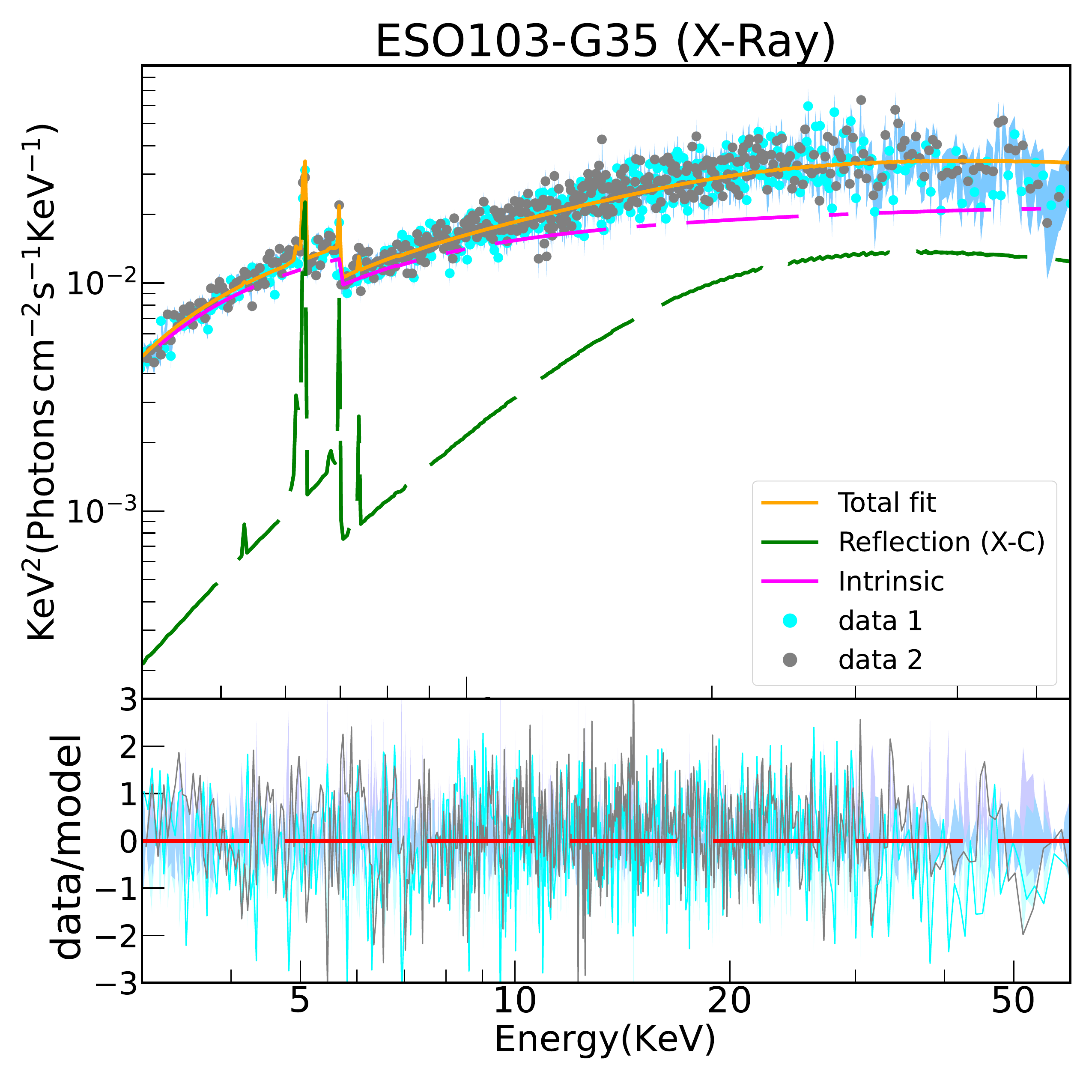}
    \includegraphics[width=0.48\columnwidth]{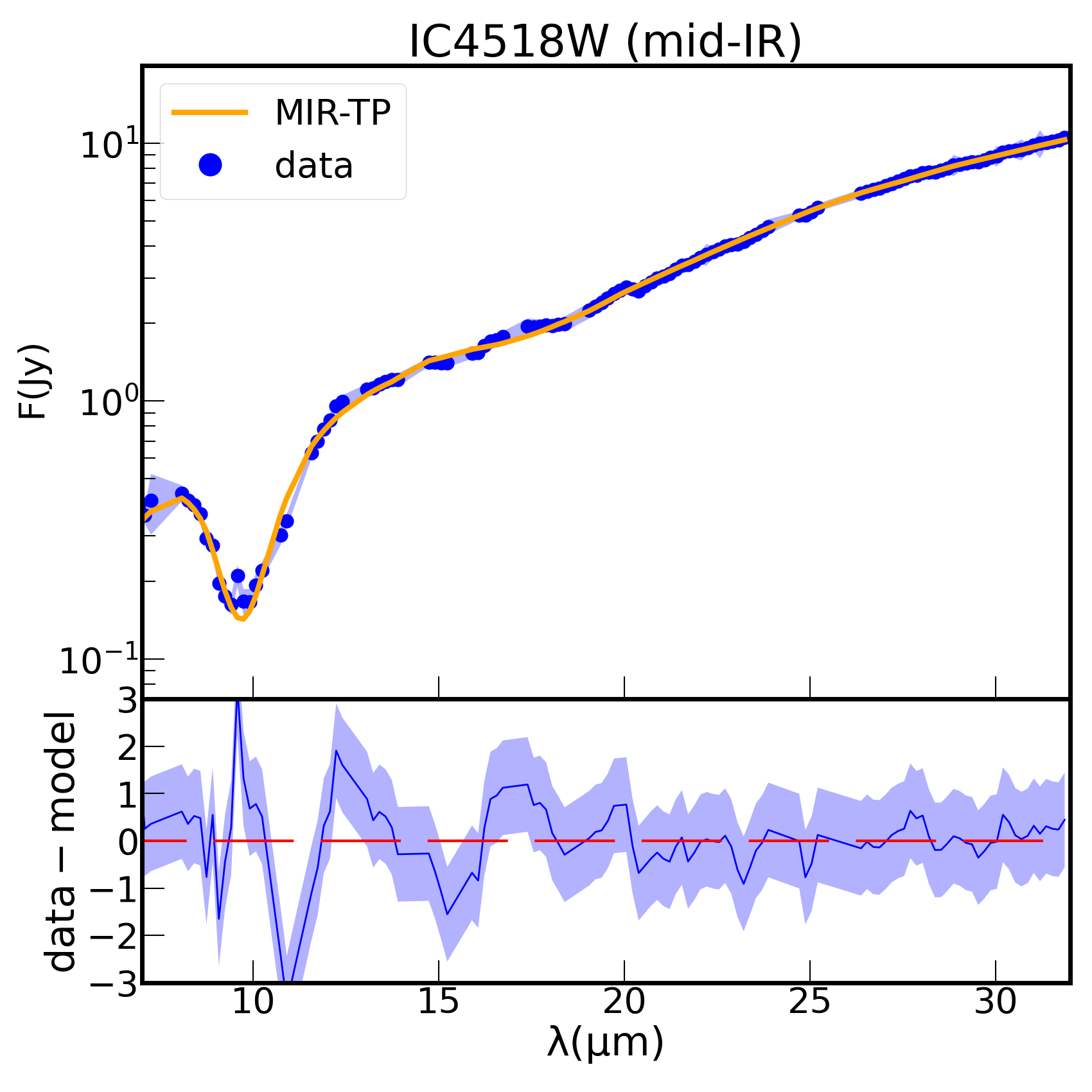}
    \includegraphics[width=0.48\columnwidth]{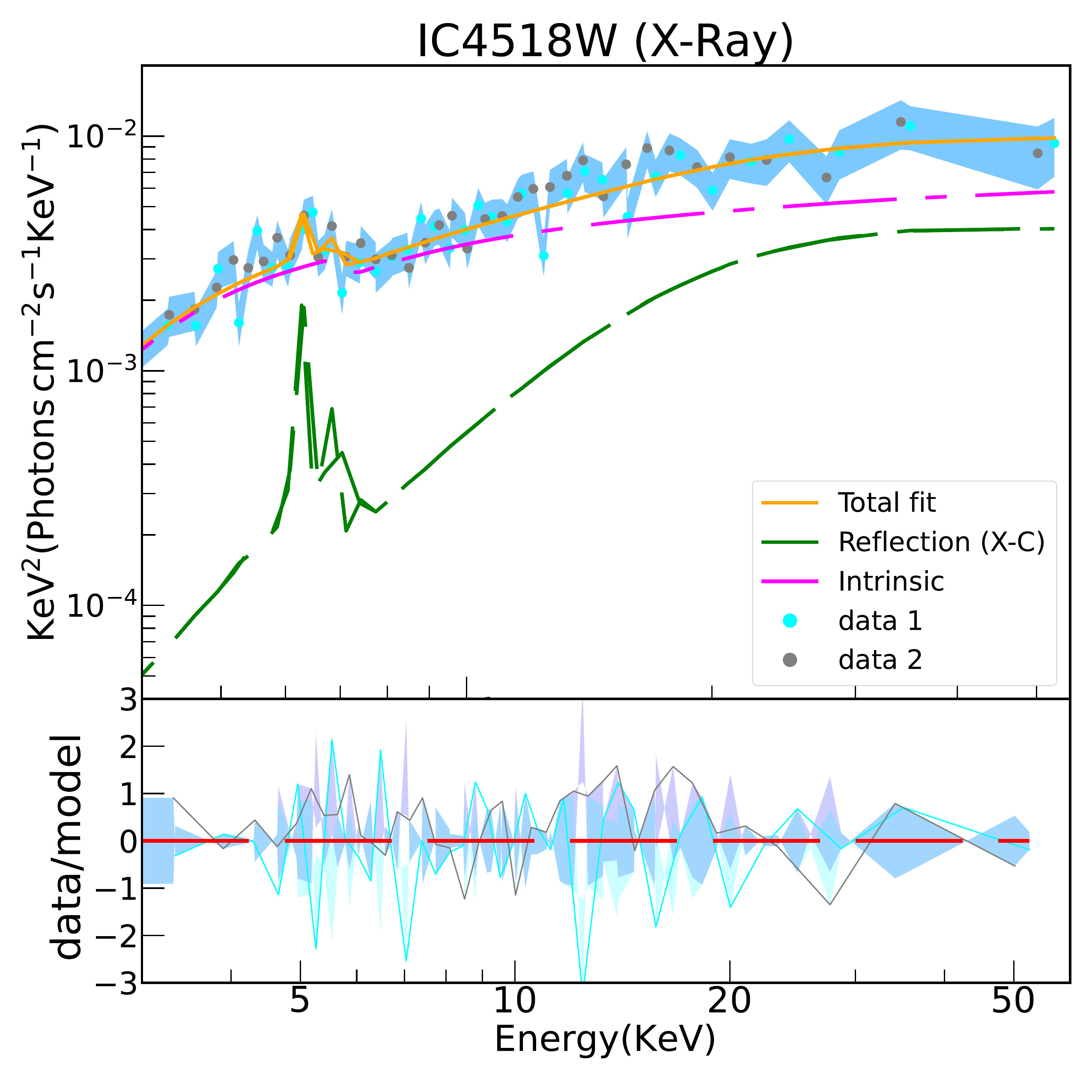}
    \caption{Spectral fits of NGC\,7213, NGC\,6300, J10594361+6504063, ESO103-G35, and IC4518W. The orange solid lines are the best fit obtained from the X-S/MIR-C, X-S/MIR-TP, X-C/MIR-S, X-C/MIR-C, and X-C/MIR-TP baseline models. For each galaxy, the \emph{Spitzer} spectrum is shown with blue points (left panel), and the \emph{NuSTAR} spectra are displayed with blue and cyan points (right panels). The magenta and green dashed lines show the absorbed power law and the reprocessed components, respectively. The lower panels display the residuals between the data and the best-fit model. The data and residual errors are shown as shaded areas.}
    \label{fig:Best_SimultaneousFits}
\end{figure*}

\section{Spectral fitting}
\label{sec:specfit}
We initially fit the \emph{Spitzer} spectra of all sources using three different models: the \emph{smooth}, \emph{clumpy}, and \emph{two-phases} models. To determine the best fit for the mid-IR data, we compare the resulting $\chi^2$ values of each source using the f-test statistic. Based on these results, Column 8 of Table \ref{tab:Sample} reports the preferred mid-IR model. Subsequently, we apply the SFT to all sources. In this step, we simultaneously charge the \emph{NuSTAR} and \emph{Spitzer} spectra of each source into XSPEC software and fit both spectra using a baseline model defined through the command sequence shown in Sect. \ref{subsec:ModelsDesc} (see Eqs.\,\ref{eq1}). It is important to note that the baseline models are a combination of the best mid-IR model with one of the X-ray models (\emph{borus02} or \emph{uxclumpy}). Hence, we have two simultaneous fits to each source at this stage, which combine the \emph{borus02} or \emph{uxclumpy} with the best mid-IR model. In all our fittings, we assume that the viewing angle is the same for both mid-IR and X-ray models ($\rm{\theta_{inc} = i}$)\footnote{Note that the viewing angle is measured with respect to the polar plane. A source observed face-on has an inclination angle of $\rm{\theta_{inc} = 19^{\circ}}$, while a source observed edge-on has an inclination angle of $\rm{\theta_{inc} = 84^{\circ}}$.}. Hereafter, we refer to the combination of the \emph{borus02} and \emph{uxclumpy} models with either the \emph{smooth}, \emph{clumpy}, or \emph{two-phases} models from mid-IR as the X-S/MIR-S, X-S/MIR-C, X-S/MIR-TP, X-C/MIR-S, X-C/MIR-C, and X-C/MIR-TP baseline models, respectively. 

To choose the baseline model that best fits the \emph{Spitzer} and \emph{NuSTAR} spectra of each source, we run the following analysis:

\begin{enumerate}

\item We link the LOS column density ($\rm{N_{H_{los}}}$ parameter) of the {\sc zphabs} component to that of the reflection models ($\rm{N_{H_{tor}}}$ parameter). We consider all the fits with $\chi/d.o.f. < 1.2$ as a plausible solution for each source\footnote{We chose the $\chi^2/d.o.f. < 1.2$ criterion to mitigate the underestimated errors of the \emph{Spitzer}/IR spectra (See appendix B in \cite{Gonzalez-Martin23}).}. Subsequently, we use the minimum Akaike criteria to select the best baseline model, and all the fits with an Akaike probability below 0.1\% are considered equally probable fits \citep[as in][]{Emmanoulopoulos16}.

\item When the \emph{borus02} model is part of the baseline model selected above, we also test unlinking the $\rm{N_H}$ parameters of the LOS from that of the reflection model\footnote{Only \emph{borus02} allows to untie the LOS $\rm{N_H}$ from that of the reflection. \emph{uxclumpy} model was computed assuming the same $\rm{N_H}$ for both components.}. We use f-test statistics to choose the simplest baseline model that describes the data.

\item After selecting the best baseline model, we test the match of the half-opening angle ($\theta_{tor}$) from the X-ray model to the value of the torus angular width ($\sigma_{tor}$) of the mid-IR model. The link between these parameters depends on the definition of the half-opening and torus angular width angles ($\rm{\theta_{tor} = \sigma_{tor}}$). In Fig.\,\ref{fig:Comp_ChiReduced} (left), we compare the $\chi^{2}/d.o.f$ obtained before and after linking these parameters. In Sect.\,\ref{subsec:bestfit}, we will provide a detailed analysis and comparison of the effect of linking and unlinking these parameters.
\end{enumerate}

As an extra test, we also run some fits using a different baseline model for the mid-IR spectra, but they are always statistically worse than that obtained using the best fit at mid-IR. The results of these tests are compared using the Akaike criteria. Indeed, the model selection at mid-IR strongly relies on its spectral features \citep{Gonzalez-Martin19a}. This is not the case for the X-ray spectra, where we find that the best models obtained from the individual fit at X-rays might change from that obtained when the SFT is applied. This is because X-ray data alone cannot fully restrict the parameters of the models \citep{Saha22}. Therefore, it is important to systematically test both \emph{borus02} and \emph{uxclumpy} models for the simultaneous fit to ensure we use the best model combination for each object.

\section{Results}
\label{sec:Results}

In this section, we show the results obtained after applying the simultaneous fitting technique and apply the tests explained in the previous section. In subsections \ref{subsec:bestfit} and \ref{subsection:ParamConstrains}, we present the results obtained before and after linking the half-opening angle from the X-ray model to the same value of the torus angular width of the mid-IR model. We use statistical tests to compare the final statistics and the possible parameter constraints of both fits to each source in these two subsections. Meanwhile, in the Subsect. \ref{subsec:DerivedParam}, we show the values of several physical parameters such as mass, covering factor, and outer radii of the torus derived from the parameters of the baseline models. In subsection \ref{subsec:Results_torusProperties}, we show the results obtained from comparing the torus properties of Sy1 and Sy2.

\begin{figure*}
    \includegraphics[width=0.67\columnwidth, clip, trim={6 14 10 28}]{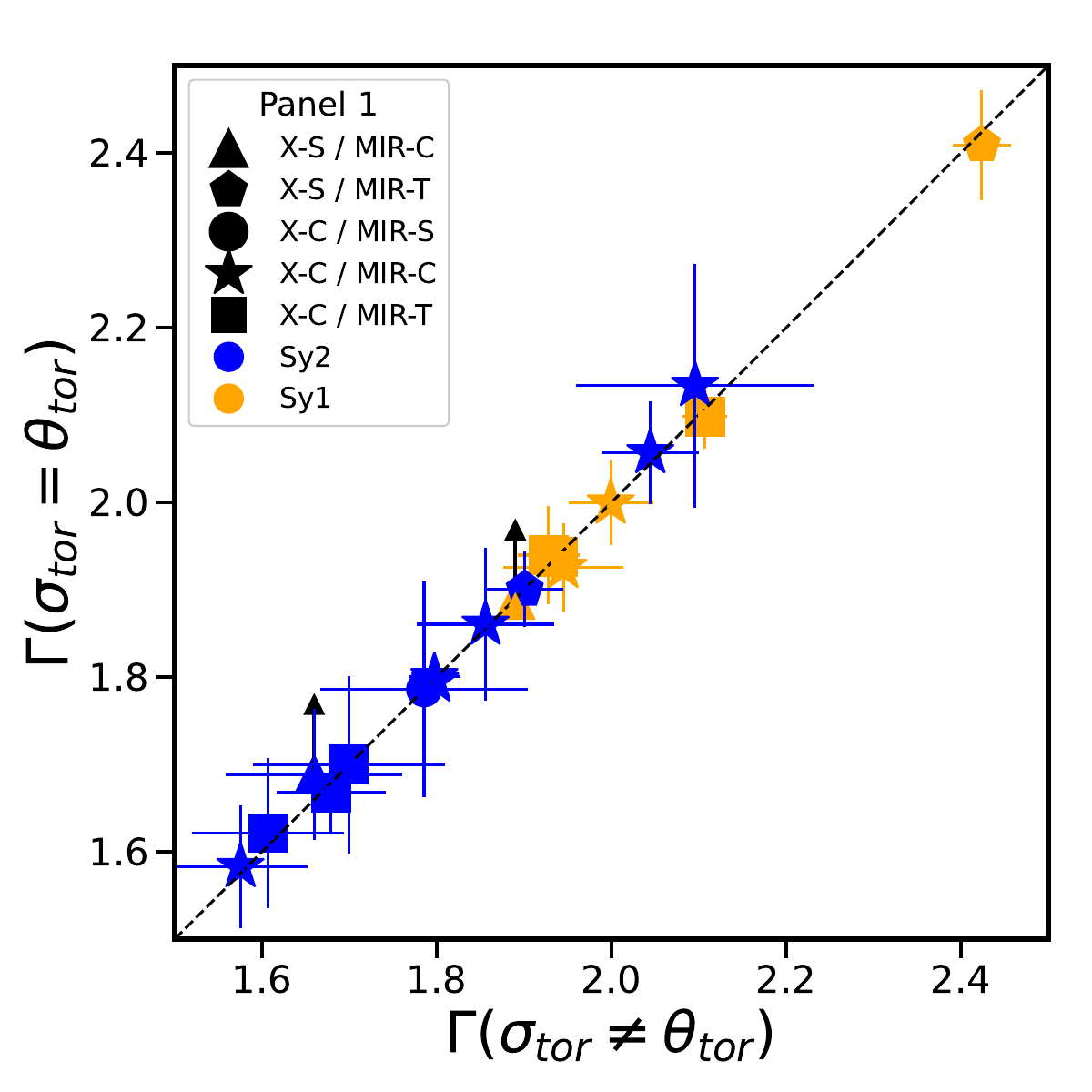}
    \includegraphics[width=0.67\columnwidth, clip, trim={6 14 10 28}]{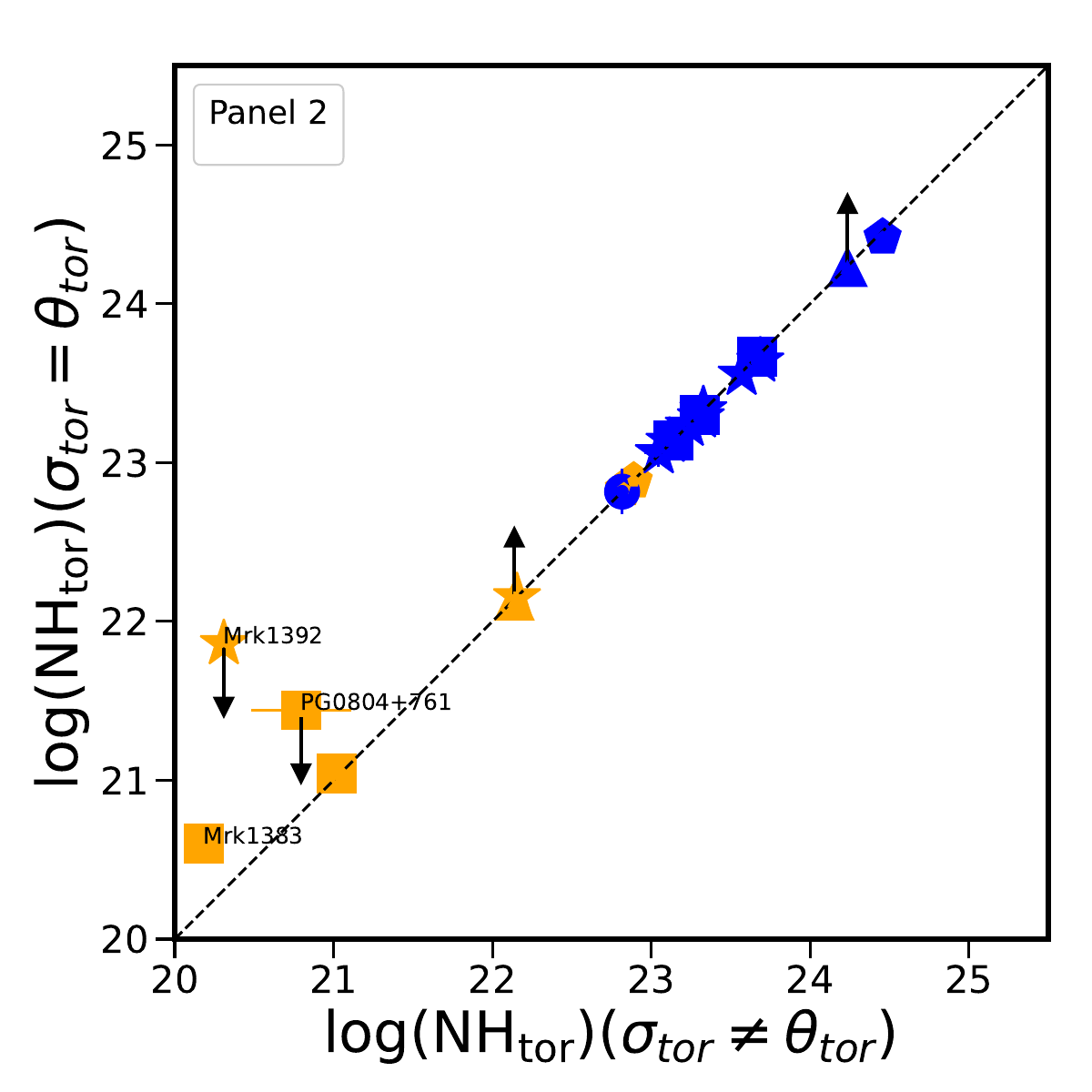}
    \includegraphics[width=0.67\columnwidth, clip, trim={6 14 10 28}]{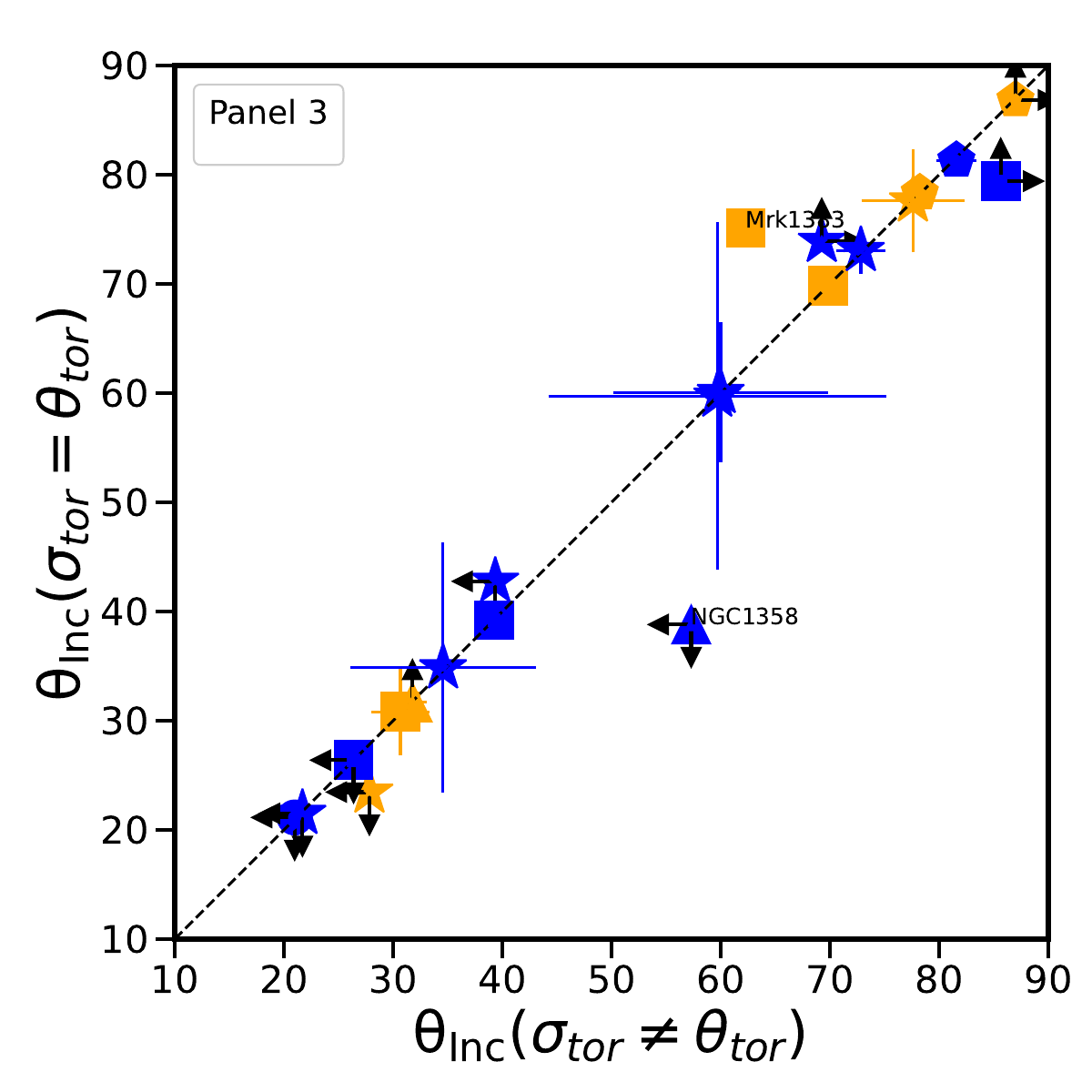}
    \includegraphics[width=0.67\columnwidth, clip, trim={6 14 10 28}]{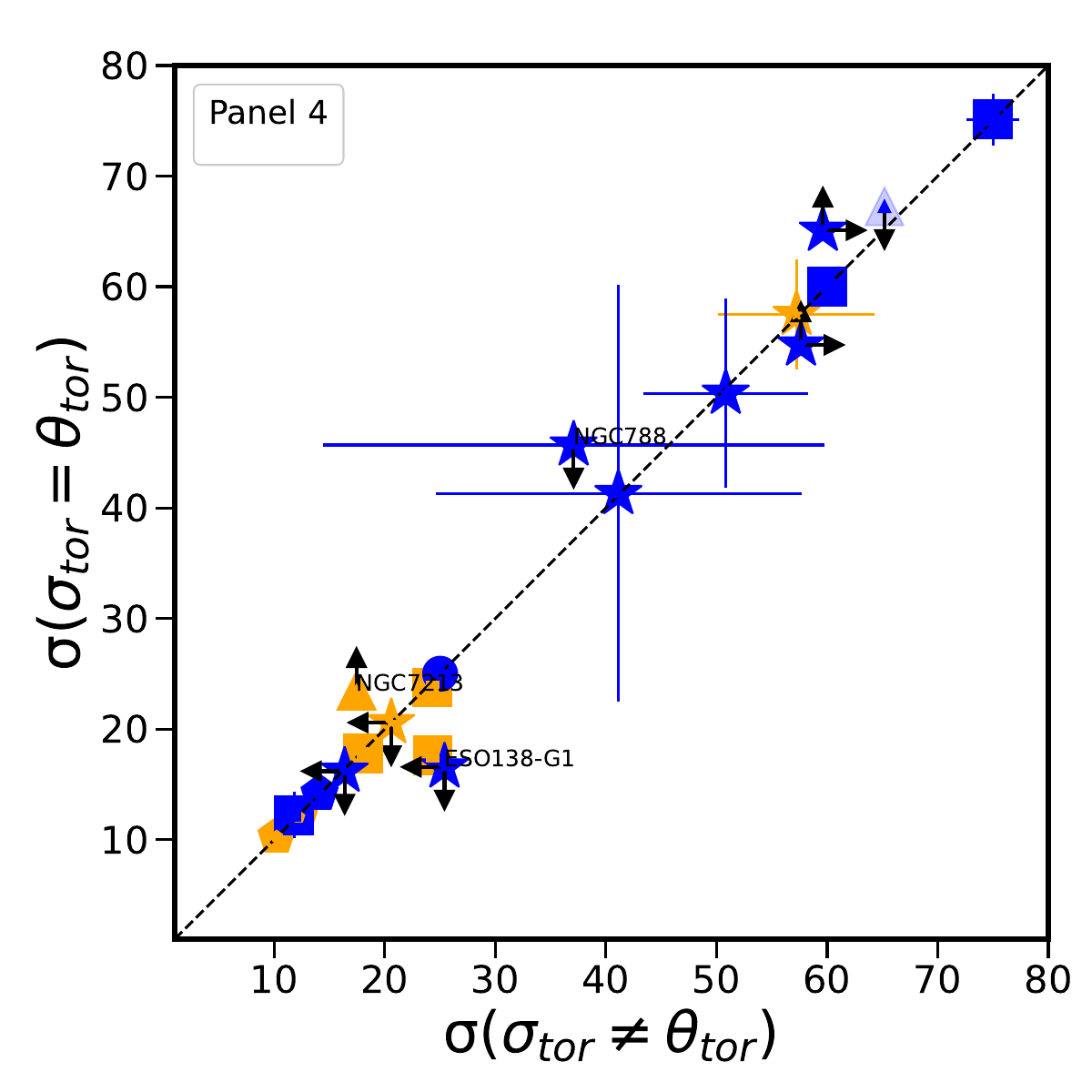}
    \includegraphics[width=0.67\columnwidth, clip, trim={6 14 10 28}]{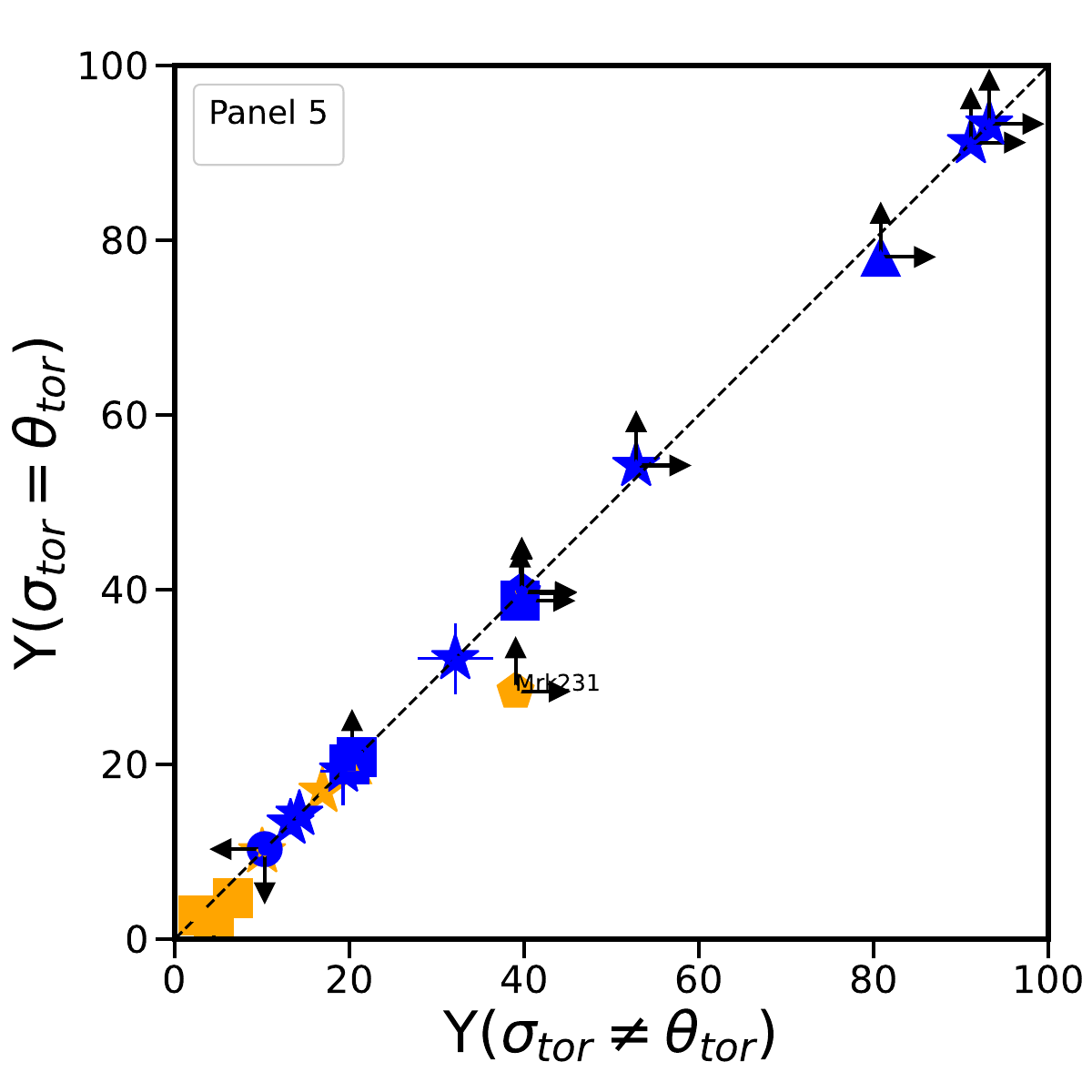}
    \includegraphics[width=0.67\columnwidth, clip, trim={6 14 10 28}]{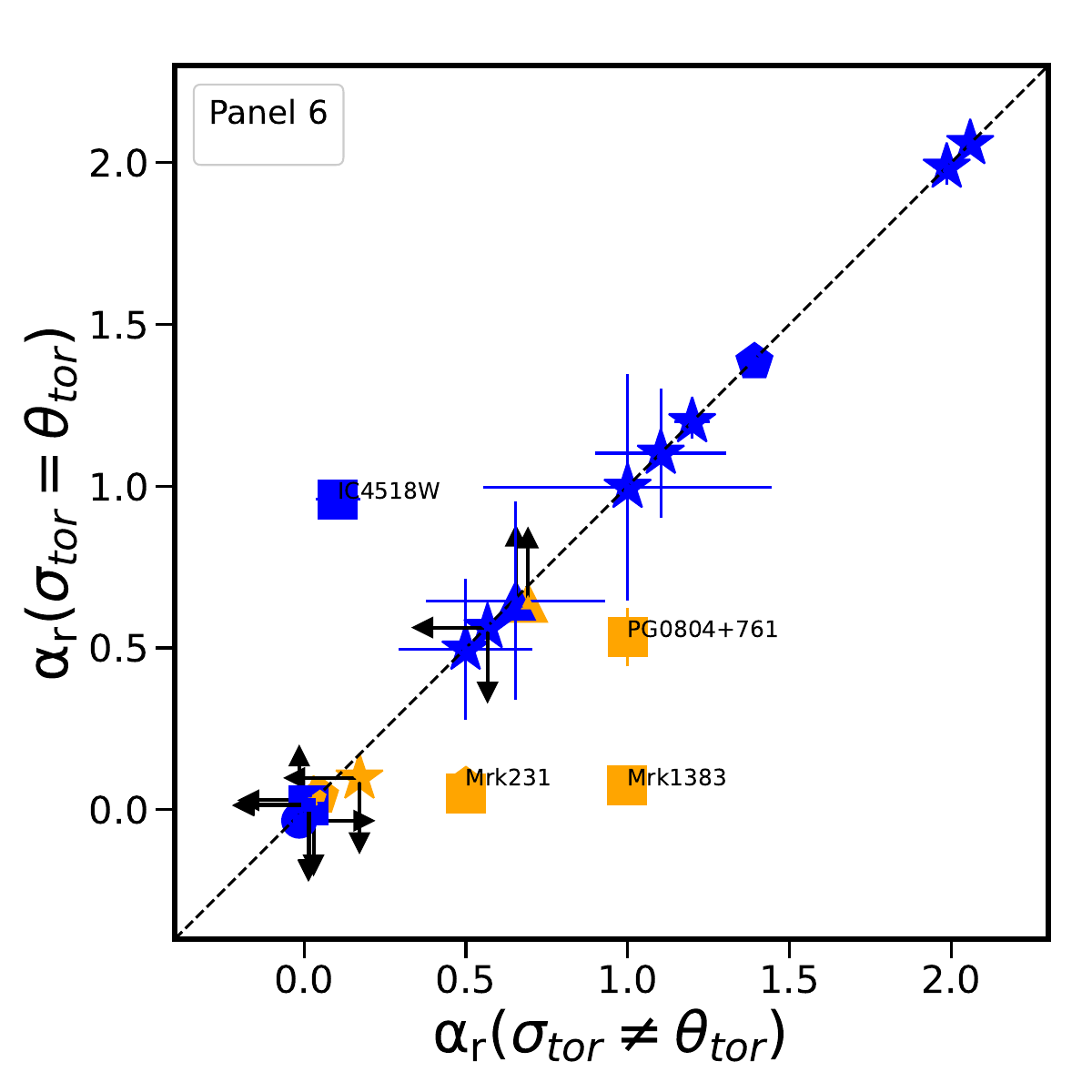}
    \includegraphics[width=0.67\columnwidth, clip, trim={6 14 19 28}]{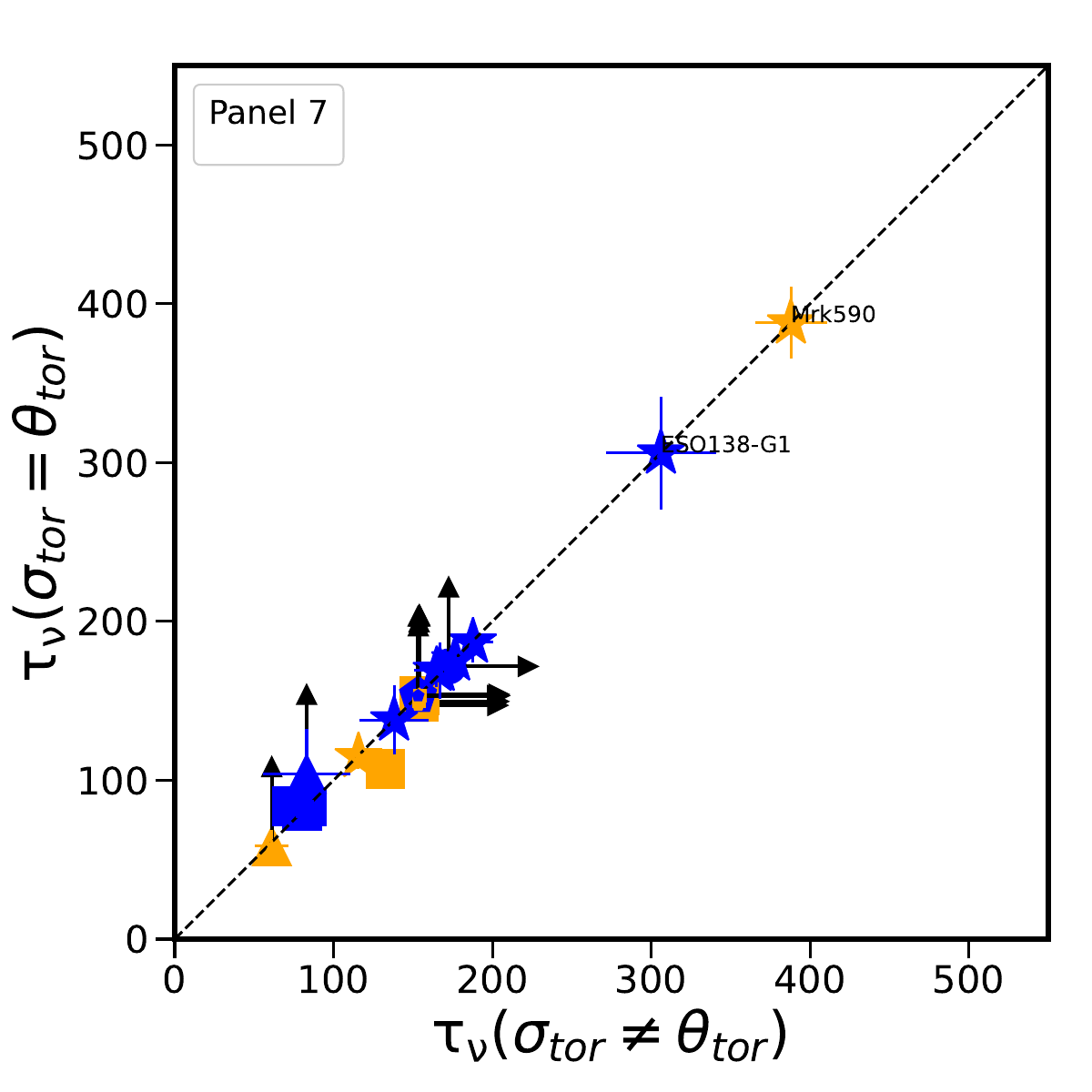}
    \includegraphics[width=0.67\columnwidth, clip, trim={6 14 10 28}]{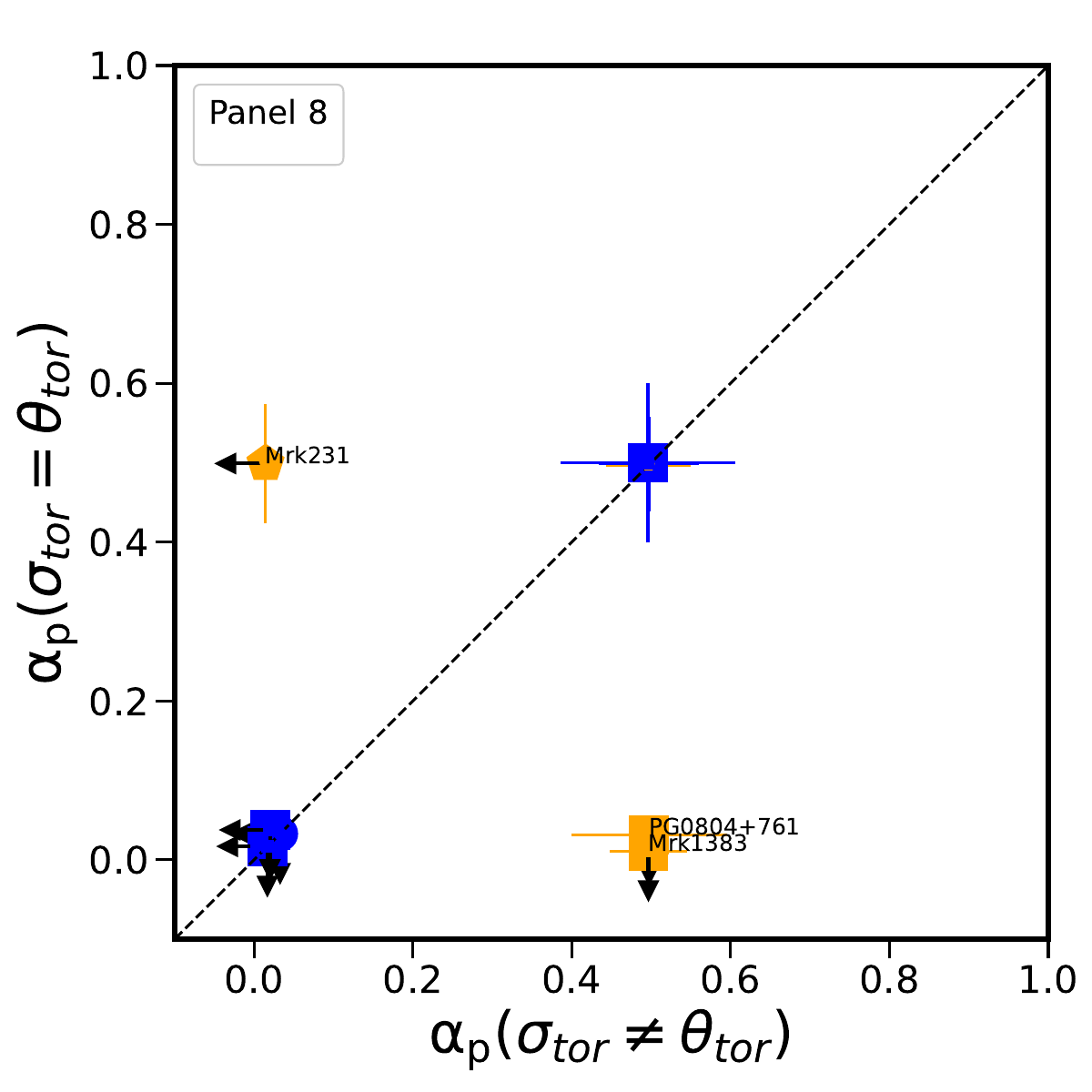}
    \includegraphics[width=0.67\columnwidth, clip, trim={6 14 10 28}]{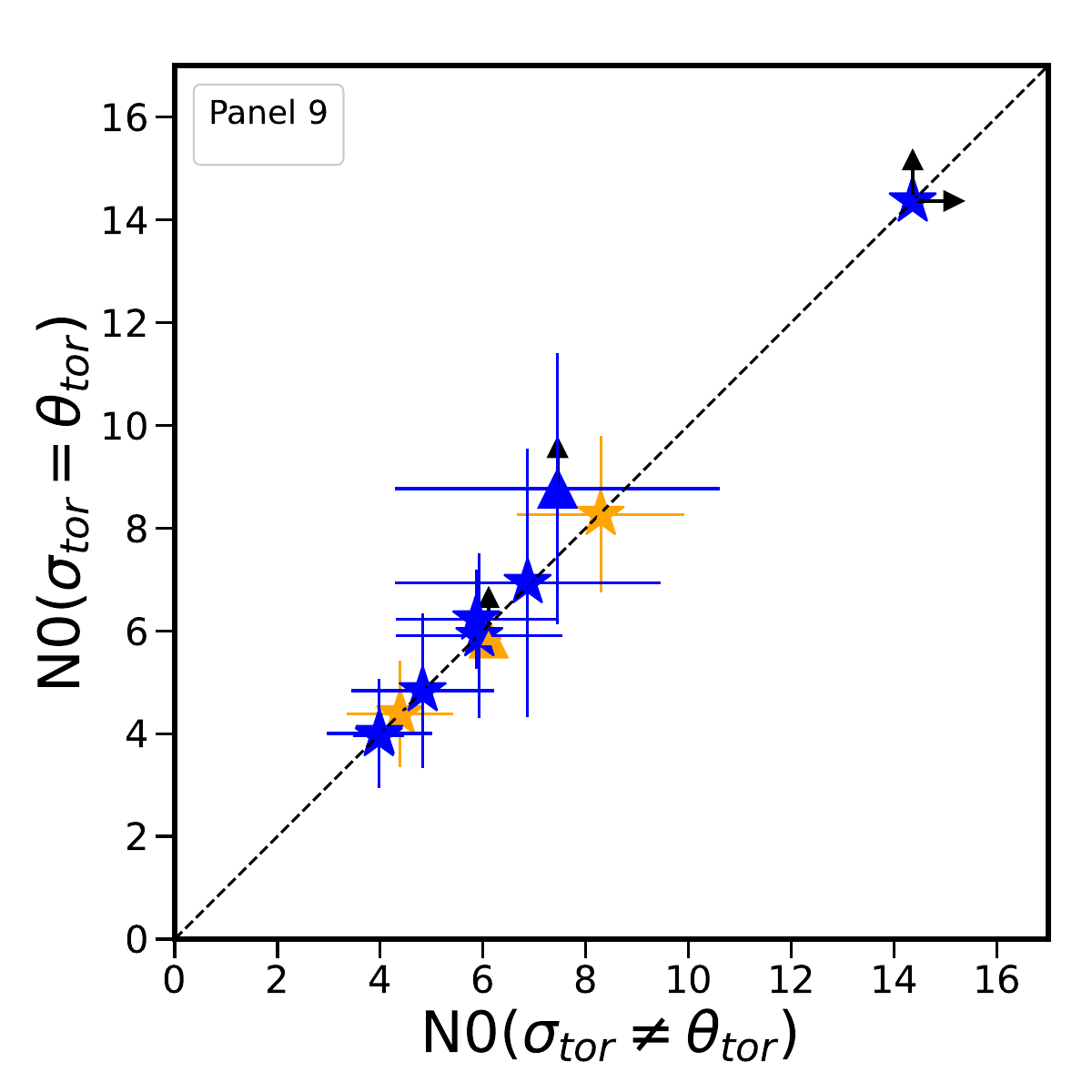}
    \caption{Comparison between parameters before and after linking the half-opening angles ($\rm{\theta_{tor}}$) and torus angular width ($\rm{\sigma_{tor}}$). Orange (blue) symbols are Sy1 (Sy2), and their shape denotes the best baseline model for each of them (see legend in panel 1). The lower and upper limit values are shown with black arrows.}
    \label{fig:Comparison}
\end{figure*}

\begin{table*}
   \caption{Simultaneous fit results ($\theta_{tor} \neq \sigma_{tor}$).}
   \label{Tab:ParametersValues}
   \centering
   \setlength{\tabcolsep}{1pt}
   \begin{footnotesize}
    \renewcommand{\arraystretch}{0.9}
    \setlength{\tabcolsep}{3pt}
    \begin{tabular}{llc|cccc|ccccccc}
        \hline \hline
        Objname & Baseline &  $\chi_{r}^2$ &  \multicolumn{4}{c}{X-ray parameters} & \multicolumn{7}{c}{IR parameters}\\
                      & Model &  &  $\Gamma$ & $log(N_{H_{tor}})$ & $\theta_{tor}$ & $\theta_{inc}$ & $\sigma_{tor}$ & Y & $\rm{\alpha_r}$ & $\rm{\alpha_p}$ & N0 &  $\tau_{\nu}$ &  log(Psize) \\
                      &  &  &  & $\rm{cm^{-2}}$ & $^{\circ}$ & $^{\circ}$ & $^{\circ}$ &  &  & & $\#$ &  &  $\mu m$ \\
                    (1) & (2) & (3) & (4) & (5) & (6) & (7) & (8) & (9) & (10) & (11) & (12) & (13) & (14)  \\
\hline \hline
    \multicolumn{14}{c}{Seyfert 1}  \\
\hline \hline
 Mrk590 & X-C/MIR-C & 0.81 &  2.0$_{1.94}^{2.04}$    &     22.16$_{22.13}^{22.24}$    &     $<$17.1   &     77.6$_{72.19}^{81.63}$    &     $<$20.58   &     10.0$_{9.9}^{10.4}$    &     $<$0.03   &     --    &     4.4$_{3.4}^{5.5}$    &     387.9$_{262.0}^{533.7}$    &     --  \\[1pt] 
PG0804+761 & X-C/MIR-TP & 0.88 &  1.93$_{1.89}^{1.96}$    &     20.8$_{20.7}^{21.3}$    &     $<$79.1   &     30.67$_{26.66}^{31.96}$    &     24.31$_{24.09}^{24.67}$    &     6.7$_{6.4}^{6.9}$    &     1.0$_{0.92}^{1.02}$    &     0.5$_{0.32}^{0.52}$    &     --    &     $>$153.9    \\[1pt] 
IRAS11119+3257 & X-S/MIR-TP & 0.92 &  2.42$_{2.4}^{2.46}$    &     22.89$_{22.83}^{22.94}$    &    $<$27.8  &     78.23$_{78.11}^{78.33}$    &     12.23$_{12.12}^{12.27}$    &     $>$39.7   &     0.05$_{0.04}^{0.07}$    &     0.5$_{0.4}^{0.51}$    &     --    &     $>$154.0   &     0.0$_{0.01}^{0.0}$ \\[1pt] 
Mrk231 & X-S/MIR-TP & 1.19 &  1.47$_{1.4}^{1.53}$    &     22.83$_{22.74}^{22.91}$    &     23.6*    &     $>$86.99   &     10.21$_{10.19}^{10.24}$    &     $>$39.0   &     0.5$_{0.44}^{0.52}$    &     $<$0.01   &     --    &     $>$153.3   &     0.19$_{0.16}^{0.21}$ \\[1pt] 
Mrk1383 & X-C/MIR-TP & 1.08 &  2.11$_{2.09}^{2.14}$    &     20.18$_{20.16}^{20.2}$    &     $<$13.7   &     62.29$_{62.05}^{62.54}$    &     24.34$_{24.12}^{24.72}$    &     4.5$_{4.3}^{4.6}$    &     1.0$_{0.98}^{1.01}$    &     0.5$_{0.41}^{0.51}$    &     --    &     $>$154.3   &     0.6$_{0.62}^{0.6}$ \\[1pt] 
Mrk1392 & X-C/MIR-C & 0.92 &  1.94$_{1.87}^{2.01}$    &     20.31$_{20.25}^{20.35}$    &     10.0$_{1.9}^{15.3}$    &     $<$27.83   &     57.23$_{49.32}^{63.45}$    &     16.9$_{16.4}^{18.4}$    &     $<$0.17   &     --    &     8.3$_{7.1}^{10.3}$    &     115.7$_{90.3}^{178.6}$    &     -- \\[1pt] 
ESO141-G055 & X-C/MIR-TP & 1.04 &  1.94$_{1.93}^{1.95}$    &     21.02$_{20.93}^{21.12}$    &     15.5$_{13.3}^{17.3}$    &     69.86$_{69.71}^{70.01}$    &     18.06$_{17.92}^{18.48}$    &     2.7$_{2.6}^{2.7}$    &     0.5$_{0.44}^{0.51}$    &     $>$1.46   &     --    &     132.6$_{130.0}^{134.9}$    &     1.3$_{1.31}^{1.29}$ \\[1pt] 
NGC7213 & X-S/MIR-C & 1.08 &  1.89$_{1.88}^{1.89}$    &     22.14$_{22.1}^{22.17}$    &     23.5$_{22.8}^{24.2}$   &     31.77$_{30.23}^{32.92}$    &     17.44$_{17.37}^{17.5}$    &     20.3$_{20.1}^{20.5}$    &     0.69$_{0.68}^{0.7}$    &     --    &     6.1$_{6.1}^{6.6}$    &     $<$66.1   &     -- \\[1pt] 
\hline \hline
     \multicolumn{14}{c}{Seyfert 2}  \\[1.2pt]
\hline \hline
UM146 & X-C/MIR-C & 0.66 &  2.10$_{1.97}^{2.24}$    &     23.05$_{22.96}^{23.13}$    &     27.8*    &     $<$39.35   &     $>$57.63   &     $>$91.1   &     1.2$_{1.14}^{1.25}$    &     --    &     4.0$_{3.3}^{5.4}$    &     138.1$_{93.3}^{258.8}$    &     -- \\[1pt] 
NGC788 & X-C/MIR-C & 0.88 &  1.58$_{1.49}^{1.64}$    &     23.69$_{23.62}^{23.75}$    &     $>$10.0   &     $>$69.25   &     37.08$_{19.05}^{64.44}$    &     13.3$_{11.7}^{14.4}$    &     1.0$_{0.31}^{1.2}$    &     --    &     5.9$_{4.8}^{8.0}$    &     164.9$_{112.6}^{269.1}$    &     -- \\[1pt] 
NGC1358 & X-S/MIR-C & 0.67 &  1.66$_{1.57}^{1.77}$    &     24.24$_{24.2}^{24.27}$    &     $<$57.8   &     $<$57.3   &     65.19*    &     $>$80.8   &     0.65$_{0.31}^{0.86}$    &     --    &     7.4$_{5.4}^{11.8}$    &     $<$249.6   &     -- \\[1pt] 
Mrk78 & X-C/MIR-C & 1.01 &  1.44$_{1.35}^{1.54}$    &     23.57$_{23.51}^{23.62}$    &     $>$9.0   &     59.68$_{39.09}^{70.01}$    &     41.12$_{29.86}^{62.95}$    &     32.1$_{26.9}^{35.6}$    &     1.1$_{0.81}^{1.22}$    &     --    &     5.9$_{4.2}^{7.5}$    &     187.8$_{118.1}^{280.6}$    &     -- \\[1pt] 
Mrk1210 & X-C/MIR-TP & 0.93 &  1.68$_{1.62}^{1.74}$    &     23.3$_{23.28}^{23.32}$    &     80.0*    &     39.24$_{38.44}^{41.26}$    &     60.0$_{59.31}^{60.56}$    &     20.0$_{19.8}^{20.2}$    &     $<$0.01   &     0.5$_{0.31}^{0.53}$    &     --    &     80.3$_{76.5}^{83.8}$    &     0.11$_{0.05}^{0.17}$ \\[1pt] 
J105943+65040 & X-C/MIR-S & 0.44 &  1.79$_{1.65}^{1.89}$    &     22.82$_{22.7}^{22.93}$    &     27.9*    &     $<$20.98   &     25.0$_{24.23}^{26.12}$    &     $<$10.3   &     $>$0.02   &     $<$0.03   &     --    &     $>$172.5   &     -- \\[1pt] 
NGC4939 & X-C/MIR-TP & 0.98 &  1.61$_{1.54}^{1.72}$    &     23.67$_{23.63}^{23.73}$    &     23.0*    &     $<$26.38   &     75.0$_{72.23}^{77.02}$    &     20.8$_{20.4}^{21.3}$    &     $<$0.01   &     $<$0.02   &     --    &     83.3$_{77.0}^{85.7}$    &     $>$1.0\\[1pt] 
IC4518W & X-C/MIR-TP & 0.92 &  1.7$_{1.59}^{1.81}$    &     23.14$_{23.07}^{23.2}$    &     10.0*    &     $>$85.64   &     11.81$_{11.31}^{13.01}$    &     $>$39.5   &     0.10$_{0.06}^{0.20}$    &     $<$0.02   &     --    &     74.0$_{69.0}^{79.9}$    &     0.12$_{0.14}^{0.09}$ \\[1pt] 
ESO138-G1 & X-C/MIR-C & 0.94 &  2.04$_{1.99}^{2.1}$    &     23.31$_{23.24}^{23.37}$    &     $<$9.7   &     60.0$_{44.34}^{64.06}$    &     $<$25.4   &     $>$52.8   &     1.99$_{1.94}^{2.02}$    &     --    &     4.8$_{3.6}^{6.3}$    &     306.1$_{169.6}^{503.3}$    &     -- \\[1pt] 
NGC6300 & X-S*/MIR-TP & 0.89 &  1.9$_{1.85}^{1.94}$    &     24.46$_{24.38}^{24.5}$    &     12.7$_{11.5}^{15.8}$    &     81.58$_{79.44}^{83.13}$    &     14.1$_{13.67}^{14.28}$    &     $>$39.7   &     1.39$_{1.35}^{1.43}$    &     0.5$_{0.39}^{0.52}$    &     --    &     $>$153.2   &     $>$1.0\\[1pt] 
ESO103-G35 & X-C/MIR-C & 1.04 &  1.8$_{1.79}^{1.81}$    &     23.24$_{23.23}^{23.25}$    &     $>$76.3   &     34.57$_{26.56}^{43.61}$    &     $>$59.61   &     14.3$_{13.8}^{14.8}$    &     0.5$_{0.17}^{0.58}$    &     --    &     4.0$_{3.8}^{4.8}$    &     176.5$_{156.9}^{228.6}$    &     -- \\[1pt] 
IC5063 & X-C/MIR-C & 0.86 &  1.8$_{1.77}^{1.83}$    &     23.33$_{23.31}^{23.35}$    &     $<$10.3   &     72.83$_{69.59}^{74.04}$    &     $<$16.36   &     $>$93.2   &     2.06$_{2.03}^{2.09}$    &     --    &     $>$14.4   &     $>$735.3   &     -- \\[1pt] 
PKS2356-61 & X-C/MIR-C & 0.67 &  1.86$_{1.78}^{1.94}$    &     23.12$_{23.05}^{23.18}$    &     $>$38.0   &     $<$21.7   &     50.83$_{46.03}^{60.85}$    &     19.3$_{18.3}^{23.4}$    &     $<$0.57   &     --    &     6.9$_{4.6}^{9.7}$    &     166.8$_{96.5}^{309.7}$    &     -- \\[1pt] 
    \hline \hline
    \end{tabular}
    \end{footnotesize}
    \tablefoot{Column(1): Source name. Column (2): Baseline model used (Combination X-ray model/IR model). The asterisk next to model indicates that $\rm{N_{H_{tor}} \leq  N_{H_{los}}}$. Column(3): Reduced $\chi^2$ ($\chi^2/d.o.f.$). Columns(4-7): Final reflection parameter values per model (see Table \ref{tab:XrayParameters}). Columns (8-14): Final dust parameter values per model (see Table \ref{tab:IRparameters}). The asterisk next to the value indicates that the parameter is not constrained. The confidence range of error calculated here is 1$\rm{\sigma}$. For each parameter, we report the value obtained from the best fit, accompanied by the minimum and maximum values calculated at +/- 1$\sigma$.}
\end{table*}

\begin{table*}
\caption{Simultaneous fit results ($\rm{\sigma_{tor} = \theta_{tor}}$).}
\label{Tab:ParametersValues_link}
\centering
\begin{scriptsize}
\renewcommand{\arraystretch}{0.9}
\setlength{\tabcolsep}{2pt}
\begin{tabular}{llc|ccc|ccccccc|cccc}
        \hline \hline
     Objname & Baseline &  $\chi_{r}^2$ &  \multicolumn{3}{c|}{X-ray parameters} & \multicolumn{7}{c|}{IR parameters} & \multicolumn{4}{c}{Derived quantities} \\
     & model &  &  $\Gamma$ & $log(N_{H_{tor}})$ & $\theta_{inc}$ & $\sigma_{tor}$ & Y & $\rm{\alpha_r}$ & $\rm{\alpha_p}$ & N0 &  $\tau_{\nu}$ &  $\rm{log(P_{size})}$ &$\rm{E_{B - v}}$ & $\rm{R_{out}}$ & $\rm{log(M_{dust})}$ & $CF$ \\
      &  &  &  & $\rm{cm^{-2}}$ & $^{\circ}$ & $^{\circ}$ &  &  & & $\#$ &  &  $\mu m$& & pc & $\rm{M_{\odot}}$ & \\
     (1) & (2) & (3) & (4) & (5) & (6) & (7) & (8) & (9) & (10) & (11) & (12) & (13) & (14) & (15) & (16) & (17)\\
\hline \hline
    \multicolumn{17}{c}{Seyfert 1}  \\
\hline \hline
Mrk590 & X-C/MIR-C & 0.85 &  2.00$_{1.96}^{2.06}$    &     22.25$_{22.19}^{22.3}$    &     78.16$_{71.63}^{79.59}$    &     $<$19.41   &     13.6$_{12.64}^{14.66}$    &     $<$0.03   &     --    &     $>$13.1   &     $>$146.22   &     -- & 2.14$_{1.53}^{2.47}$ & $\rm{6.24_{5.33}^{7.16}}$ & $\rm{6.43_{6.32}^{6.53}}$ & $\rm{0.38_{0.30}^{0.46}}$\\[1pt] 
PG0804+761 & X-C/MIR-TP & 0.88 &  1.94$_{1.88}^{1.99}$    &     $<$21.44   &     30.82$_{25.07}^{33.07}$    &     23.76$_{23.44}^{24.17}$    &     4.67$_{4.46}^{4.90}$    &     0.53$_{0.44}^{0.62}$    &     $<$0.03   &     --    &     $>$149.65   &     0.29$_{0.31}^{0.24}$ &  < 0.22  & $\rm{3.52_{2.52}^{4.52}}$ & $\rm{3.91*}$ & $\rm{0.40_{0.39}^{0.41}}$\\[1pt] 
I11119+3257 & X-S/MIR-TP & 0.91 &  2.41$_{2.36}^{2.49}$    &     22.88$_{22.82}^{22.94}$    &     78.34$_{78.3}^{78.51}$    &     12.05$_{11.92}^{12.10}$    &     $>$39.63   &     0.04$_{0.03}^{0.06}$    &     0.5$_{0.38}^{0.51}$    &     --    &     $>$154.12   &     0.0$_{0.01}^{0.01}$ &  $6.89_{6.75}^{7.02}$  & $\rm{1.44_{1.37}^{1.50}}$ & $\rm{4.05_{3.93}^{4.16}}$ & $\rm{0.21_{0.20}^{0.22}}$\\[1pt] 
Mrk231 & X-S/MIR-TP & 1.17 &  1.47$_{1.43}^{1.52}$    &     22.84$_{22.79}^{22.88}$    &     $>$86.85   &     10.29$_{10.21}^{10.56}$    &     $>$28.35   &     0.07$_{0.04}^{0.12}$    &     0.5$_{0.39}^{0.54}$    &     --    &     $>$147.18   &     0.0$_{0.0}^{0.03}$ & 5.83$_{5.61}^{6.16}$ & $\rm{2.12_{1.47}^{2.78}}$ & $\rm{4.62_{4.40}^{4.84}}$ & $\rm{0.18_{0.17}^{0.19}}$\\[1pt] 
Mrk1383 & X-C/MIR-TP & 0.99 &  2.10$_{2.05}^{2.12}$    &     20.6$_{20.58}^{20.66}$    &     75.13$_{74.83}^{75.30}$    &     17.6$_{17.23}^{17.86}$    &     $<$2.03   &     0.07$_{0.05}^{0.14}$    &     $<$0.01   &     --    &     $>$153.31   &     0.99$_{1.01}^{0.98}$ & <0.33 & $\rm{1.49_{0.38}^{2.60}}$ & $\rm{3.94_{3.92}^{3.96}}$ & $\rm{0.30_{0.29}^{0.31}}$\\[1pt] 
Mrk1392 & X-C/MIR-C & 0.92 &  1.93$_{1.89}^{1.99}$    &     $<$21.86   &     $<$23.46   &     57.5$_{51.07}^{60.97}$    &     16.91$_{16.50}^{18.07}$    &     $<$0.1   &     --    &     8.27$_{7.08}^{10.13}$    &     114.92$_{92.38}^{182.98}$    &     -- & 2.03$_{1.71}^{2.38}$ & $\rm{2.19_{1.59}^{2.80}}$ & $\rm{5.95_{5.84}^{6.07}}$ & $\rm{0.95_{0.92}^{0.97}}$\\[1pt] 
ESO141-G055 & X-C/MIR-TP & 1.03 &  1.94$_{1.93}^{1.95}$    &     21.04$_{20.9}^{21.13}$    &     69.86$_{69.57}^{70.20}$    &     17.79$_{17.57}^{18.13}$    &     2.72$_{2.68}^{2.76}$    &     0.05$_{0.02}^{0.1}$    &     0.5$_{0.42}^{0.87}$    &     --    &     107.07$_{105.06}^{109.07}$    &     1.6$_{1.62}^{1.58}$ & $<0.21 $  & $\rm{0.86_{0.40}^{1.33}}$ & $\rm{5.99_{5.89}^{6.09}}$ & $\rm{0.30_{0.31}^{0.32}}$ \\[1pt] 
NGC7213 & X-S/MIR-C & 1.08 &  1.89$_{1.88}^{1.89}$    &     22.13$_{22.05}^{22.16}$    &     31.67$_{29.84}^{32.61}$    &     23.46$_{23.33}^{23.60}$    &     20.02$_{18.33}^{21.70}$    &     0.64$_{0.63}^{0.65}$    &     --    &     5.85$_{5.62}^{5.89}$    &     $<$59.32   & -- & 4.81$_{4.74}^{4.87}$ & $\rm{0.49_{0.32}^{0.67}}$ & $\rm{3.07_{2.98}^{3.16}}$ & $\rm{0.54_{0.52}^{0.56}}$\\[1pt]
\hline \hline
    \multicolumn{16}{c}{Seyfert 2}  \\
\hline \hline
UM146 & X-C/MIR-C & 0.66 &  2.13$_{1.98}^{2.26}$    &     23.06$_{22.97}^{23.15}$    &     $<$42.77   &     $>$54.73   &     $>$91.18   &     1.2$_{1.15}^{1.25}$    &     --    &     4.0$_{3.33}^{5.46}$    &     138.0$_{94.97}^{263.22}$    &     -- & 3.92$_{3.35}^{4.47}$ & $\rm{2.67_{2.25}^{3.10}}$ & $\rm{4.66_{4.56}^{4.76}}$ & $\rm{>0.84}$\\[1pt] 
NGC788 & X-C/MIR-C & 0.88 &  1.58$_{1.50}^{1.64}$    &     23.64$_{23.59}^{23.69}$    &     $>$73.9   &     $<$45.69   &     13.28$_{11.82}^{14.49}$    &     1.0$_{0.52}^{1.22}$    &     --    &     6.23$_{5.46}^{7.39}$    &     169.2$_{127.43}^{237.26}$    &  -- & 1.69$_{1.41}^{1.95}$ & $\rm{0.74_{0.30}^{1.19}}$ & $\rm{3.94_{3.75}^{4.12}}$ & $\rm{>0.84}$ \\[1pt] 
NGC1358 & X-S/MIR-C & 0.66 &  1.69$_{1.63}^{1.78}$    &     24.23$_{24.21}^{24.27}$    &     $<$38.81   &     $<$67.26   &     $>$78.09   &     0.65$_{0.27}^{0.88}$    &     --    &     8.77$_{6.16}^{11.45}$    &     $<$260.87   & -- & $ > 7.95$  & $\rm{7.67_{5.11}^{10.23}}$ & $\rm{6.48_{6.25}^{6.71}}$ & $\rm{>0.84}$ \\[1pt]
Mrk78 & X-C/MIR-C & 1.01 &  1.40$_{1.32}^{1.52}$    &     23.56$_{23.51}^{23.62}$    &     59.73$_{40.97}^{72.83}$    &     41.29$_{26.98}^{64.67}$    &     32.12$_{27.45}^{35.55}$    &     1.1$_{0.82}^{1.22}$    &     --    &     5.9$_{4.25}^{7.47}$    &     187.01$_{118.36}^{280.95}$    &     -- & 4.92$_{4.59}^{5.26}$  & $\rm{3.78_{2.35}^{5.20}}$ & $\rm{5.22_{5.04}^{5.40}}$ & $\rm{>0.86}$ \\[1pt] 
Mrk1210 & X-C/MIR-TP & 0.92 &  1.67$_{1.63}^{1.72}$    &     23.3$_{23.28}^{23.32}$    &     39.21$_{38.52}^{40.59}$    &     60.02$_{59.38}^{60.59}$    &     20.0$_{19.76}^{20.21}$  &   $<$0.01   &  0.5$_{0.33}^{0.53}$    &  --  &     80.5$_{76.54}^{83.83}$  &  0.11$_{0.05}^{0.17}$ & $< 0.07$
 & $\rm{1.84_{1.55}^{2.12}}$ & $\rm{4.86_{4.72}^{4.99}}$ & $\rm{0.87_{0.86}^{0.88}}$\\[1pt] 
J105943+65040 & X-C/MIR-S & 0.44 &  1.79$_{1.65}^{1.90}$    &     22.82$_{22.65}^{22.94}$    &     $<$21.14   &     25.0$_{23.53}^{26.1}$    &     $<$10.3   &     $>$0.03   &     $<$0.03   &     --    &     $>$171.67   & -- & $3.80_{3.39}^{4.38}$ & $\rm{2.79_{2.39}^{3.19}}$ & $\rm{>7.54}$ & $\rm{0.42_{0.40}^{0.44}}$ \\[1pt]  
NGC4939 & X-C/MIR-TP & 0.98 &  1.62$_{1.56}^{1.73}$    &     23.67$_{23.62}^{23.71}$    &     $<$26.39   &     75.12$_{72.21}^{76.94}$    &     20.85$_{20.39}^{21.32}$    &     $<$0.01   &     $<$0.02   &     --    &     83.3$_{76.93}^{85.81}$    &     $>$1.0& $<0.07$ & $\rm{0.78_{0.64}^{0.92}}$ & $\rm{1.69*}$ & $\rm{0.97_{0.95}^{0.98}}$\\[1pt] 
IC4518W & X-C/MIR-TP & 0.8 &  1.70$_{1.59}^{1.79}$    &     23.14$_{23.07}^{23.2}$    &     $>$79.41   &     12.22$_{11.63}^{15.82}$    &     $>$38.72   &     0.96$_{0.9}^{1.01}$    &     $<$0.04   &     --    &     83.44$_{81.76}^{89.03}$    &     $>$1.0 & $12.65_{12.25}^{13.08}$ & $\rm{3.0_{2.8}^{3.2}}$ & $\rm{2.02*}$ & $\rm{0.21_{0.20}^{0.27}}$\\[1pt] 
ESO138-G1 & X-C/MIR-C & 0.95 &  2.06$_{2.00}^{2.11}$    &     23.29$_{23.23}^{23.34}$    &     60.06$_{54.92}^{67.73}$    &     $<$16.56   &     $>$54.22   &     1.99$_{1.92}^{2.03}$    &     --    &     4.83$_{3.63}^{6.64}$    &  306.01$_{170.92}^{529.97}$    &  -- & 4.32$_{3.94}^{4.50}$ & $\rm{1.3_{0.6}^{2.0}}$ & $\rm{2.71_{2.60}^{2.82}}$ & $\rm{0.35_{0.31}^{0.40}}$ \\[1pt] 
NGC6300 & X-S/MIR-TP & 0.89 &  1.90$_{1.86}^{1.94}$    &     24.42$_{24.37}^{24.47}$    &     81.31$_{80.79}^{82.33}$    &     14.17$_{13.55}^{14.34}$    &     $>$39.73   &     1.38$_{1.35}^{1.42}$    &     0.5$_{0.4}^{0.52}$    &     --    &     $>$153.23   &     $>$1.0& $13.44_{13.12}^{13.76} $ & $\rm{1.03_{0.99}^{1.08}}$ & $\rm{3.45_{3.34}^{3.55}}$ & $\rm{0.24_{0.23}^{0.25}}$\\[1pt] 
ESO103-G35 & X-C/MIR-C & 1.04 &  1.80$_{1.78}^{1.81}$    &     23.24$_{23.23}^{23.24}$    &     34.84$_{26.58}^{49.49}$    &     $>$65.1   &     14.28$_{13.8}^{14.71}$    &     0.5$_{0.14}^{0.58}$    &     --    &     3.96$_{3.7}^{4.28}$    &     175.82$_{153.66}^{204.31}$    &     -- & $6.72_{6.52}^{6.90}$ & $\rm{1.90_{1.39}^{2.41}}$ & $\rm{5.21_{5.11}^{5.31}}$ & $\rm{0.90_{0.88}^{0.91}}$\\[1pt] 
IC5063 & X-C/MIR-C & 0.87 &  1.80$_{1.78}^{1.83}$    &     23.33$_{23.31}^{23.35}$    &     73.06$_{69.82}^{74.06}$    &     $<$16.19   &     $>$93.31   &     2.06$_{2.03}^{2.09}$    &     --    &     $>$14.36   &     $>$735.26   &     -- & $3.12_{2.94}^{3.28}$ & $\rm{5.85_{4.88}^{6.83}}$ & $\rm{3.94_{3.88}^{4.00}}$ & $\rm{0.43_{0.41}^{0.46}}$\\[1pt] 
PKS2356-61 & X-C/MIR-C & 0.67 &  1.86$_{1.78}^{1.95}$    &     23.13$_{23.07}^{23.20}$    &     $<$21.51   &     50.37$_{41.06}^{58.14}$    &     19.24$_{15.47}^{23.33}$    &     $<$0.56   &     --    &     6.94$_{4.65}^{9.88}$    &     169.26$_{98.16}^{332.3}$    &     -- & $6.09_{5.65}^{6.52}$ & $\rm{8.45_{4.95}^{11.95}}$ & $\rm{6.71_{6.55}^{6.86}}$ & $\rm{0.95_{0.92}^{0.97}}$\\[1pt]  
    \hline \hline
    \end{tabular}
    \end{scriptsize}
    \tablefoot{Column(1): Source name. Column (2): Baseline model used. Column(3): Reduced $\chi^2$ ($\chi^2/d.o.f.$). Columns(4-6): Final reflection parameter values per model. Columns (7-13): Final dust parameter values per model. Column (14): Color excess for the foreground extinction. The asterisk next to the value indicates that the parameter is not constrained. For each parameter, we report the value obtained from the best fit, accompanied by the minimum and maximum values calculated at +/- 1$\sigma$. The outer radius is calculated as $\rm{R_{out} = Y*R_{in} = Y*1.3*(L_{bol}/10^{46})^{1/2}}$.}
\end{table*} 

\subsection{The best baseline model}
\label{subsec:bestfit}
Figure\,\ref{fig:Comp_ChiReduced} (left panel) shows the $\chi^{2}/d.o.f.$ values obtained before and after linking the half-opening angle of the X-ray model and the torus angular width of the IR models to the same value for each AGN. Most of the sources have good statistics when their \emph{NuSTAR} and \emph{Spitzer} spectra are fitted with the best baseline model chosen ($\chi^{2}/d.o.f.$ < 1.2). Only three sources (NGC\,4507, NGC\,1052, and J05081967+1721483) show relatively poor spectral-fitting statistics (1.2 < $\chi^{2}/d.o.f.$ < 1.4) due to their mid-IR spectra, which are not well fit with the baseline model tested. We discard these three sources for the rest of the analysis and will study them in detail in future work. The spectral fits of these sources are shown in Figs. \ref{fig:Fits_2}-\ref{fig:Fits_4} in appendix \ref{app:Spectra_fits}).

In Figure\,\ref{fig:Comp_ChiReduced} (left panel), we also identify with different colors and symbols the five resulting combinations of X-ray and mid-IR models that best fit the spectra of sources: X-S/MIR-C (green triangles), X-S/MIR-TP (pink pentagons), X-C/MIR-S (cyan circle), X-C/MIR-C (orange stars), and X-C/MIR-TP (purple squares). Inside the shaded area, the X-C/MIR-C and X-C/MIR-TP baseline models are preferred for most sources: nine (two Sy1 and seven Sy2) and six (three Sy1 and three Sy2), respectively. The X-C/MIR-S baseline model is only preferred for one source: J10594361+6504063. NGC\,7213 and NGC\,1358 prefer the X-S/MIR-C baseline model. Meanwhile, the X-S/MIR-TP baseline model is preferred by three sources: IRAS\,11119+3257, Mrk\,231, and NGC6300. In column\,2 of Table\,\ref{Tab:ParametersValues}, we reported the baseline model chosen for each source.

For the seven sources that prefer the \emph{borus02} model at X-ray, we can also explore the possibility that the column density parameters from the LOS and the reflection component differ. To test this hypothesis, we fit these sources with a baseline model where the column densities of the line-of-sight absorption and X-ray model are different and independent \citep[see][]{Balokovic18}. We then use an f-test to compare the chi-reduced values of these fits with the obtained values when the column densities of both components are the same. We find that only NGC\,6300 prefers a baseline model where the column density of these two components is unlinked (that is $\rm{N_{H_{tor}} \neq N_{H_{los}}}$).

Finally, we use the f-test to explore whether the half-opening angle from X-ray and torus angular width from mid-IR models could be linked at the same value as a possible indicator of the same structure producing both components. Indeed, among the 21 AGNs (eight Sy1 and thirteen Sy2) with good statistics ($\rm{\chi^2/d.o.f. < 1.2}$), only Mrk\,590 preferred a baseline model without linking the half-opening angle and torus angular width (this object is slightly over the 1:1 diagonal line in the left panel of Fig.\,\ref{fig:Comp_ChiReduced}). As examples, the final fits of five sources, one for each resulting baseline model, are shown in Fig.\,\ref{fig:Best_SimultaneousFits}. The residuals (ratio between data and model) are in the range of [-3 and 3], with the largest departures at mid-IR wavelengths at the 9.7$\rm{\mu m}$ silicate features.

To summarize, 95\% of the sources in our sample favor a simple baseline model where the half-opening angle and torus angular width parameters are linked to the same value. This result suggests that for most of our sources, the mid-IR continuum emission and the reflection component of X-ray stem from the same structure \citep{Esparza-Arredondo19}.

\subsection{Constraints on the model parameters}
\label{subsection:ParamConstrains}
The SFT is an excellent tool to constrain most of the physical parameters of the obscuring material because it combines information from both X-ray and mid-IR wavelength ranges. To further explore this improvement, in each panel of Fig.\,\ref{fig:Comparison}, we compare the values of parameters before and after linking the half-opening angle and torus angle width. We use different symbols and colors to identify the baseline models chosen and the type of source, respectively. Additionally, in Tables\,\ref{Tab:ParametersValues} and \ref{Tab:ParametersValues_link}, we report the final best model, $\chi^{2}/d.o.f.$ obtained, and the parameters before and after linking the half opening angles, respectively. Quoted errors are estimated as one standard deviation.

Concerning the X-ray parameters, we find that the $\Gamma$ and $\rm{N_{H_{tor}}}$ parameters are constrained in most of the sources independently of whether the half-opening angle and torus angular width are linked or not, except by the column density of Mrk\,1392 which is a lower limit when $\rm{\theta_{tor} = \sigma_{tor}}$. The viewing angle $\theta_{inc}$ parameter is constrained in six out of eight Sy1 and six out of the 13 Sy2 before and after linking the half-opening angle and torus angular width. 

Among the mid-IR parameters, we find that, before linking the half-opening angle and torus angular width, the $\rm{\sigma_{tor}}$ parameter is constrained in 14 among the 20 sources with good statistics, and the parameter is unrestricted for only NGC\,1358 (marked as asterisks in Col.\,8 of Table\,\ref{Tab:ParametersValues}). Interestingly, the unconstrained value for NGC\,1358 results in an upper limit when we link the half-opening angle and torus angular width.

Before linking the half-opening angle and torus angular width, the parameter more difficult to constrain is $\theta_{tor}$ for most sources; it is only constrained to three Sy1 and one Sy2 (see Column 6 in Table \ref{Tab:ParametersValues}). Indeed, in six sources, the half-opening angle could be any value (that is, not even upper limits are obtained). This is fixed by linking it to the mid-IR torus angular width ($\rm{\sigma_{tor}}$). As mentioned at the end of our previous subsection, this link is statistically preferred by most sources in our sample. However, it is worth exploring if the half-opening angle and torus angular width obtained prior to this link are consistent with it. In Fig.\,\ref{fig:Comp_ChiReduced} (right), we compare the $\rm{\theta_{tor}}$ and $\rm{\sigma_{tor}}$ values obtained when we only link the viewing angles (that is both parameters are allowed to vary independently). The seven sources (NGC\,4939, UM\,146, Mrk\,1210, NGC1358, J10594361+6504063, Mrk\,231, and IC4518W) where one of the half-opening parameters is unrestricted are shown with grey marks. We find 12 where the $\rm{\sigma_{tor}}$ and $\rm{\theta_{tor}}$ values are similar. Indeed, the values of both parameters are very different only for three sources (Mrk\,1392, UM\,146, and NGC\,4939). 

The size of the dusty torus is parameterized through the $Y$ parameter, which is obtained from the mid-IR baseline models (X-ray models do not depend on the outer radius of the torus). This parameter is constrained in 11 out of the 20 AGNs, independent of whether the half-opening angle and torus angular width are linked or not. Finally, the slope of the radial density profile ($\rm{\alpha_r}$) and optical depth ($\rm{\tau_{\nu}}$) are constrained in 15 and 11 sources, respectively. Regardless of whether the half-opening angle and torus angular width are linked or not (see panels 6-7 in Fig.\,\ref{fig:Comparison}).

\begin{figure*}[ht]
\centering
\includegraphics[width=1.8\columnwidth]{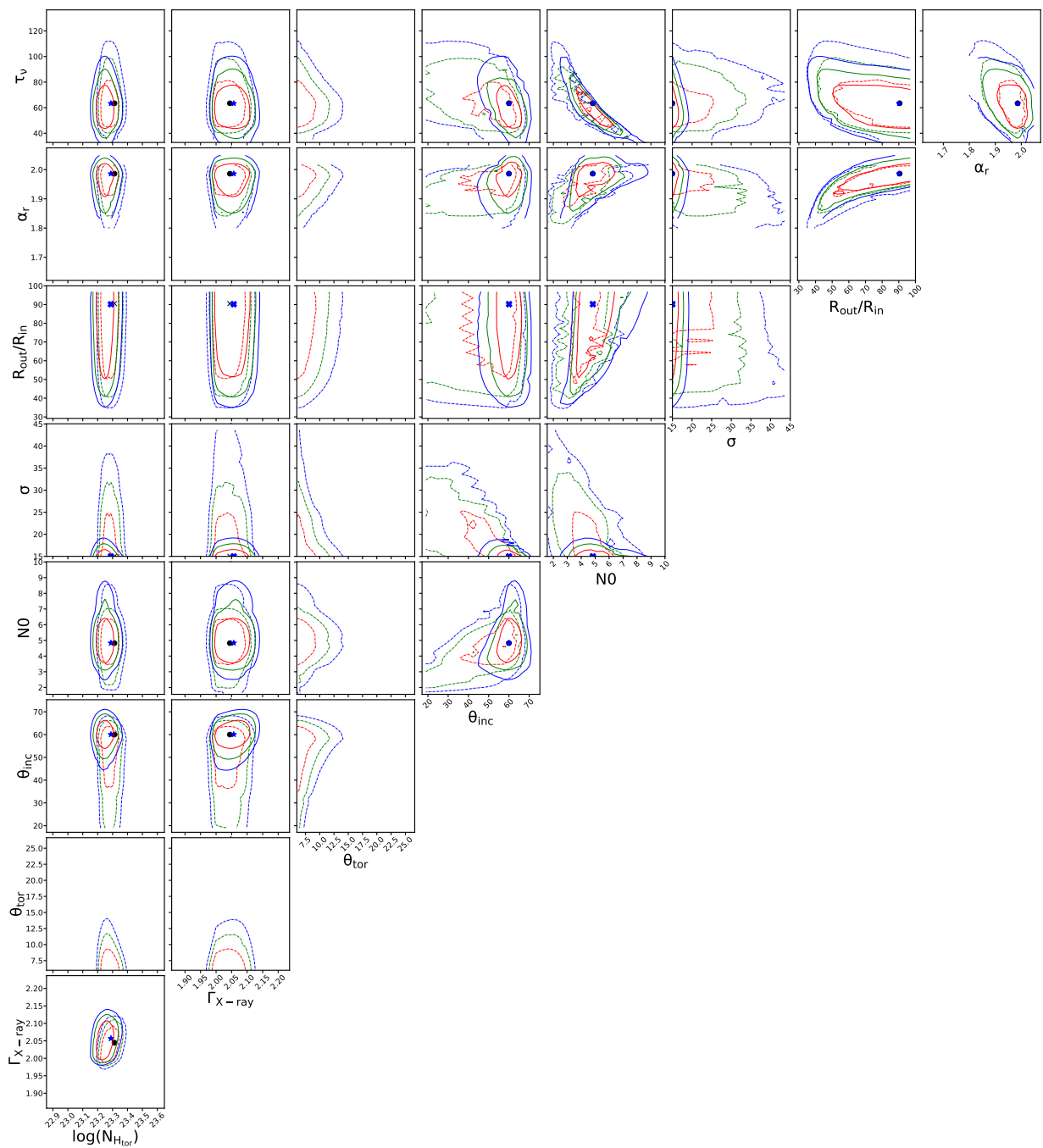}
\caption{Two-dimensional $\rm{\Delta \chi^2}$ contours for the resulting free parameters when we used the X-C/MIR-C baseline model before (dotted lines) and after (solid lines) linking the half-opening angle and torus angular width to fit ESO\,138-G01 spectra. The red, green, and blue (dotted and solid) contours mark the $\rm{1\sigma}$, $\rm{2\sigma}$, and $\rm{3\sigma}$ errors, respectively. The black circles and blue stars are the resulting values for each parameter before and after linking the half-opening angle and torus angular width, respectively.}
\label{fig:Contours_ESO138-G01}
\end{figure*}

As reported in our previous work, this technique of linking the half-opening angle and torus angular width is an excellent help to constrain the parameters further and break possible degeneration \citep{Esparza-Arredondo19}. The analysis presented in this work confirms this result. To illustrate the improvement on the parameter restriction, Fig.\,\ref{fig:Contours_ESO138-G01} shows the two-dimensional $\chi^{2}$ distribution for each free parameter before and after linking the half-opening angle and torus angular width with dotted and solid lines, respectively, for ESO\,138-G01. The black points and blue stars show the values reported in Table \ref{Tab:ParametersValues} and \ref{Tab:ParametersValues_link}, respectively\footnote{Note that these values are located inside the 1$sigma$ contour and do not necessarily have to be in its centers.}. Figures \ref{fig:Contours_Mrk590} to \ref{fig:Contours_PKS2356-61} in Appendix \ref{sec:app2} show the results for all the objects in the sample.

For most of the sources, we find that parameters from the baseline models where $\rm{\sigma_{tor}  = \theta_{tor}}$ are better constrained at all levels. We can observe this result in the case of ESO\,138-G01, where the viewing angles and torus angular width are much better restricted after assuming this link between parameters. The $\theta_{tor}$ parameter is the most difficult to constrain for most of the sources, even at the 1-$\rm{ \sigma}$ level\footnote{If the 3$\rm{\sigma}$, 2$\rm{\sigma}$, or 1$\rm{\sigma}$ contours cover a third of the total range of each parameter, then we consider that parameter is well constrained at the respective level.}. However, the $\rm{\sigma_{tor}}$ parameter is constrained in most of the sources at 2$\rm{\sigma}$ level, except for NGC\,788, NGC\,1358, Mrk\,78, and PKS\,2356-61. According to the f-test, Mrk\,590 is the only source that prefers a baseline model without the half-opening angle and torus angular width linked (see previous subsection). We also find that the $\theta_{inc}$ and $Y$ parameters are not constrained at any level for three (UM\,146, NGC\,1358, and Mrk\,78) and two (IC\,4518W, ESO\,138-G1) sources, respectively. NGC\,1358 is the source with fewer parameters constrained at the 2$\rm{\sigma}$ level ($\rm{\tau}$, $\rm{Y}$, $\rm{\Gamma}$, and $\rm{N_{H_{tor}}}$ are constrained). Furthermore, the $\rm{\theta_{inc}}$ parameter of ESO\,103-G35 could have two opposite values.

\subsection{Dust quantities derived from mid-IR parameters}
\label{subsec:DerivedParam}
The outer radius of the torus ($\rm{R_{out}}$), the total AGN dust mass ($\rm{M_{dust}}$), and the covering factor ($\emph{Cf}$) are pivotal IR quantities that hold significant physical significance \citep[][for a review]{RamosAlmeida17}. To obtain these important values, we derive them from the posterior distributions of the parameters involved\footnote{We calculate the probability distribution function (pdf) for each quantity from the pdfs of the individual parameters involved. We use the STEPPAR command from XSPEC and divide the parameter space of each one into 30 equally spaced bins to obtain the individual PDFs.}. In Table \ref{Tab:ParametersValues_link}, we report the final values for each of these quantities and their corresponding confidence interval. Determining the outer radius involves the Y parameter of each model and assumes the connection between the dust sublimation radius and AGN bolometric luminosity to derive the inner radius. Regardless of the baseline model chosen, the outer radius within our sample spans a range of values from 0.5\,pc to 16\,pc. Approximately 80\% of our sample shows $\rm{R_{out} < 5\,pc}$. NGC\,4507 has the highest $\rm{R_{out}}$, while NGC\,7213 hosts the smallest value. When considering a \emph{clumpy} distribution of dust, the outer radius spans a broad range from 0.49\,pc to 8.45\,pc, except for NGC\,4507. In contrast, the range narrows for sources favoring a \emph{two-phases} distribution of the dust, ranging from 0.78\,pc to 3.52\,pc. The $\rm{R_{out}}$ for 2MASX\,J10594361+6504063 (the only one that prefers a \emph{smooth} model at mid-IR) agrees with the previously mentioned ranges.

To calculate the $\rm{M_{dust}}$, we integrate the dust density distribution across the entire dust volume\footnote{We use the QUAD function within SCIPY (Python 3.0) to make all integration.}. We use the definition of the density distribution given in the primary papers of the mid-IR models \citep[see][]{Fritz06, Nenkova08b, Gonzalez-Martin23}. We find that the total dust masses of our sources range between $\rm{log\,(M_{dust})=[2.7-7.5]}$. NGC\,4939 and  2MASX\,J10594361+6504063 show higher and lower dust mass values, respectively. The sources that prefer baseline models with a \emph{two-phases} distribution of the dust have masses in the range  $\rm{log\,(M_{dust})=[3.4-5.9]}$. Meanwhile, the sources that prefer baseline models with a \emph{clumpy} distribution of the dust have extended masses between the range $\rm{log(M_{dust})=[2.5-6.5]}$.

The $\rm{M_{dust}}$ and $\rm{R_{out}}$ values are consistent with those found in previous works \citep[e.g.,][]{Lira13, Garcia-Bernete19, Garcia-Bernete22}. However, these measurements differ from those obtained using ALMA observations, which trace the cold dust and molecular gas distribution. For example, NGC\,6300 and NGC\,7213 are sources in common with the GATOS sample reported by \cite{Garcia-Burillo21}. They measured outer radii and masses with higher values than in this work and in previous studies where mid-IR imagining and interferometric data were used \citep[e.g.,][]{Packham05, Radomski08}. The difference may be due to the mid-IR models not representing the observed physical size of dust by ALMA for the GATOS sample or the temperature differences that give rise to the emission of cold and hot gas \citep{Garcia-Bernete22}. This result suggested that mid-IR only traces the inner parts of the torus where the dust has high temperatures and is mixed with gas traced through X-ray emission \citep[see][]{Lopez-Rodriguez18, Alonso-Herrero21, Nikutta21}.

Finally, the calculation of the covering factor involves computing the unity minus the escape probability, that is, $ \int{e^{-\tau_{\nu}(lso)} \,Cos \, \sigma_{tor}\, d\sigma_{tor}}$ where $\rm{\tau_{\nu}(lso)}$ is the line-of-sight opacity. This opacity is computed from the equatorial opacity and the density distribution for \emph{smooth} models and from the distribution of clouds for \emph{clumpy} models. In the case of the \emph{two-phases} model, the $\emph{Cf}$ only depends on the $\rm{sigma}$ value. We find that $\emph{Cf}$ values are in a broad range of $\emph{Cf} = $[0.18-0.90]. Except for the $\emph{Cf}$ of Mrk\,1392, all Sy1s cover a range between [0.18:54]. Meanwhile, the Sy2s cover a range between [0.21:0.97]. A detailed analysis and discussion of these results is presented in Sect.\,\ref{subsec:CF_AGNaccretion}.

\begin{figure*}[ht]
    \centering
    \includegraphics[width=2.\columnwidth, trim={0 0 0 0},clip]{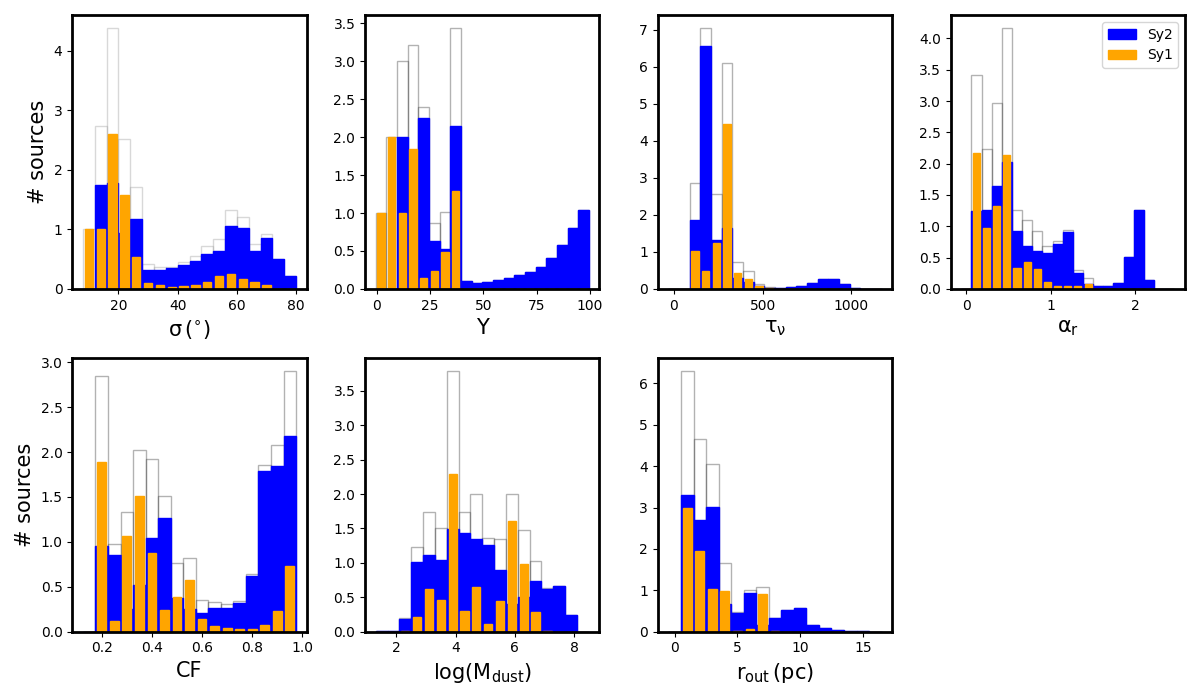}
    \caption{Posterior distributions (PDs) of mid-IR torus model parameters for all AGNs in our sample with good statistics. The top row shows (from left to right) the torus angular width ($\rm{\sigma = \theta_{tor} }$), the ratio between the outer and inner torus (\emph{Y}), the optical depth ($\rm{\tau_{\nu}}$), and the slope of the radial distribution of dust ($\rm{\alpha_r}$). The bottom row shows the distribution of derived parameters (from left to right): covering factor (\emph{Cf}), total dust mass ($\rm{M_{dust}}$), and the outer radius of the torus ($\rm{r_{out}}$). The orange and blue histograms correspond to Sy1 and Sy2, respectively. The white bars show the total combined distributions of Sy1 and Sy2.}
    \label{fig:MIRparamdistrib}
\end{figure*}

\subsection{Torus properties and AGN type}
\label{subsec:Results_torusProperties}
\begin{figure}[!t]
    \centering
    \includegraphics[width=0.8\columnwidth, trim={30 20 27 34},clip]{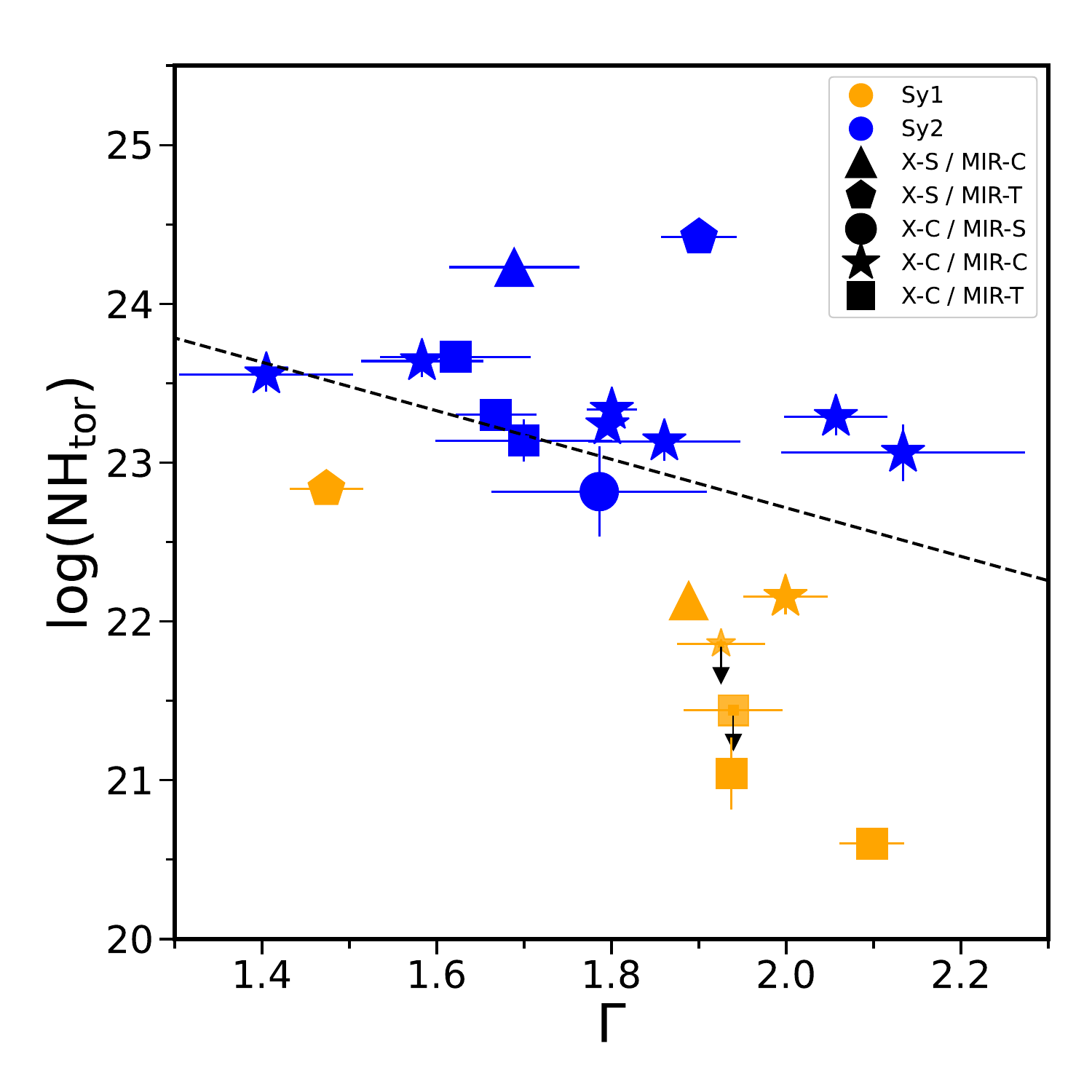}
     \caption{Column density of the reflector ($\rm{N_{H_{tor}}}$) versus the photon index ($\rm{\Gamma}$). Sy1s are shown in orange, and Sy2s are shown in blue. Different symbols are shown according to the best models used for each object (see text). }
    \label{fig:NHvsGamma}
\end{figure}

It is well established in the literature that the classical unification model is too simple to explain the properties and differences of Sy1s and Sy2s \citep[see][for a review]{RamosAlmeida17}. Therefore, Sy2s are not only edge-on views of Sy1s, as the unification ideas propose \citep{Urry95}. We find that most Sy1s (5 out of 8) have an inclination angle greater than $60^{\circ}$, suggesting an edge-on view system. However, the torus angular widths of these sources are small ($< 20^{\circ}$). This indicates that their classification as type 1 sources is due to the torus not intercepting the line of sight. Meanwhile, the $\theta_{inc}$ of Sy2s cover a wide range [21$^{\circ}$:82$^{\circ}$] (see panel 3 of Fig.\,\ref{fig:Comparison}). More than a decade ago already, pioneer works by \citet{RamosAlmeida11} and \citet{Alonso-Herrero11} using SED fitting at mid-IR wavelengths found intrinsic differences between Sy1 and Sy2 galaxies, besides the viewing angle. They found that the accretion disk of Sy1s is intrinsically less covered due to a relatively thinner torus than that found in Sy2s. Using a volume-limited sample of nearby AGNs with sub-arcsecond angular resolution spectroscopy \citet{Garcia-Bernete19} confirmed these results.

Our results further confirm these differences. To illustrate it, Fig.\,\ref{fig:MIRparamdistrib} shows the posterior distributions (PDs) for each mid-IR parameter across our AGNs sample, which includes both Sy1 and Sy2 populations. The histograms represent the combined PDs for all sources normalized by the area rather than individual PDs, providing a global view of the distribution of each parameter within the sample. The color coding highlights the differences between Sy1 and Sy2, facilitating a comparison of trends between the two types. The angular sizes, $\sigma_{tor}$, of Sy1 galaxies (orange) cover a narrow range between $\rm{[10^{\circ}-24^{\circ}]}$ (except for the case of Mrk\,1392). Meanwhile, the $\rm{\sigma_{tor}}$ values of Sy2 galaxies (blue) are grouped into two peaks; one is consistent with Sy1s while the other peak has a maximum at $\rm{60^{\circ}}$. We confirm these results by comparing the posterior distributions using the Kolmogorov$-$Smirnov (K-S) test\footnote{The null hypothesis is rejected if the p-value is lower than 0.05. Therefore, the PDs that we are comparing are different.}. The comparison between the Sy1 and Sy2 PDs shows that their distributions are different (p-value $= 0.002$). We also find clear differences in the covering factors. Except for Mrk\,1392, all sources classified as Sy1s are consistent with $\rm{\emph{Cf} < 0.5}$ and 11 out of the 13 Sy2s with good fits are consistent with $\rm{\emph{Cf} > 0.5}$ (exceptions are IC\,4518W and NGC\,6300). The comparison of \emph{Cf} distributions of Sy1 and Sy2 through the K-S test confirms that the PDs are different (p-value$= 0.04$). The average covering factor of Compton-thin obscured AGNs analyzed by \citet{Zhao21} at X-rays is $\rm{\emph{Cf} \sim 0.67}$, consistent with our results of Sy2 galaxies. \cite{Garcia-Bernete19} found that the $Cf$ obtained though only mid-IR fits depends on the torus dust distribution and geometry assumed by the model. They observed that using the \emph{two-phases} models provides large values of covering factor (Cf > 0.6), while the clumpy models tend to favor intermediate values. Although most baseline models chosen by our sources incorporate both mid-IR distributions inside the SFT, we did not observe this distinction in our sample. Instead, we found that the values of $Cf$ cover similar ranges, irrespective of the chosen baseline model.

Recently, \citet{Kawamuro24} used CIGALE code to model the full SED of the Sy1 Pox52, finding a torus with a small value of the angle between the equatorial plane and the edge of the torus ($\sigma_{tor}$), consistent with our results. No statistical differences are found regarding the total dust mass (p-value$ = 0.13$ from K-S test) and, interestingly, among the five AGNs with $\rm{R_{out} > 5\,pc}$, four of them are classified as Sy2s, although AGNs with $\rm{R_{out} < 5\,pc}$ are found for both Sy1 and Sy2 galaxies. The comparison of PDs of the $\rm{R_{out}}$ parameters of Sy1 and Sy2 reveals a p-value$ = 0.04$, which implies that they are different distributions. 

According to K-S test, with the exception of $\rm{\alpha_r}$ and $\rm{M_{dust}}$, all parameter distributions differ between Sy1 and Sy2 ($\rm{\alpha_r}$: \, p-value = 0.09, and $\rm{M_{dust}}$: p-value = 0.13). Additionally, the PDs of Sy1 PDs differ from those of the total sample (p-value$ < 0.04$), while the PDs of Sy2 are similar to those of the total sample (p-value$ >0.27$). Consequently, the mid-IR properties of the dust in Sy1s suggest that they belong to a well-defined group of objects with low covering factors and narrow half-opening angles, which increases the probability of the central photons escaping. In contrast, Sy2s are a mixed bag with highly-covered objects, although the half-opening angle could be as low as that found for Sy1s.

X-ray parameters also reinforce this behavior when comparing Sy1s and Sy2s properties. To illustrate it, we show in Fig.\,\ref{fig:NHvsGamma} the column density of the reflection component, $\rm{N_{H_{tor}}}$, versus the photon index, $\rm{\Gamma}$, of the intrinsic continuum (associated with the X-ray corona). Most Sy1s show $\rm{N_{H_{tor}}<10^{22-23}\,cm^{-2}}$ ($\rm{20.6 < N_{H_{tor}} < 22.9}$) while Sy2s show larger values ($\rm{23.0 < N_{H_{tor}} < 24.5}$). This is consistent with the X-rays classification of type-1/type-2 AGNs \citep{Matt00}. \citet{Zhao21} investigated a sample of 93 obscured and nearby AGNs using only X-ray \emph{NuSTAR} observations and the \emph{smooth} torus model. They found that Compton-thin and Compton-thick AGNs may harbor similar tori, whose average column density is nearly Compton thick \citep[see also][]{Buchner19}. This is consistent with our results for Sy2 galaxies. The photon index is also consistent with the canonical value of $\rm{\Gamma= [1.8-2.1]}$ for most Sy1s, while Sy2s show values as low as $\rm{\Gamma= 1.5}$. Eleven out of the 13 Sy2 have $\rm{\Gamma < 1.9}$ and all Sy1s have $\rm{\Gamma > 1.9}$ (see panels 1 and 2 of Fig.\,\ref{fig:Comparison}). These two quantities anti-correlate with each other when we exclude the objects where an X-C/MIR-C baseline is preferred. Except for three sources (NGC\,6300, ESO\,138-G1, and UM\,146), we observe that objects with a steeper X-ray intrinsic continuum are less obscured than those showing a flatter X-ray intrinsic continuum. This behavior is beyond the natural degeneration of photon index and column density, which would imprint a positive correlation between these two quantities. 

\section{Discussion}
\label{sec:Disc}
\subsection{Distribution of the dust and gas}
One difference between the first SED fitting results \citep{RamosAlmeida09, RamosAlmeida11, Alonso-Herrero11, Martinez-Paredes17, Garcia-Bernete19} and those presented in this analysis is that they were based on a single SED library of model, consisting of a \emph{clumpy} distribution of dust in a torus-like geometry. In the last few years, several studies have tested and compared models that propose new geometries and dust distributions available, including a \emph{disk+wind} geometry \citep[e.g.,][]{Gonzalez-Martin19b, Martinez-Paredes21, Garcia-Bernete22}. These works found that the \emph{disc+wind} model is more suitable for reproducing the SEDs of type-1 and more luminous AGNs, whilst \emph{clumpy} torus models are preferred for less-luminous and type-2 AGNs.
However, recently \cite{Gonzalez-Martin23} found that the \emph{two-phases} torus model can reproduce the 85\%-88\% of the spectra of a sample of 68 nearby and luminous AGNs. As we mentioned in Sect.\,\ref{sec:sample}, we only found four Sy1 (RBS0770, PG1211+143, RBS\,1125, and Mrk\,1393) from a previous sample of 36 AGNs that fitted better with the \emph{disk+wind} model at mid-IR wavelengths. We are not including the analyses of these sources in this work because of the lack of this geometry at X-ray wavelengths. However, we agree that fully understanding AGNs obscuration requires looking beyond the geometry of the dust distribution.

Besides its geometry, the distribution of dust and gas, either \emph{clumpy} or \emph{smooth}, has also been discussed \citep{RamosAlmeida17}. At mid-IR, the \emph{smooth} distribution is not preferred for almost any Seyfert \citep{Garcia-Bernete22}, but it is preferred for more extremely obscured objects \citep[e.g.,][]{Garcia-Bernete22b, Efstathiou22}. This result is confirmed in our analysis where the \emph{smooth} distribution is preferred at mid-IR only for the Sy2 IRAS\,J105943+65040 (see Table\,\ref{Tab:ParametersValues_link}). The 52\% and 43\% of our sample require either a \emph{clumpy} or a \emph{two-phases} distribution of dust (11 and 9 AGNs, respectively). Most Sy1s prefer the \emph{two-phases} distribution (5 out of 8 Sy1), while 8 out of the 13 Sy2s with good fits require the \emph{clumpy} distribution of dust. Using full SED analysis and the CIGALE code to disentangle host and AGN contributions, \citet{Kawamuro24} found that a \emph{two-phases} torus better describes the emission for the Sy1 galaxy Pox\,52 than a \emph{smooth} torus model, also consistent with our results.

At X-rays, the distribution of gas is still controversial, partially because of the lack of good data sets with enough information to distinguish between models \citep{Saha22}. Despite that, recently, \citet{Sengupta23} as part of the effort of the Clemson-INAF group \citep{Marchesi18, Zhao21, Torres-Alba21, Traina21} to classify Compton-thick AGN candidates, found that 78\% of the sources show a LOS column density different from that of the reflector, which they interpret as a preferred \emph{clumpy} distribution of the material. The \emph{clumpy} distribution of gas is preferred in 85\% of the Sy2s analyzed here (only five AGNs prefer the \emph{borus02} model).

Now, if we consider gas and dust, we find a wide variety of distribution combinations that do not seem to be associated with the AGN type:

\begin{itemize}
    \item[$\circ$] \emph{Clumpy} distributions of both dust and gas (X-C/MIR-C baseline models): 2 Sy1s and 7 Sy2s. 
    \item[$\circ$] \emph{Clumpy} gas distribution and \emph{two-phases} dust distribution (X-C/MIR-TP baseline model): 3 Sy1s and 3 Sy2s.
    \item[$\circ$] \emph{Smooth} gas distribution and \emph{two-phases} dust distribution (X-S/MIR-TP): 2 Sy1s and 1 Sy2.
    \item[$\circ$] \emph{Smooth} gas distribution and \emph{clumpy} dust distribution (X-S/MIR-C): 1 Sy1 and 1 Sy2.
\end{itemize}

These results highlight the complexity of the obscurer when considering multiwavelength information \citep{Esparza-Arredondo21}. 

\cite{Laha20} investigated a sample of 20 Sy2s and found that 13/20 showed no significant variability in LOS column density, suggesting a \emph{smooth} distant gas distribution. In our previous work, the \emph{smooth} distributions of gas seemed to be preferred for AGNs without variable absorption, while the \emph{clumpy} distribution of gas is needed when the object shows signs of LOS column density variability \citep{Esparza-Arredondo21}. However, this was based on independent SED fitting of mid-IR and X-rays. Moreover, the \emph{two-phases} model of the dust was not considered in our previous work. In the present work, where a baseline model is fitted to both wavelengths simultaneously, this result is not confirmed. \citep{Laha20} torus models incorporating both clouds or over-dense regions (two-phases distribution) should account for LOS column densities as low as few $\rm{10^{21} cm^{-2}}$. We speculate that including the \emph{two-phases} model at mid-IR (without a counterpart at X-rays) in our current study might have weakened the link between column density variability and the chosen distribution of gas/dust. Perhaps new \emph{two-phases} torus models at X-rays are needed, which could be done soon thanks to the extension of the radiative transfer code SKIRT that now accounts for the physics behind the X-ray reflection component \citep{VanderMeulen23}.

\subsection{Dust and gas connection}
\begin{figure}[!t]
    \centering
    \includegraphics[width=1.0\columnwidth, trim={0 0 27 34},clip]{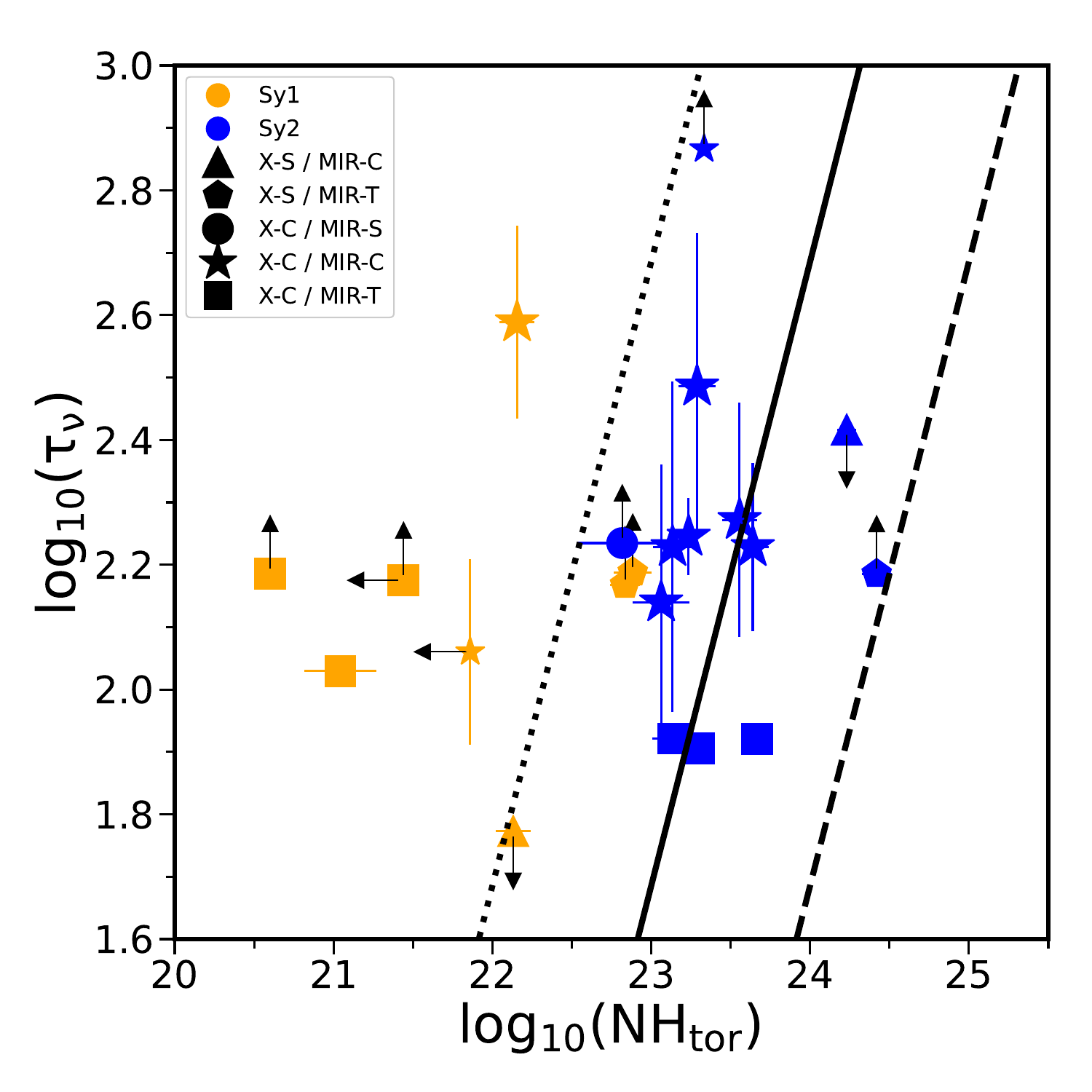}
    \caption{Optical depth of the dust versus hydrogen column density of the gas. Sy1s are shown in orange and Sy2s are shown in blue. Symbols refer to the best baseline model per object. The dotted, solid, and dashed black lines show 10, 1, and 0.1 times the Galactic dust-to-gas ratio, respectively. Objects with upper/lower limits are shown with smaller symbols for clarity.}
   \label{fig:compDeriveParam}
\end{figure}

For decades, the relationship between gas and dust in different objects in the universe has been studied, and it is known that the properties of both materials are related \citep[e.g., the metallicity or the grain growth][]{Hensley23}. In the case of the galaxies, this connection is supported by several scaling relations, for example, between the dust mass, total gas mass, stellar mass, and star-formation at different redshifts \citep[e.g.,][]{Orellana17, Popping23}. In this work, we explored the possible connection between the column density of the gas ($\rm{N_{H_{tor}}}$) and the opacity $\tau_{\nu}$ of the dust. This relationship has been broadly studied in our Galaxy through different methods and considering the possible bias in the observations given rise to different quotient values between both variables \citep[e.g.,][]{Reina73, Gorenstein75}. This work assumes the galactic dust-to-gas ratio presented by \citet{Draine03} and given as: $\rm{N_{H}/A_v = 1.9 \times 10^{21} [cm^{-2} mg^{-1}]}$ \citep{Watson11}. As is expected, for the sources that are further away from us, this relationship is affected due to the interstellar medium between the observer and the source \citep[e.g.,][]{Kahre18}. In the case of AGN, \citet{Maiolino01} find that this relationship is slightly above the Galactic dust-to-gas ratio.

Figure\,\ref{fig:compDeriveParam} compares the $\rm{N_{H_{tor}}}$ and $\rm{\tau_{\nu}}$ values derived from the final best fits of each source. The dotted, solid, and dashed black lines show 10, 1, and 0.1 times the Galactic dust-to-gas ratio, respectively. Most Sy2 galaxies in our sample are consistent with the galactic values. Some of these galaxies show an increase in the $\rm{N_{H_{tor}}}$ value, which could be explained as due to dust-free neutral gas within the sublimation radius. Meanwhile, the Sy1 galaxies show slightly lower values for the galactic dust-to-gas ratio. These values agree with previous results \citep{Burtscher16, Huang11}. However, the low dust-to-gas ratio seen in Sy1s is hard to explain. Of course, variable absorption could be a reason. However, among the two Sy1s with restricted measurements of both $\rm{N_H}$ and $\rm{\tau_{V}}$, Mrk\,590 is not variable \citep{Laha20}. Another possibility is that the gas within the sublimation radius has been expelled from the nucleus for these Sy1s throughout AGN winds and jets \citep[see][]{Garcia-Burillo21}.

\subsection{Torus versus AGN luminosity}
\begin{figure*}[ht]
    \centering
    \includegraphics[width=0.67\columnwidth, trim={10 10 0 0},clip]{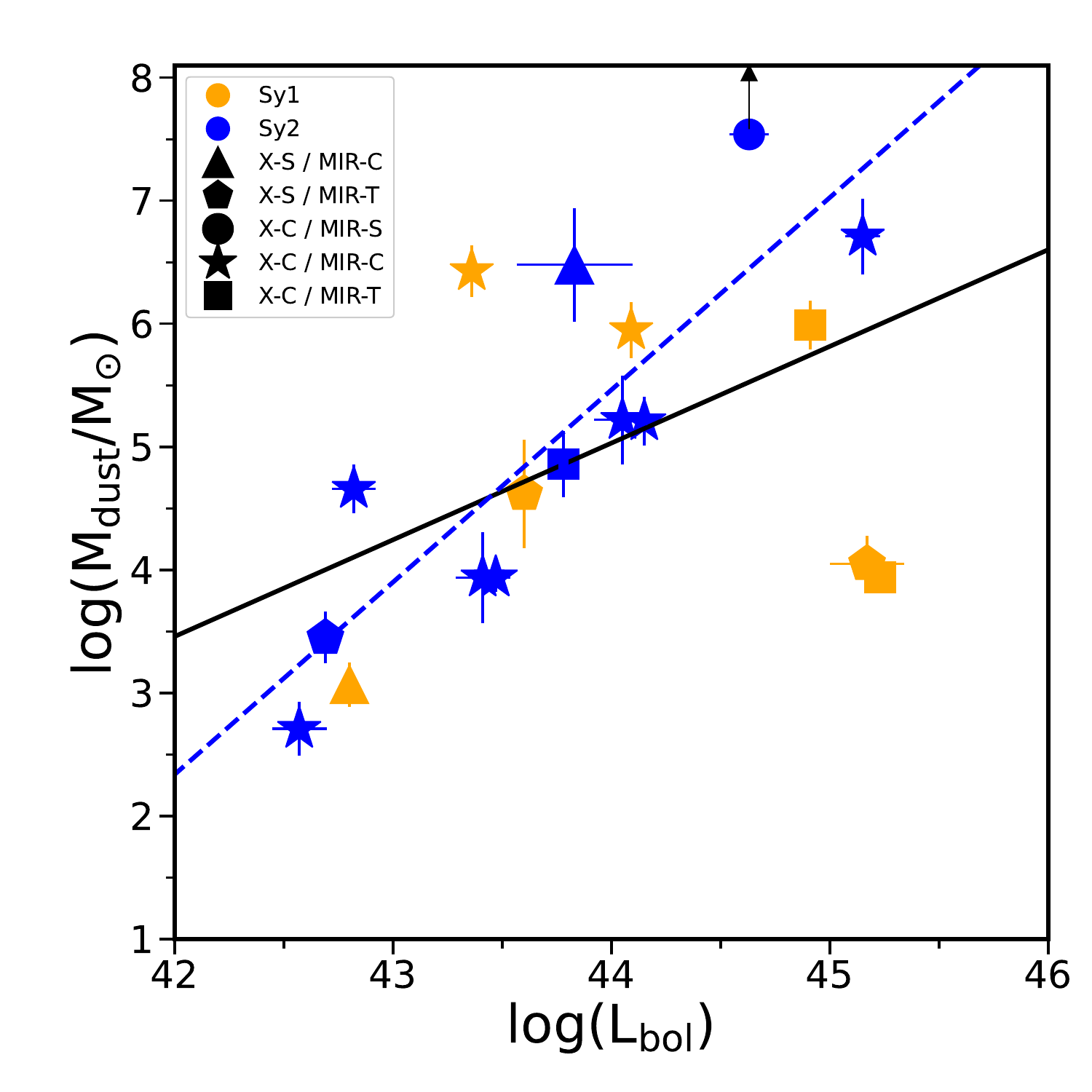}
    \includegraphics[width=0.67\columnwidth, trim={10 10 0 0},clip]{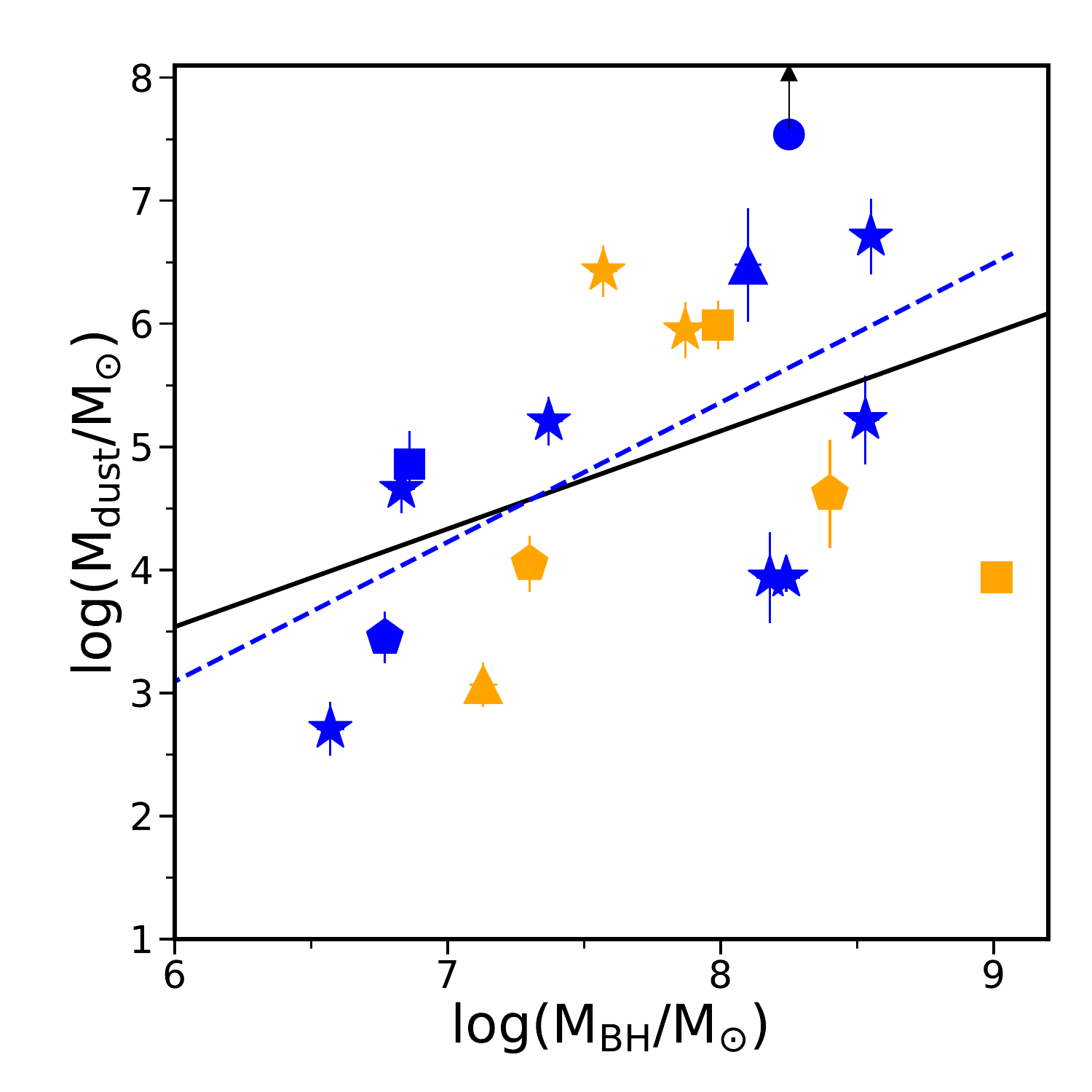} 
    \includegraphics[width=0.67\columnwidth, trim={10 10 0 0},clip]{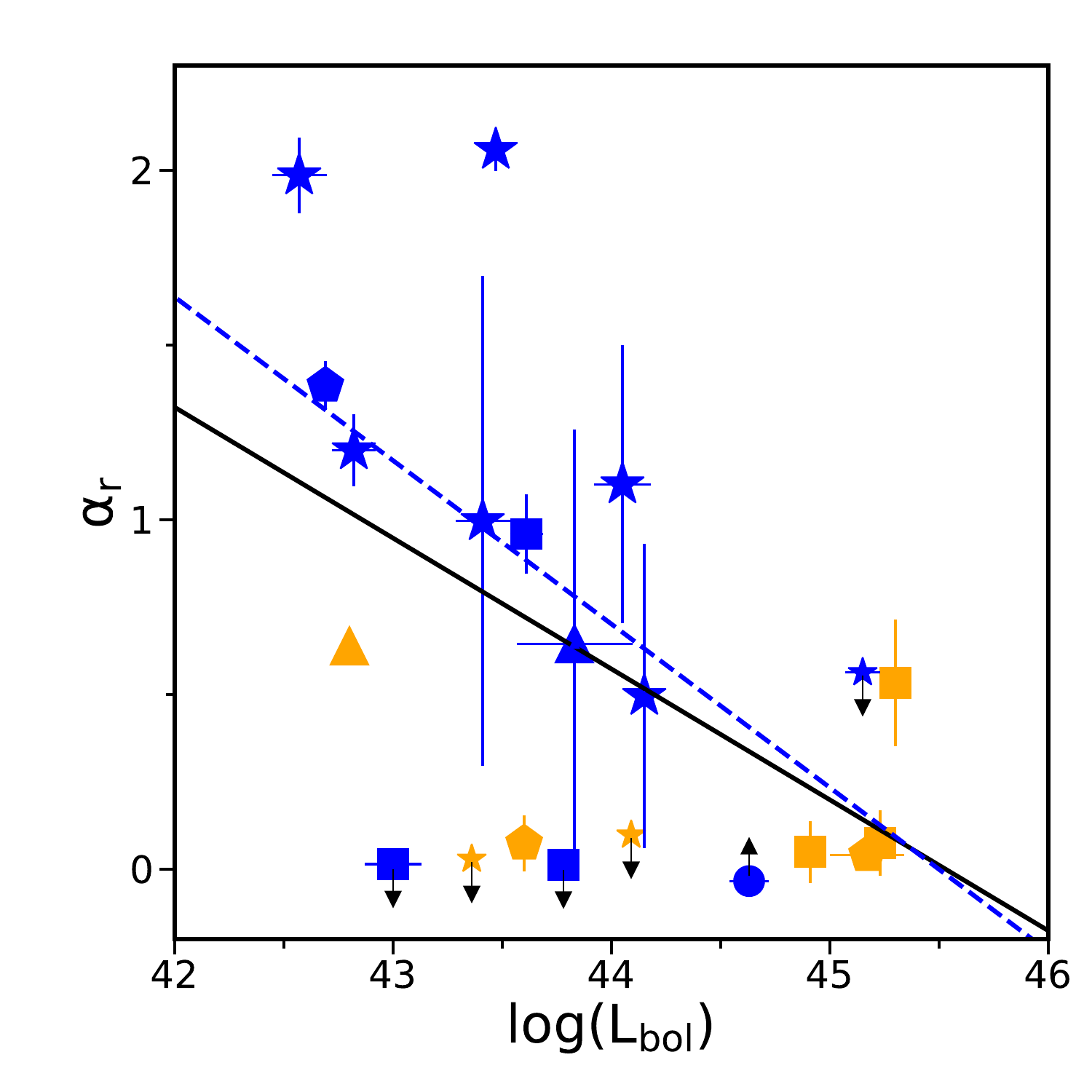}
    \caption{First panel: Dependence of the torus total dust mass. Middle panel: Total dust mass versus the SMBH mass. Last panel: Slope of the radial distribution. Sy1s are shown in orange, and Sy2s are shown in blue. Symbols refer to the best baseline model found for each object (see legend in Fig.\,\ref{fig:Comp_ChiReduced}). The solid black and dashed blue lines show the best-fit linear relation between these two quantities considering all sources and only the Sy2, respectively. Objects with upper/lower limits are shown with smaller symbols for clarity.}
    \label{fig:Lbol_vs_TParam}
\end{figure*}

Intrinsic differences in the torus have also been claimed for high- versus low-luminosity AGNs \citep{Gonzalez-Martin17}. \citet{Gonzalez-Martin19b} found a larger percentage of best fits for the \emph{disk+wind} model for high-luminosity objects \citep[see also][]{Martinez-Paredes21}, while a larger number of low-luminosity AGNs were better fitted to the \emph{clumpy} torus model \citep[see also][]{Garcia-Bernete22}. Furthermore, \citet{Garcia-Bernete22} also showed that structural parameters, such as the outer radius of the torus, might depend on the luminosity of the source. To explore this, we delve into the possible relationships between the $\rm{L_{bol}}$ and the parameters characterizing dusty gas torus, as derived from our final fits. In Table\,\ref{tab:Sample}, we report the bolometric luminosity,  Eddington ratio, $\rm{L_{bol}/L_{Edd}}$, and the black hole mass, $\rm{M_{BH}}$, of each source.

We find a possible positive correlation between the total dust mass of the torus ($\rm{log_{10}(M_{dust})}$) and the bolometric luminosity ($\rm{log_{10}(L_{bol})}$), as depicted in Figure\,\ref{fig:Lbol_vs_TParam} (left panel). Considering all sources, this relation is not statistically robust, yielding a Pearson correlation coefficient\footnote{$R^2=1$ means zero dispersion.} (R) of $\rm{0.51}$ and a determination coefficient ($R^2$) of $\rm{0.26}$. However, considering only the galaxies classified as Sy2, the correlation becomes more robust ($\rm{R=0.86}$ and $\rm{R^{2}=0.73}$). This result might show that the AGN luminosity is regulated by the amount of material available for AGN feeding or that the amount of AGN heated dust (and therefore emitting at mid-IR) is linked to the AGN luminosity. Although the accretion rates are affected by large uncertainties since no correlation is found between dust mass and the Eddington ratio ($\rm{R=0.17}$), the most natural explanation is that the total dust mass emitting in the mid-IR is directly linked to the AGN power heating the dust. Therefore, the current efficiency ($\rm{\lambda_{edd}}$) of the accretion process would not be linked to the reservoir of material at parsec scales, which is expected considering the timescale for the torus to fall into the accretion disk (to nurture the AGN) is much longer than the timescales for the accretion disk to change its state. In the case of NGC\,1068, the timescale estimate for the torus fade away is 1-4 Myr \citep{Garcia-Burillo19}. Meanwhile, the accretion disk is estimated to vary in timescales of hours \citep{Hawkins07}. Interestingly, although with larger dispersion ($\rm{R=0.40}$ and $R^2 =0.15$, see Figure\,\ref{fig:Lbol_vs_TParam} middle panel), a similar relationship is found for $\rm{log_{10}(M_{dust})}$ and  $\rm{log_{10} M_{BH}}$. This relationship suggests that a more massive SMBH is able to drag more dust from the circumnuclear medium into the AGN.

To study the robustness of the above relationship, we also explore if this dependence is due to the parameters involved in computing the total $\rm{log_{10}(M_{dust})}$, such as $\rm{\sigma_{tor}}$, $Y$, $\rm{\alpha_r}$, $\rm{\tau_{\nu}}$, and $\rm{N_0}$. No clear relationship is found between $\rm{L_{bol}}$ and neither $\tau_{\nu}$, $\sigma_{tor}$ nor $N0$ parameters. A robust link emerges between the bolometric luminosity and the slope of radial density distribution ($\rm{R=-0.7}$ and $\rm{R^2 = 0.48}$, right panel of Fig.\,\ref{fig:Lbol_vs_TParam}). For Sy2 galaxies, high $\alpha_r$ values are associated with low $\rm{log_{10}(L_{bol})}$ values, indicating that the obscuring material is distributed closer to the nucleus in intermediate luminosity sources ($\rm{\alpha_r > 1}$ and $\rm{log_{10}(L_{bol}) < 43.5}$ erg/s); meanwhile, the high luminosity sources tend to have a distribution of material farther extended (relatively flat) from the nucleus ($\rm{\alpha_r < 1}$ and $\rm{log_{10}(L_{bol}) < 44}$ erg/s). Furthermore, most Sy1 tend to have lower $\rm{\alpha_r}$ values. It is important to notice that, even if limits are not considered in our statistical tests, four objects (NGC\,4939, Mrk\,590 and Mrk\,1210, and Mrk\,1392) are hard to reconcile within the general trend.

\citet{Gonzalez-Martin17} found evidence that the outer radius of the torus also decreases with the AGN bolometric luminosity for low-luminosity AGNs ($\rm{log_{10} L_{bol} < 42 }$ erg/s). This behavior was confirmed for intermediate and high luminosity AGNs ($\rm{41.7 < log_{10} L_{bol} < 44.7}$ erg/s) by \citet{Garcia-Bernete22}. However, the latter authors do not find a clear dependence between $Y$ and $\rm{L_{bol}}$, suggesting that the correlation between torus size and $\rm{L_{bol}}$ might be caused, at least in part, by the sublimation radius, which is dependent on the $\rm{L_{bol}}$. Our outer radii ($\rm{R_{out} \sim 1-9\,pc}$) are in agreement with the range of mid-IR torus sizes estimated in previous works for nearby AGNs using data obtained through MIDI/VLT \citep[e.g.,][]{Tristram09, Burtscher13}. However, a clear linear dependence is not observed, partially due to the limited number of objects and the large amount of upper/lower limits in our results. It is important to remark that the SFT at both mid-IR and X-ray wavelengths does not help to restrict the outer radius of the torus because the X-ray reflection component does not depend on this parameter \citep[most of the emission comes from the inner region of the torus][]{VanderMeulen23}. Therefore, this parameter probably needs near-IR observations in order to be further constrained, as shown by \citet{RamosAlmeida14}.

\subsection{Covering factor as a function of AGN accretion state}
\label{subsec:CF_AGNaccretion}
The covering factor is an effective parameter for understanding the role of the obscuring material in the context of the AGN feedback mechanism \citep[][for a review]{RamosAlmeida17}. It is defined as the fraction of the sky obscured by the dust associated with the torus. There are different ways to measure this quantity, and several assumptions are present in all cases. Frequently, the $\emph{Cf}$ is defined as the relationship between the nuclear-infrared luminosity and the bolometric AGN luminosity \citep{Maiolino07, Treister08, Gu13, Toba21}. This method assumes that the infrared luminosity is dominated by the dust located within the torus and also assumes that both luminosities are isotropic \citep[e.g.,][]{Stalevski16, Ralowski23}. The $\emph{Cf}$ can also be measured by counting the number of obscured sources at X-rays or the number of type-2 AGNs in volume-limited samples \citep{Ueda03, Ricci23}.

In this work, the $\emph{Cf}$ parameter is derived from parameters obtained through mid-IR fit, so it is mostly associated with the dust-obscuring material (reported in Table\,\ref{Tab:ParametersValues_link}, Col.\,16). We find values between 0.15 to 1 (including error bars; see also the bottom-left panel in Fig.\,\ref{fig:MIRparamdistrib}). The lack of values below $\rm{\emph{Cf}<0.15}$ is intrinsic to the modeling since all the models assume a torus angular size of at least $\rm{\sigma_{tor}=10^{\circ}}$ \citep[see also][]{Gonzalez-Martin19a}. We performed a general analysis on the derived $\emph{Cf}$ in objects in common with the previous analysis, finding that the $\emph{Cf}$ strongly depends on the method (and its assumptions). This is shown by \citet{Gonzalez-Martin19a} where even using the SED technique the \emph{Cf} depends on the model assumed. Therefore, it is important to reinforce that these $\emph{Cfs}$ are associated with dust, and they assume a torus-like geometry.

In the last decades, several works have revealed that there is an anticorrelation between the $\emph{Cf}$ of AGNs and the accretion rate \citep{Zhuang18, Ezhikode17, Toba21} or the luminosity \citep{Gu13}. This relationship is interpreted by the fact that the inner radius of the obscuring material increases with the incident luminosity of the AGN \citep{Ricci17}. \citet{Naddaf24} find that the covering factor is intrinsically linked to the mass, accretion rate, and metallicity of the clouds under a scenario in which the clouds in the outer, less ionized part of the BLR are launched by the radiation pressure acting on dust. Recently, \cite{Ricci23} compared the relationship between the $\rm{L_{bol}}$ and the covering factor obtained from the IR luminosity versus the \emph{Cf} from X-rays. They found a slightly less steep trend calculated through IR measurements compared with the trend found with the X-ray.

\begin{figure}[!t]
    \centering
    \includegraphics[width=1.0\columnwidth, trim={15 0 10 0},clip]{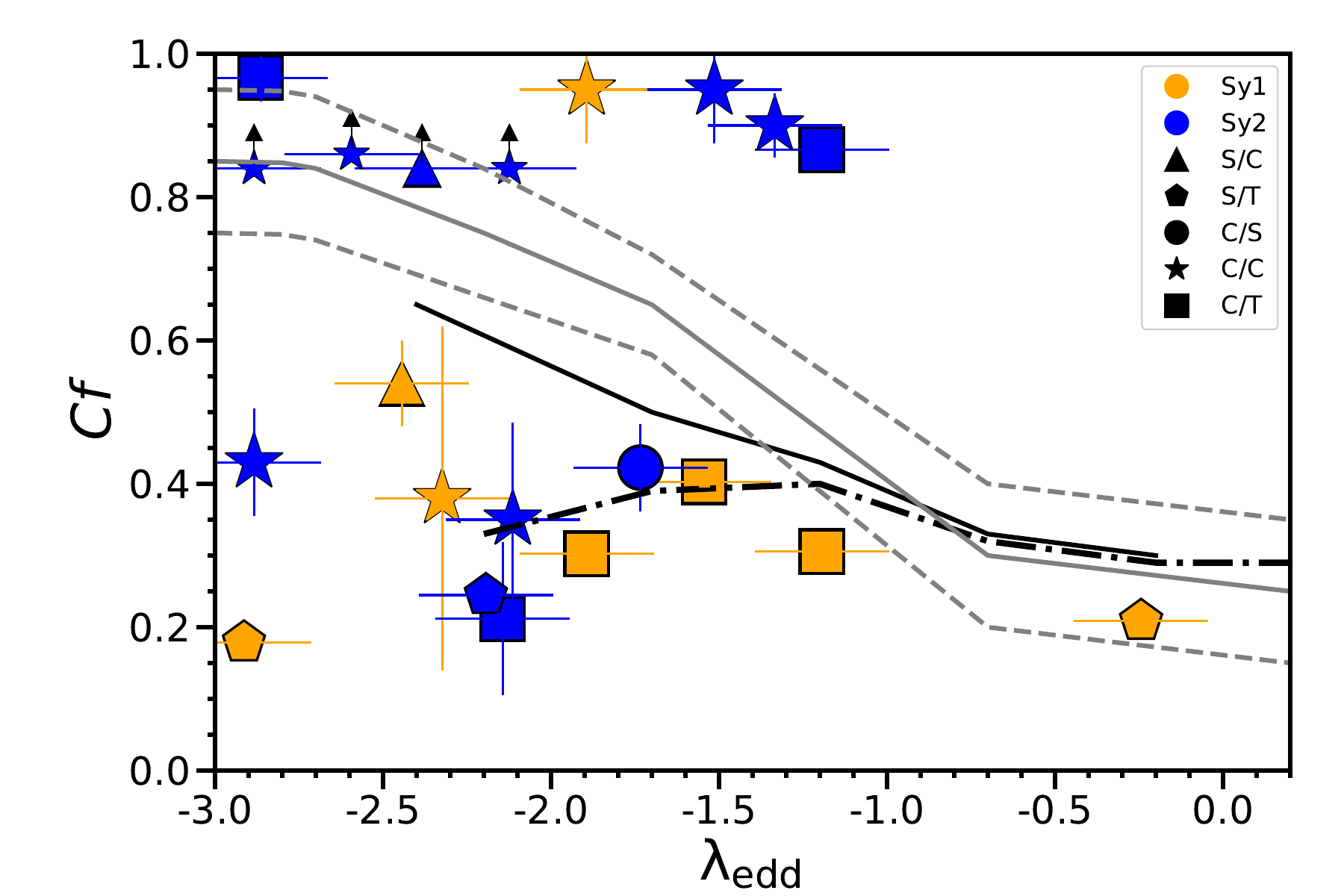}
    \caption{Eddington ratio against covering factor (\emph{Cf}). The solid and dashed grey lines show the trend found by \cite{Ricci22} and the $1\sigma$ uncertainty around it, respectively, when the covering factor of gas is inferred from the fraction of obscured sources using X-ray observations. The solid and dashed-dotted black lines show the trends that follow the median \emph{Cf}, derived through IR data, obtained for obscured and unobscured, respectively, sources from the sample of \cite{Ricci23}.}
    \label{fig:Cf_vs_lambedd}
\end{figure}
\noindent

Figure\,\ref{fig:Cf_vs_lambedd} shows the covering factor values derived here versus the Eddington ratio. The solid and dashed grey lines show the trends, using X-ray data (dusty gas and dust-free) to derive $\emph{Cf}$, found by \citet{Ricci22} for a sample of obscured AGNs. The area between dashed grey lines represents the 1$\sigma$ uncertainty around it. The solid and dashed-dotted black lines show the trends that follow the median \emph{Cf}s obtained for obscured and unobscured sources from the sample of \citet{Ricci23}, derived through IR (dusty-gas) and using the correction of \cite{Vasudevan07}, respectively. All covering factors derived from our sample are inside the range of values that they reported. Objects with $\rm{\emph{Cf}<0.6}$ in our analysis seem consistent with the $\emph{Cf}$ estimated from mid-IR by them. The $\emph{Cf}s$ from X-ray are always above this trend (grey lines), indicating that the sources contain a significant contribution from gas. This is consistent with the work done by \cite{Ichikawa19}, where they found that the $\emph{Cf}$ obtained from X-ray observations (gas) always exceeds the average $\emph{Cf}$ of the dust.

In an object-by-object comparison, except for Mrk\,1392 and NGC\,7213, all the sources classified as Sy1 of our sample appear to follow the trend found by \cite{Ricci23} for unobscured sources (black dashed-dotted line) together with four Sy2 (ESO138-G1, 2MASXJ10594361, NGC6300, and IC4518W). These four Sy2 sources show $Cfs$ and $\sigma$ values that do not clearly distinguish them from Sy1s. However, the hydrogen column densities from X-rays are higher than those of Sy1. Therefore, the obscuration of these sources could be due to the distribution or density of the gas and dust. For example, NGC\,6300 galaxy exhibit smooth and two-phase gas and dust distributions, respectively, which could explain their higher column density ($\rm{log(N_{H_{tor}})}$ = 24.42) despite a relatively low dust mass ($\rm{log(M_{dust}) = 3.45}$). In the case of the other three Sy2, their column density values are also higher and show low dust mass values, but their gas and dust distributions are in clumps and in two phases. A possible explanation is that gas clumps intercept the line-of-sight but not dust clumps. These findings indicate that the obscuration mechanism in Sy2s is more complex. The remaining 60\% of Sy2 shows higher $Cfs$, and five are consistent with the trend obtained by X-rays. PKS\,2356-61, ESO\,103-G35, and Mrk\,1210 are the sources with $\rm{\lambda_{edd} > -1.5}$ and higher $\emph{Cf}$s that do not clearly follow the trend. Finally, IC\,5063 is the only source with lower $\emph{Cf}$ and $\lambda_{edd}$ values.

Alternatively, the distribution of our sources in the $\lambda_{edd}$ versus $Cf$ diagram may be related to past interactions between host galaxies and their neighbors. These interactions can create perturbations that influence both the black hole mass and the accretion rate \citep{Krongold02}. For example, sources with higher $\emph{Cf}, (> 0.5)$ and lower $\rm{\lambda_{edd}, (< -1.)}$ can obscure the AGNs not only due to the dust and gas content or viewing angle but also due to the relatively low power of the AGN, which is insufficient to clear the surrounding material. Conversely, sources with lower $\rm{\lambda_{edd}, (< -1.) }$ and $\emph{Cf}, (< 0.5)$ may simply lack dust and gas. Meanwhile, objects like IRAS\,11119+3257, which has higher $\rm{\lambda_{edd}, (> -1.) }$ and lower $\emph{Cf},(< 0.6)$, could represent AGNs with sufficient energy to expel surrounding dust. In future work, we will explore the environments of these galaxies to investigate this interpretation.

\begin{figure}[!t]
    \centering
    \includegraphics[width=1.0\columnwidth, trim={15 0 10 0},clip]{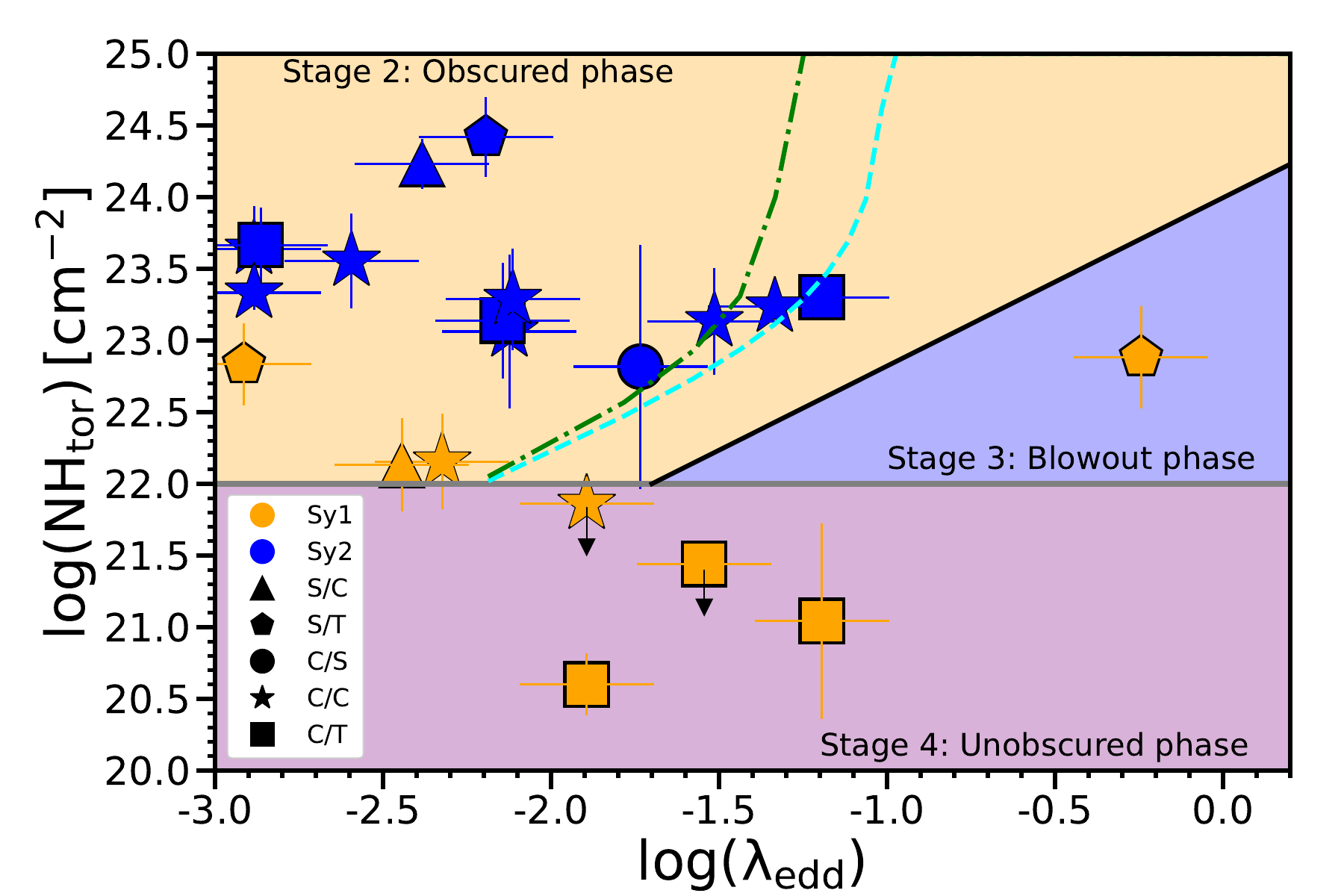}
    \caption{The $\rm{N_{H_{tor}}}$-$\lambda_{Edd}$ plane. The solid black line shows the effective Eddington limit introduced by \cite{Fabian06}. The dashed cyan and green dotted-dashed lines show this limit if the radiation pressure on dust is considered, assuming one and two times the galactic dust-gas ratio, respectively. The shaded areas show the three areas that cover the different stages according to the radiation-regulated unification model proposed by \cite{Ricci17}. The solid grey line shows the limit where the $N_{H_{tor}}$ could have an important contribution from the host galaxy. Symbols and colors follow the conventions used in the previous figures.}
    \label{fig:NH_vs_lambedd}
\end{figure}

\subsection{The column density as a function of AGN accretion state}

The $\rm{N_{H}-\lambda_{edd}}$ plane is used to explore the effect of the radiation pressure of the accretion disk on surrounding material \citep[e.g.,][]{Venanzi20, Alonso-Herrero21, Garcia-Bernete22}. In Figure \ref{fig:NH_vs_lambedd}, we present the $\rm{N_{H_{tor}}}$ against the Eddington ratio for the sources of our sample. The solid black line shows ``the effective Eddington limit'', introduced by \cite{Fabian06}. This limit considers the effect of the radiation pressure over the surrounding dust. It divides the plane into two regions known as the long-lived obscuration region and the forbidden region (or blowout region). The effective Eddington limit was reviewed considering the radiation-trapping and assuming different dust-to-gas ratios by \cite{Ishibashi18}. The radiation trapping could play an important role in the source where the obscuring material has a \emph{clumpy} distribution; in this case, the reprocessed radiation tends to leak out through lower-density channels (that is, the path of less resistance), reducing the effective optical depth. The cyan dashed line in Fig. \ref{fig:NH_vs_lambedd} shows the relationship assuming the galactic dust-to-gas ratio. \cite{Ishibashi18} found that an increase in the dust-to-gas ratio implies that the forbidden region increases, meaning the dust is short-lived at lower Eddington ratios. \cite{Ricci17} gave an evolutionary meaning to the $N_{H}-\lambda_{Edd}$ plane, where the AGN could move in this diagram during the life cycle. According to their recent work \cite{Ricci23}, there are four stages: 1) Accretion event, 2) Obscured phase, 3) Blowout phase, and 4) Unobscured phase. The radiation-regulated unification model agrees with the evolutionary sequence proposed by \cite{Krongold02} and is discussed in our previous section. 

The sources from our sample are located in the last three stages proposed by \cite{Ricci23}. IRAS11119+3257 is located in stage 3, where the \emph{Cf} and $\rm{N_{H_{tor}}}$ are consistent with a source where the accretion rate is so high that the obscuring material has been expelled. \cite{Tombesi15} reported the detection of a powerful ultra-fast outflow in the X-ray spectrum of this source using \emph{Suzaku} observations. They confirmed this result using \emph{NuSTAR} observations and reported that the wind launching radius is likely at a distance of r $\leq$ 16 $r_s$ from the central black hole and has a mass outflow rate of $\rm{\sim 0.5 - 2 \, M_{\odot} yr^{-1}}$ \citep{Tombesi17}. 

In the area defined for stage 2, we found that 76\% of our sources and 62\% are classified as Sy2, which is consistent with the prediction that obscured sources occupy this region with random inclination angles. Except for NGC\,6300 and IC\,4518W, all \emph{Cf}s are consistent with the mean value found by \cite{Ricci23} for this stage. As mentioned in the previous section, these sources might have more circumnuclear material or an AGN with lower energy incapable of removing the material. Finally, the grey solid line marks the $\rm{N_{H_{tor}} < 10^{22}}$ limit, below which absorption by extended dust lanes may become important \cite{Fabian08, Fabian09}. According to \cite{Ricci23}, in this area, there are sources in stage 4, where most of the obscuring material has been expelled, and therefore, these sources would be observed as unobscured AGNs. ES0\,141-G055, PG\,0804+761, Mrk\,1383, and Mrk\,1392 are the four Sy1 galaxies located in this region. The three sources that chose the X-C/MIR-TP baseline model have \emph{Cf}s consistent with the mean value found by \cite{Ricci23} for this region.

\section{Summary and Conclusions}
\label{sec:conclusion}
In this work, we used a simultaneous spectral fitting technique to study the properties of the dusty-gas torus in AGNs by analyzing \emph{NuSTAR} and \emph{Spitzer} spectra. Our sample consisted of 24 nearby AGNs ($\rm{z < 0.4}$) selected from our previous study \citep{Esparza-Arredondo21}: eight objects are type 1 Seyfert (Sy1), and sixteen are type 2 (Sy2). We utilized two and three X-rays and mid-IR models to fit the data, including the \emph{uxclumpy} and \emph{two-phases} models in our SFT, as an improvement over our previous work, enabling us to explore novel distributions of dust and gas.

Our technique allowed us to investigate whether the same structure could produce both X-ray reflection and mid-IR continuum. For this, we linked the half-opening angle and torus angular width values to the same value, as well as the inclination angles. This link between parameters implies fewer free parameters and, therefore, simplifies the baseline models as it guarantees that the data are not overfitted. From a technical point of view, some of our main findings are that we are capable of successfully achieving good fits and constraining most physical parameters for most of the sample sources. Furthermore, the simplified baseline models also helped break the degeneracy among some parameters, e.g., the column density and photon index.

While we acknowledge the need to enlarge the sample size for more robust results, our current findings strongly reinforce the evolution of the dusty gas around AGNs in several ways:

\begin{itemize}

    \item[$\circledast$] The mid-IR dusty torus inferred radius between 0.5 up to 16\,pc in our sample and dust mass in a range between $\rm{[10^2,10^7] M_{\odot}}$.

    \item[$\circledast$] We find a clear difference in the angular sizes of Sy1 and Sy2 galaxies. Most Sy1s show torus angular width below $\rm{\sim 25^{\circ}}$. This implies a covering factor below $\rm{\sim 0.5}$. On the other hand, Sy2s seem to be a mixed group with a broad distribution of torus angular width and covering factors. At X-rays wavelengths, the differences are even more evident; Sy1s show steep photon indices ($\rm{\Gamma>1.8}$) and low column densities of the torus ($\rm{N_{H}<10^{22}cm^{-2}}$) while Sy2s show large column densities ($\rm{N_{H}>10^{22}cm^{-2}}$) and a broader distribution of photon indices ($\rm{1.4<\Gamma<2.2}$). Therefore, Sy1s and Sy2s have different dusty-gaseous structures. 

    \item[$\circledast$] Our analysis shows that their luminosity and mass influence the structure and distribution of the dusty gas torus in AGNs. A positive correlation exists between the total dust mass and bolometric luminosity for type 2 Seyfert galaxies (Sy2). This relationship suggests that AGNs with higher luminosity might have more available dust in their torus for feeding or that AGNs power heats the dust, leading to more mid-IR emission.
    
    \item[$\circledast$] The \emph{clumpy}$-$\emph{clumpy} and \emph{clumpy}$-$\emph{two-phases} distribution is preferred for most sources (16 out of 23), although other combinations work better for some of the objects. It is important to note that the \emph{smooth} distribution of dust is preferred only for one object in this analysis. Therefore, dust is distributed in clumps or in a \emph{two-phases} medium, as suggested in previous works.
    
    \item[$\circledast$] Our best fit is obtained when the column density of the reflection component is linked to that of the line-of-sight; only NGC\,6300 prefers them not to be linked. This suggests that the gaseous counterpart of the torus is homogeneous with a flat azimuthal distribution profile. Indeed, most of our models prefer a slope of azimuthal distribution at mid-IR of $\rm{\alpha_{p}<0.5}$.
    
    \item[$\circledast$] Our sources are located in three different states of the radiation-regulated unification model, with most currently in an obscuration phase. In the context of the evolution sequence of Seyfert galaxies, these sources may contain a more significant amount of circumnuclear material, or they may have an AGN with lower energy that cannot remove the material due to different factors. In future work, we will analyze the environment of these sources to explore possible galaxy interactions, intrinsic properties, and environmental conditions.
 
 \end{itemize}

It is important to remark that, among the parent sample of 36 AGNs studied by \citet{Esparza-Arredondo21}, three sources could not be fitted with any model at X-rays, and four sources could not be fitted with any model at mid-IR. This indicates that further complexity from the models at both wavelengths is needed in a non-negligible fraction of AGNs. Additionally, four AGNs (all Sy1) prefer the disk$+$wind model at mid-IR. Even though all of these sources were excluded due to the lack of a corresponding model at X-rays, this indicates that the distribution of dust might vary for some objects. Models of X-rays with this distribution could help determine if the X-ray reflection is also produced by material associated with this disk$+$wind structure.

\begin{acknowledgements}
     The authors thank the anonymous referee for the careful reading and constructive suggestions that improved the paper. D.E.A. acknowledges financial support from MICINN through the Juan de la Cierva program and SECIHTI through the Estancias postdoctorales por México EPM(1) 2024 (CVU:592884). D.E.A. and C.R.A. acknowledge support from the Agencia Estatal de Investigaci\'on of the Ministerio de Ciencia, Innovaci\'on y Universidades (MCIU/AEI) under the grant ``Tracking active galactic nuclei feedback from parsec to kiloparsec scales'', with reference PID2022-141105NB-I00 and the European Regional Development Fund (ERDF).
    D.E.A. and B.G.L. acknowledge support from the Spanish Ministry of Science and Innovation through the Spanish State Research Agency (AEI-MCINN/10.13039/501100011033) through grants "Participation of the Instituto de Astrofísica de Canarias in the development of HARMONI: D1 and Delta-D1 phases with references PID2019-107010GB100 and PID2022-140483NB-C21 and the Severo Ochoa Program 2020-2023 (CEX2019-000920-S). 
    OGM acknowledges financial support from PAPIIT UNAM IN109123 and ”Ciencia de Frontera” CONAHCyT CF2023-G100. 
    This research is mainly funded by the UNAM PAPIIT project IN109123 and CONAHCyt Ciencia de Frontera project CF-2023-G-100 (PI OG-M). IGB acknowledges support from STFC through grants ST/S000488/1 and ST/W000903/1.
    A.A.H. acknowledges financial support from the grant PID2021-124665NB-I00, funded by the Spanish Ministry of Science and Innovation and the State Agency of Research MCIN/AEI/10.13039/501100011033 PID2021-124665NB-I00, and by ERDFA's Way of Making Europe.
    IGB acknowledges support from STFC through grants ST/S000488/1 and ST/W000903/1.
    This work made use of data from the \emph{NuSTAR} mission, a project led by CalTech, managed by JPL, and funded by NASA. We thank the \emph{NuSTAR} Operations, Software and Calibration teams for support with the execution and analysis of these observations. This research has made use of the \emph{NuSTAR} Data Analysis Software (NuSTARDAS) jointly developed by the ASI Science Data Center (ASDC, Italy) and CalTech. This work is based in part on observations made with the \emph{Spitzer} Space Telescope, operated by the Jet Propulsion Laboratory, California Institute of Technology, under a contract with NASA.
\end{acknowledgements}

\onecolumn
\begin{appendix}

\section{Equations to derive dust mass and covering factor parameters.}
We integrate the density distribution of dust over the dust volume to obtain the $\rm{M_{dust}}$. This density distribution depends on different parameters according to the baseline model. Using our notation for each parameter, the equations to compute the $\rm{M_{dust}}$ are as follows:

\begin{itemize}
        \item X-S/MIR-S, X-C/MIR-S, X-S/MIR-TP, and X-C/MIR-TP baseline models:
    \begin{equation}
            M_{dust} = C  \int_0^{\pi/2} e^{- \gamma Cos \theta} cos \theta d \theta   \int_1^{R_{out}/R_{in}} r^{\alpha_r+2} dr
    \end{equation}   
    \item X-S/MIR-C and X-C/MIR-C baseline models
    \begin{equation}
        M_{dust} = C  \int_0^{\pi/2} e^{-\theta^2/\sigma_{tor}^2} cos \theta d \theta \int_1^{R_{out}/R_{in}} r^{-\alpha_r+2} dr
    \end{equation}
    where $C = 4.* \pi*mh*N_{H}*\tau_{\nu}*R_{in}^2$; $N_{H} = 1.9\times 10^{21}$; and $mh = 1.76 \times 10^{-27}$.
\end{itemize}

We calculate the $\emph{Cf}$ as the unity minus the escape probability; using our notation, the equations used are:

\begin{itemize}
        \item X-S/MIR-S, X-C/MIR-S, X-S/MIR-TP, and X-C/MIR-TP baseline models:
    \begin{equation}
            \emph{Cf} = 1. -  \int_0^{\pi/2} e^{-\tau_{9.7} e^{\gamma Cos \theta}} cos \theta d \theta
    \end{equation}   
    \item X-S/MIR-C and X-C/MIR-C baseline models
    \begin{equation}
        \emph{Cf} = 1.-  \int_0^{\pi/2} e^{-N0 e^{\theta^2/\sigma_{tor}^2}} cos \theta d \theta
    \end{equation}
    where $C = 4.* \pi*mh*N_{H}*\tau_{\nu}*R_{in}^2$; $N_{H} = 1.9\times 10^{21}$; and $mh = 1.76 \times 10^{-27}$. 
\end{itemize}

In the case of the C-C and S-C baseline models the $\rm{M_{dust}}$ is obtained using the $N0$, $\sigma_{tor}$, $Y$, and $\tau_{\nu}$ parameters. Meanwhile, the $\rm{M_{dust}}$ for the X-S/MIR-S, X-C/MIR-S, X-S/MIR-T, and X-C/MIR-T baseline models is derived through the $\alpha_p$, $\alpha_q$, Y0, and $\tau_{\nu}$ parameters. 

\clearpage
\twocolumn

\section{Specral fits}
\label{app:Spectra_fits}
\hspace{10cm}

Final spectral fits of the sources of our sample. The orange solid lines are the best fit obtained from the X-S/MIR-C, X-S/MIR-TP, X-C/MIR-S, X-C/MIR-C, and X-C/MIR-TP baseline models. For each galaxy, the \emph{Spitzer} spectrum is shown with blue points (left panel), and the \emph{NuSTAR} spectra are displayed with blue and cyan points (right panels). The magenta and green dashed lines show the absorbed power law and the reprocessed components, respectively. The lower panels display the residuals between the data and the best-fit model. The data and residual errors are shown as shaded areas. 

\begin{figure}[!b]
    \includegraphics[width=0.48\columnwidth]{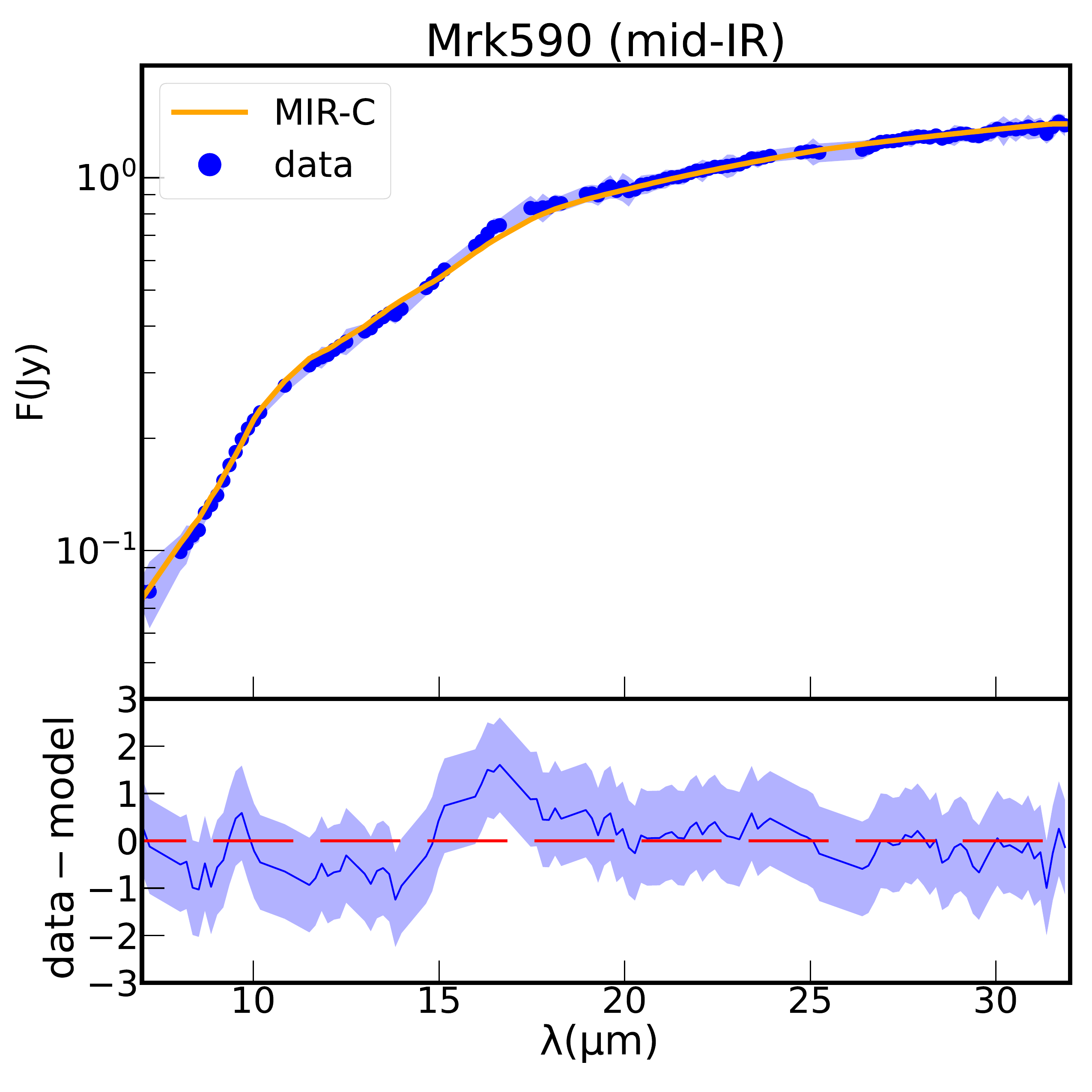}
    \includegraphics[width=0.48\columnwidth]{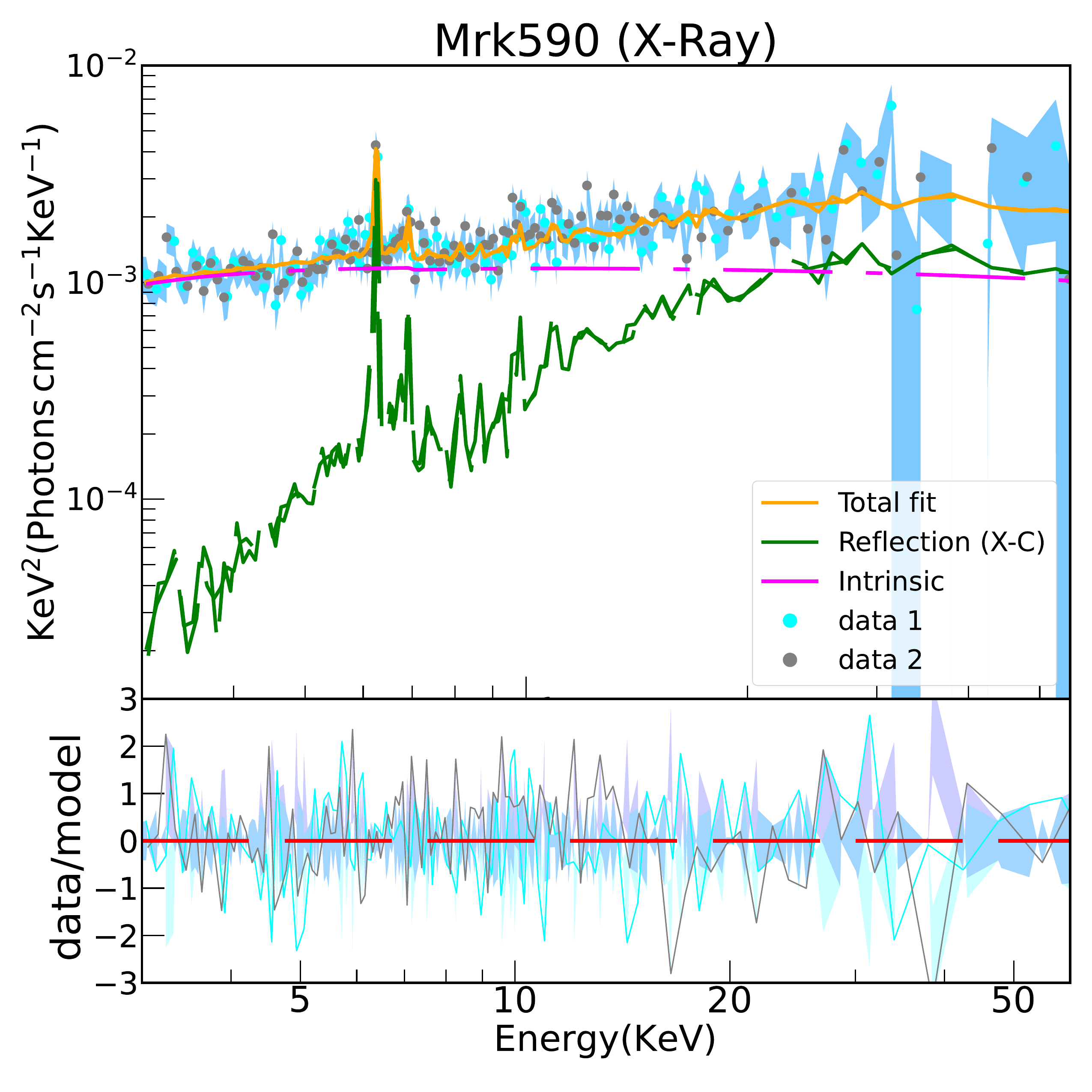}
    \includegraphics[width=0.48\columnwidth]{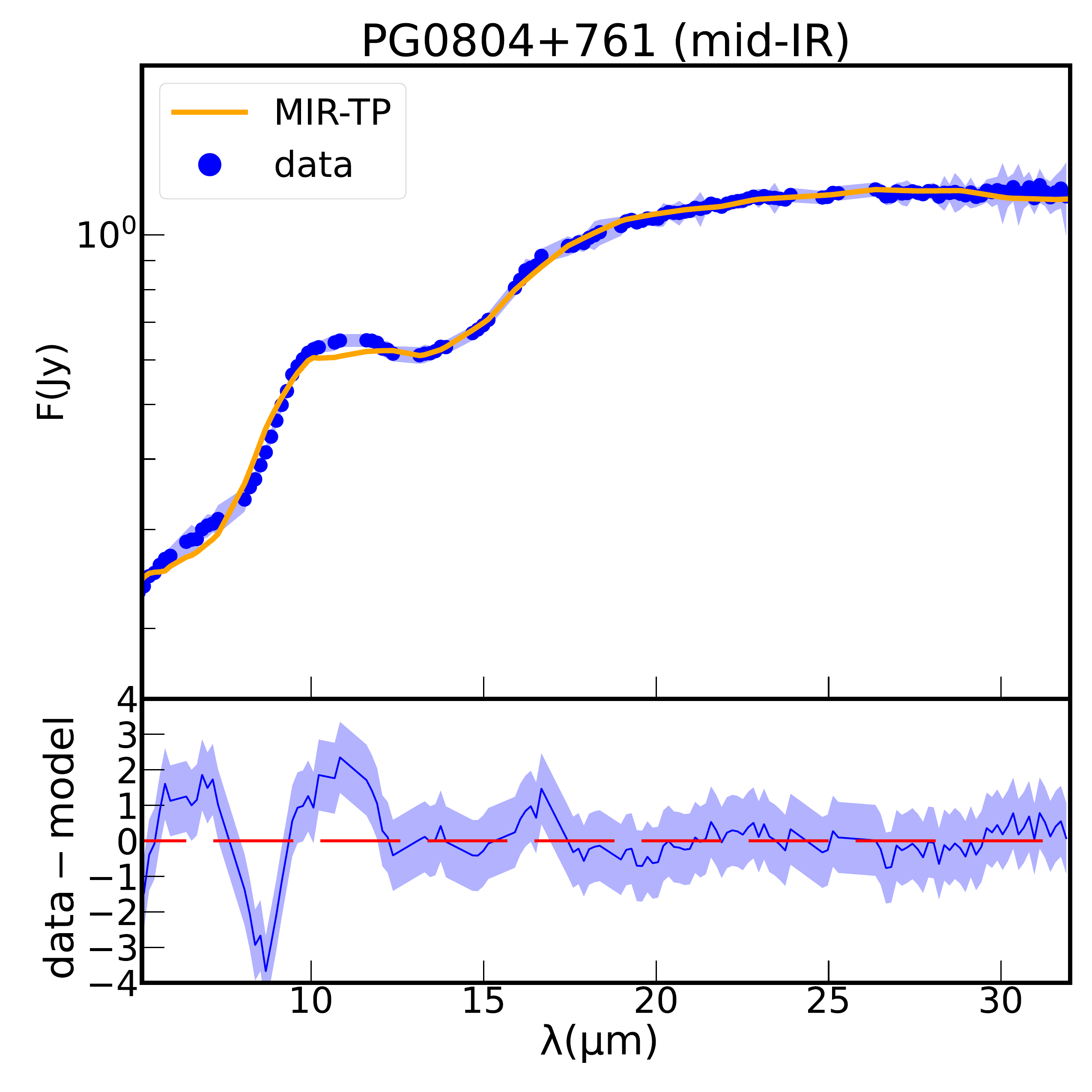}
    \includegraphics[width=0.48\columnwidth]{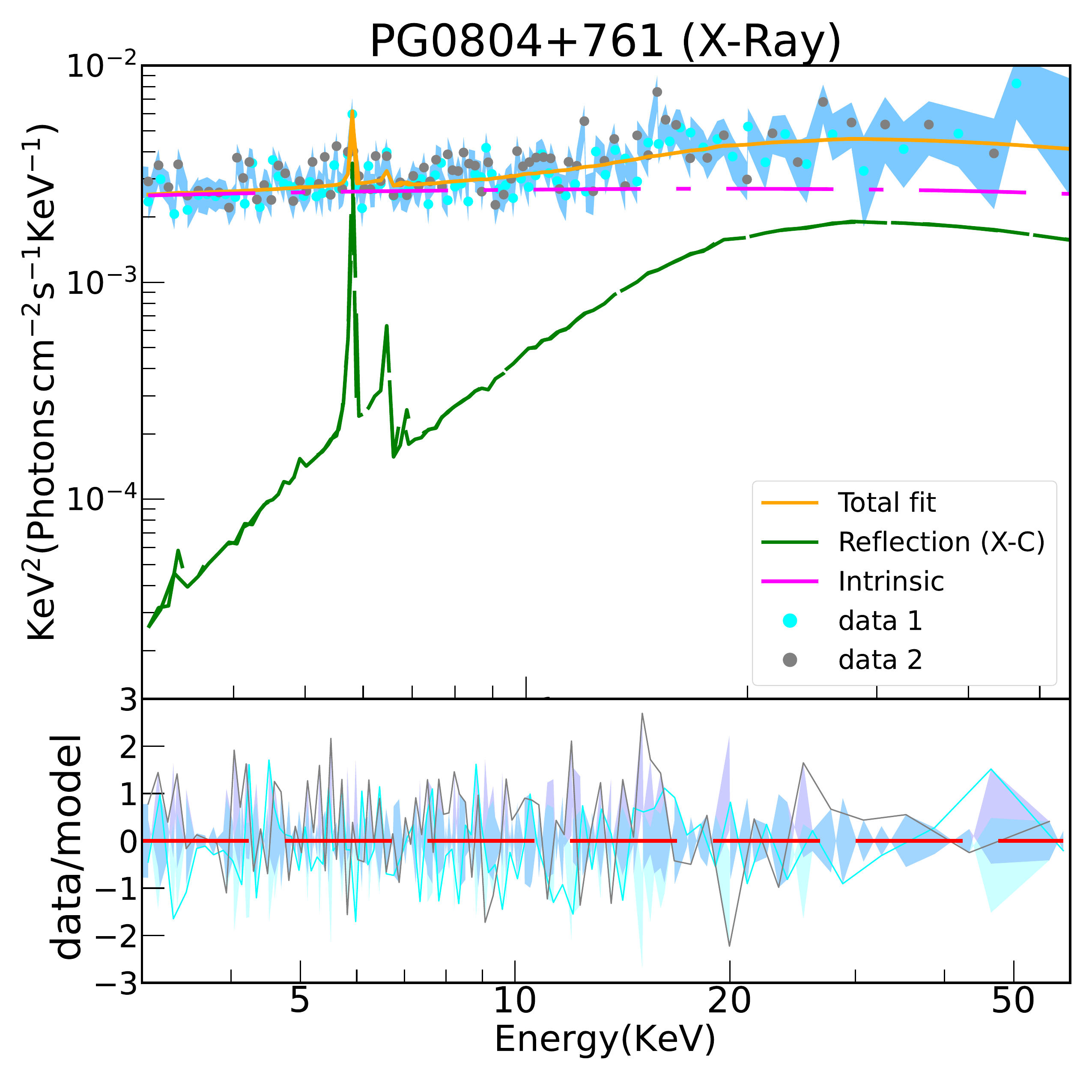}
     \includegraphics[width=0.48\columnwidth]{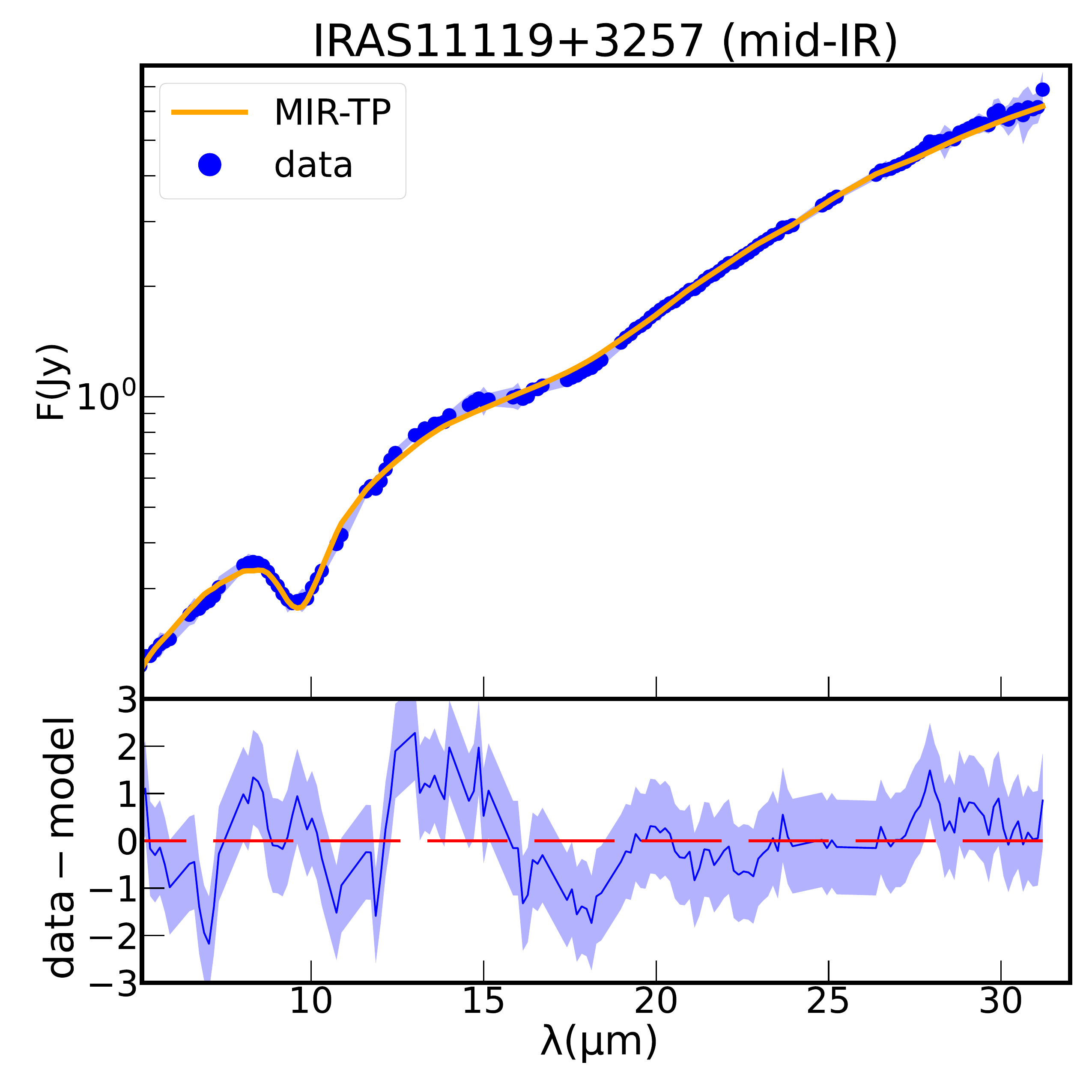}
    \includegraphics[width=0.48\columnwidth]{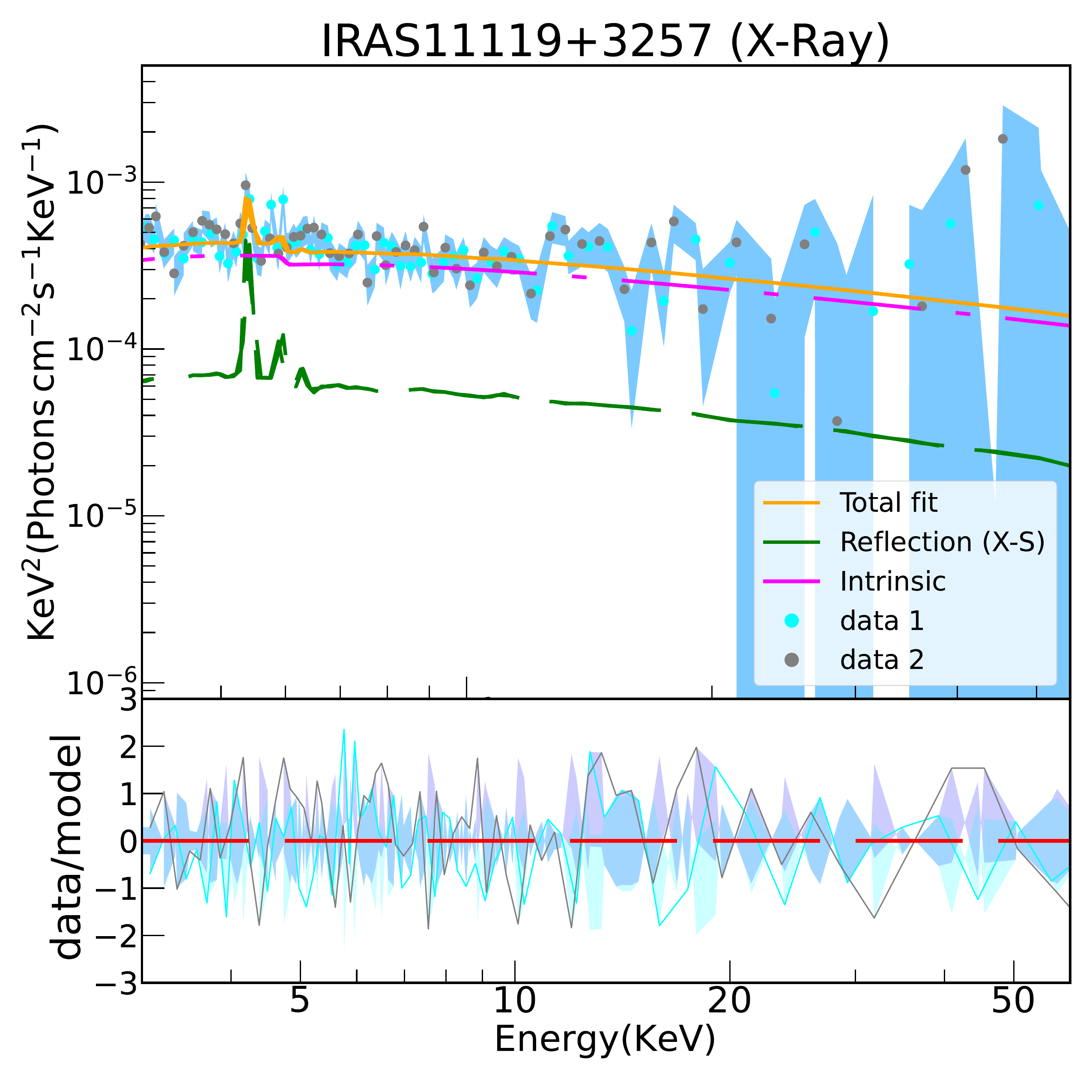}
    \includegraphics[width=0.48\columnwidth]{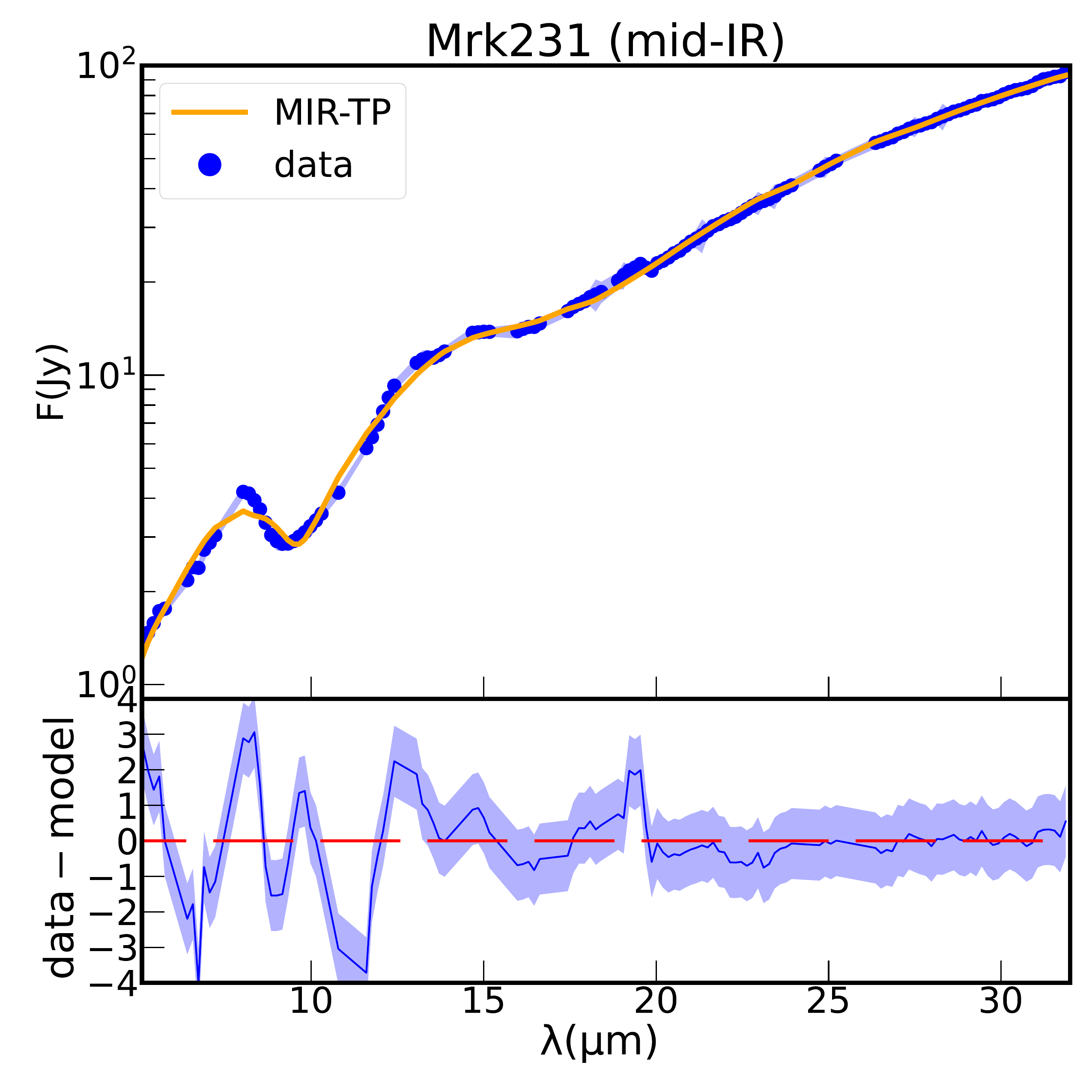}
    \includegraphics[width=0.48\columnwidth]{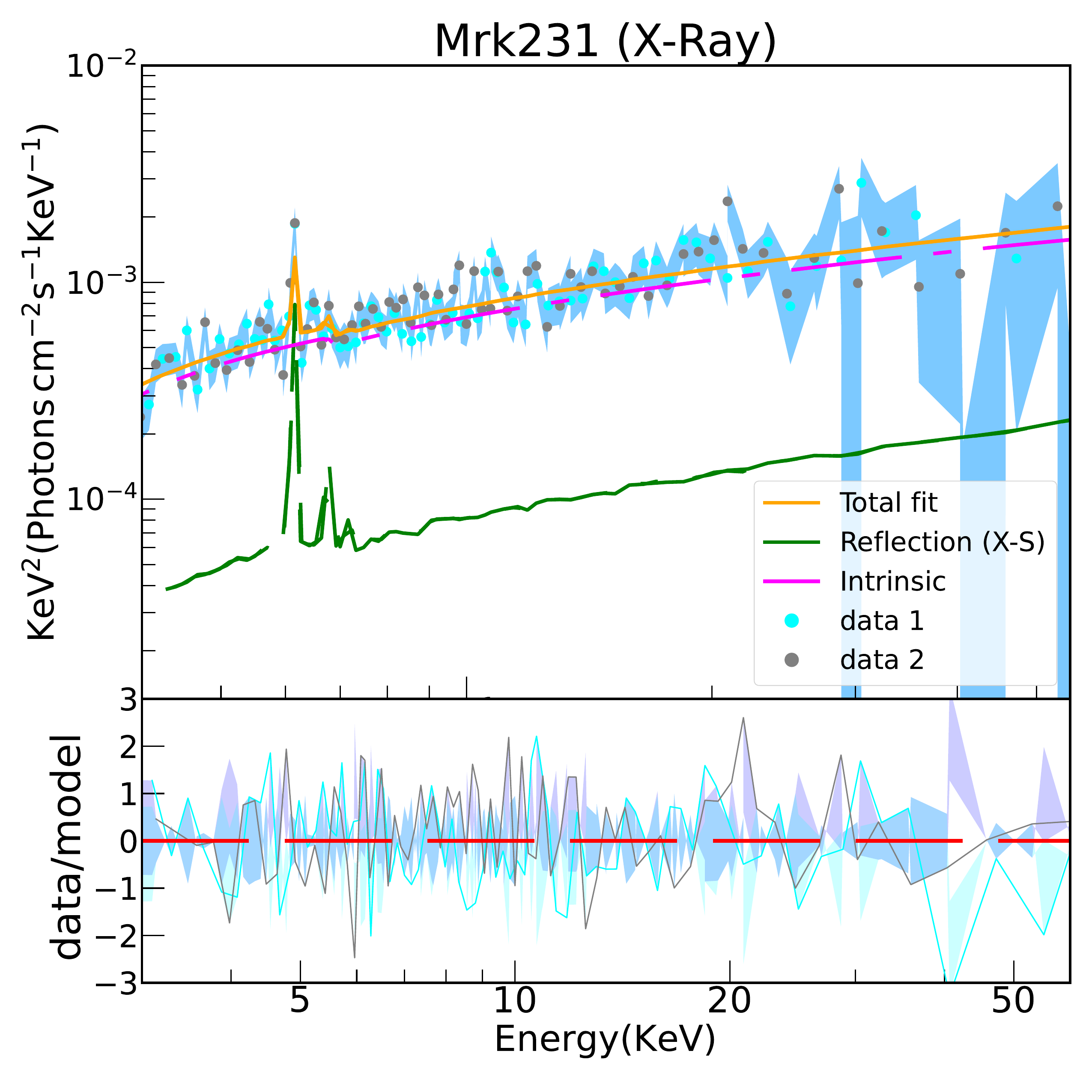}
    \caption{See the description at the beginning of this section.}
    \label{fig:Fits_1}
\end{figure}

\begin{figure}[!b]
    \includegraphics[width=0.48\columnwidth]{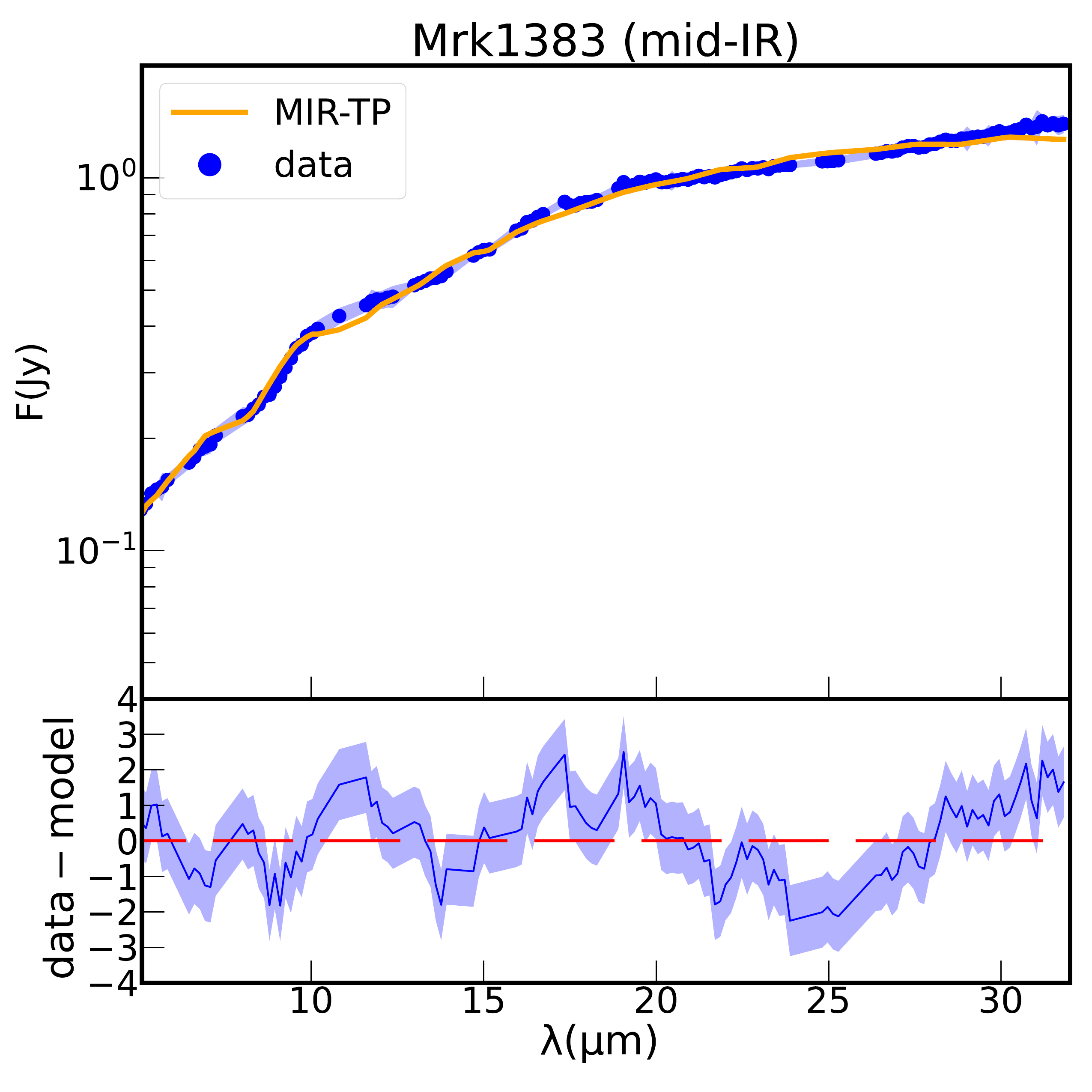}
    \includegraphics[width=0.48\columnwidth]{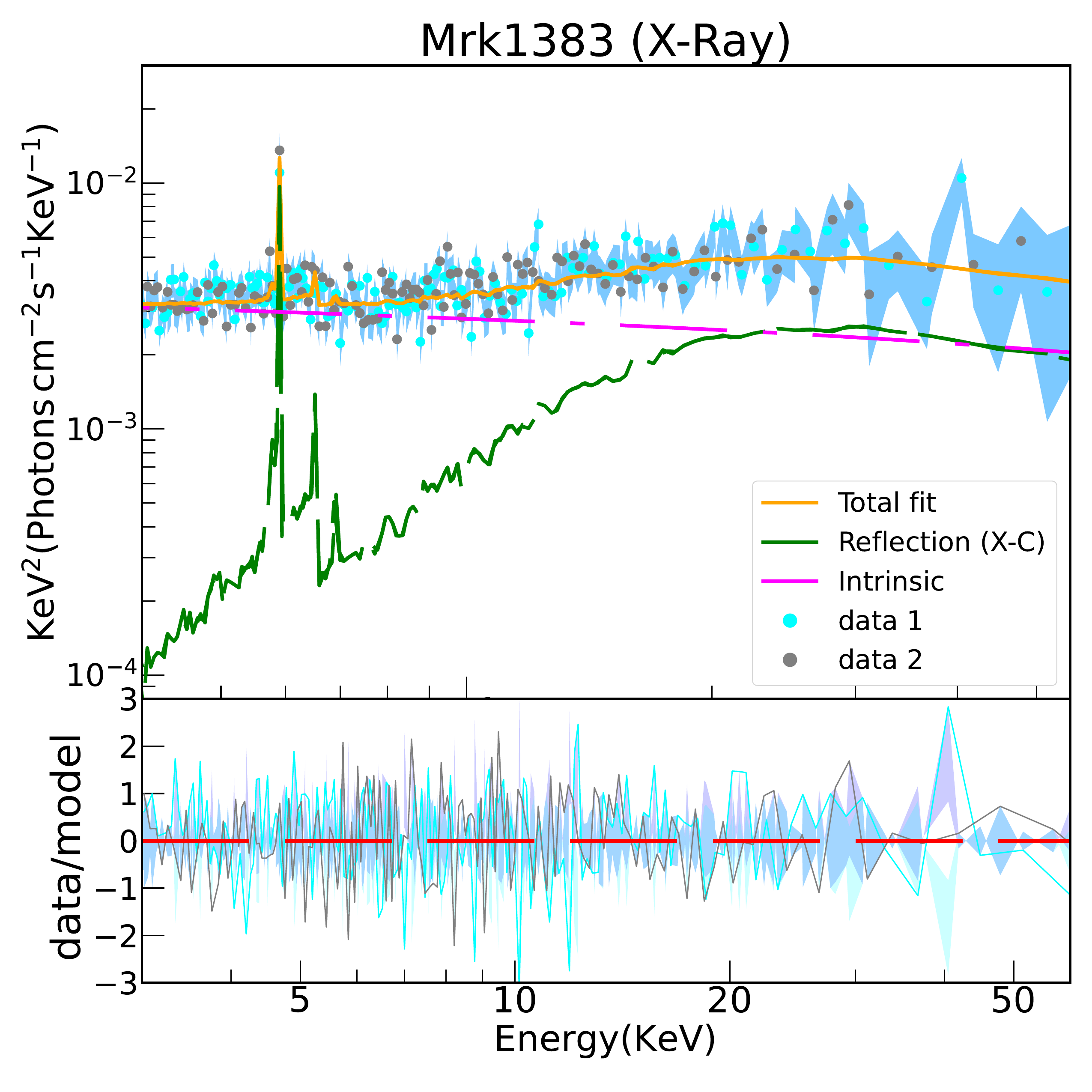}
    \includegraphics[width=0.48\columnwidth]{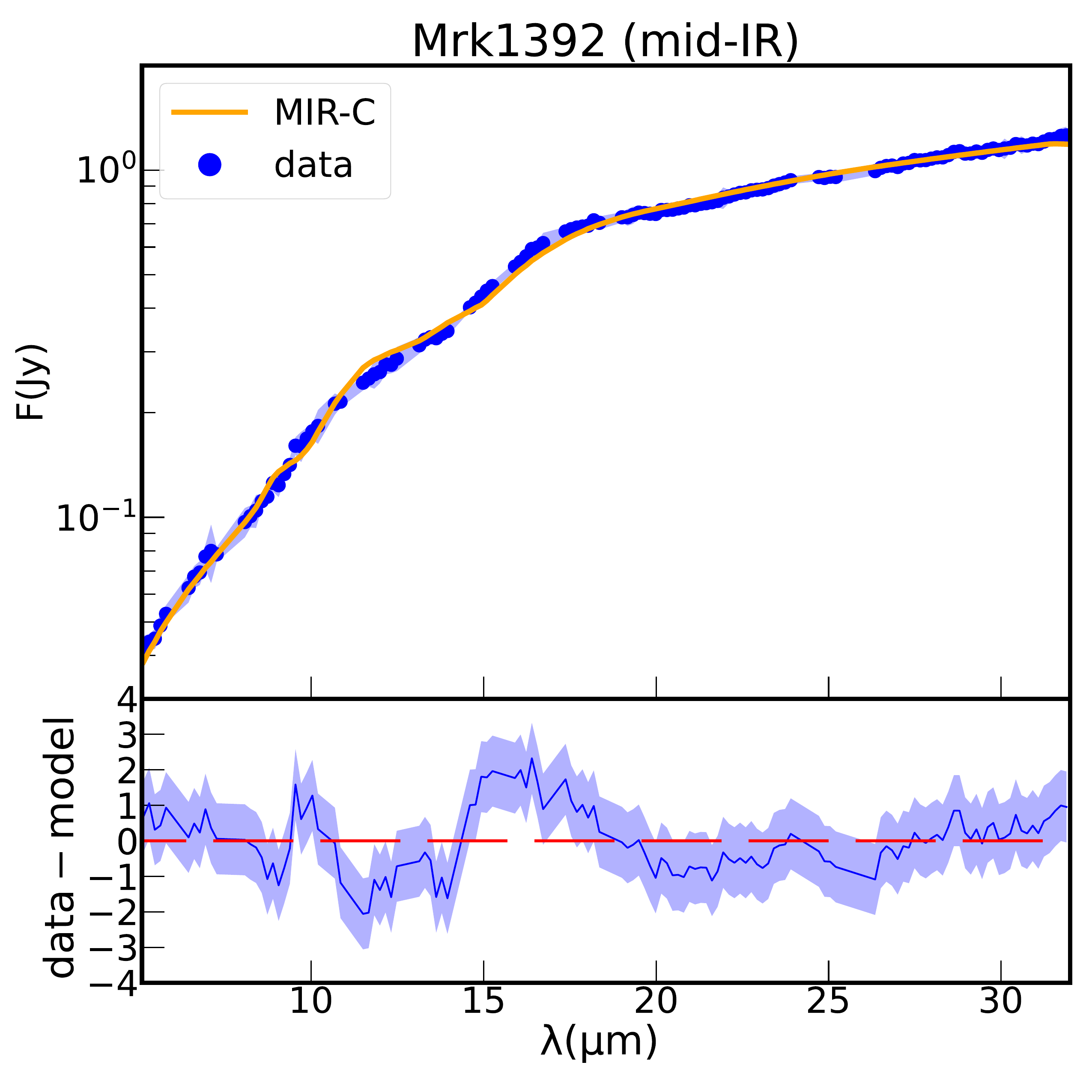}
    \includegraphics[width=0.48\columnwidth]{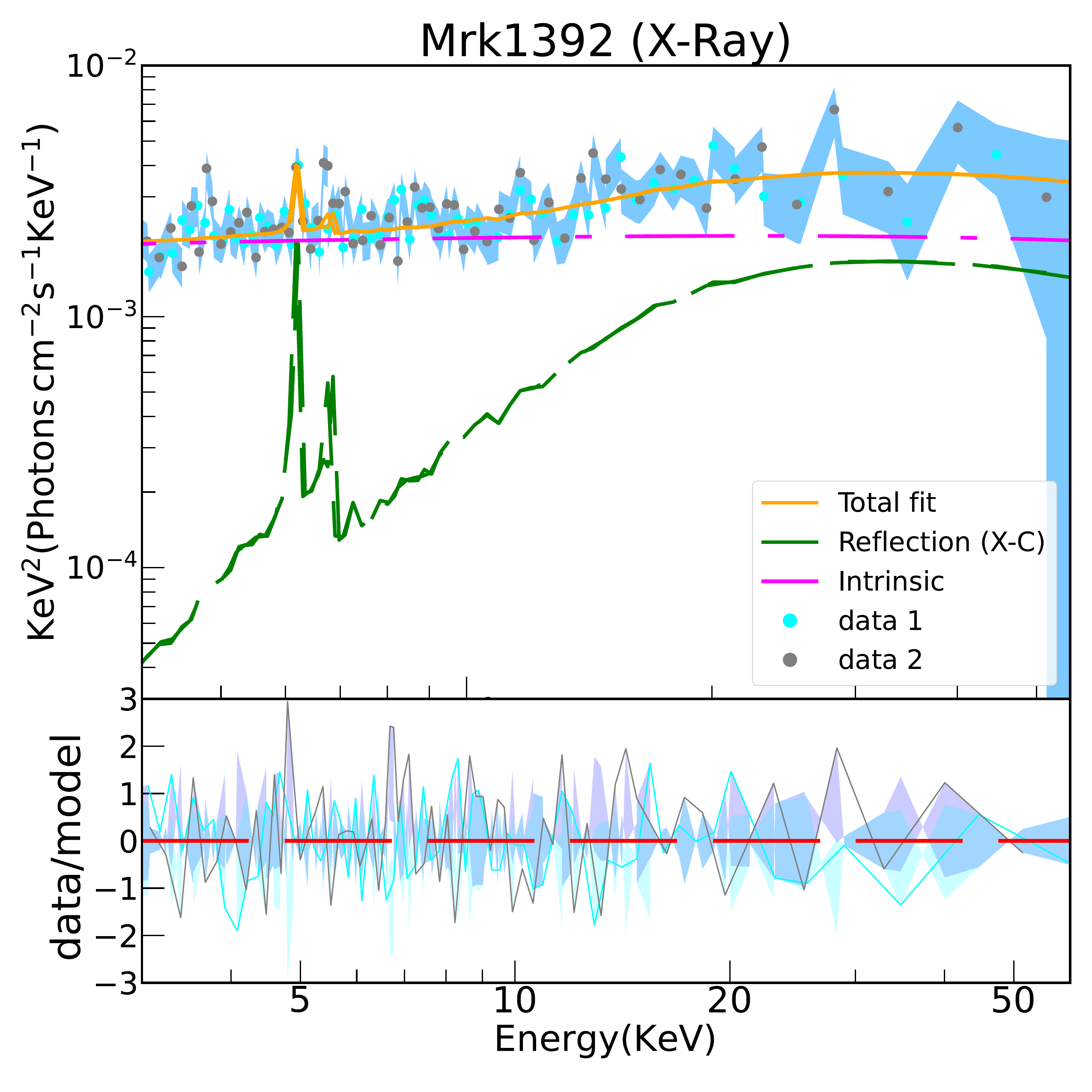}
    \includegraphics[width=0.48\columnwidth]{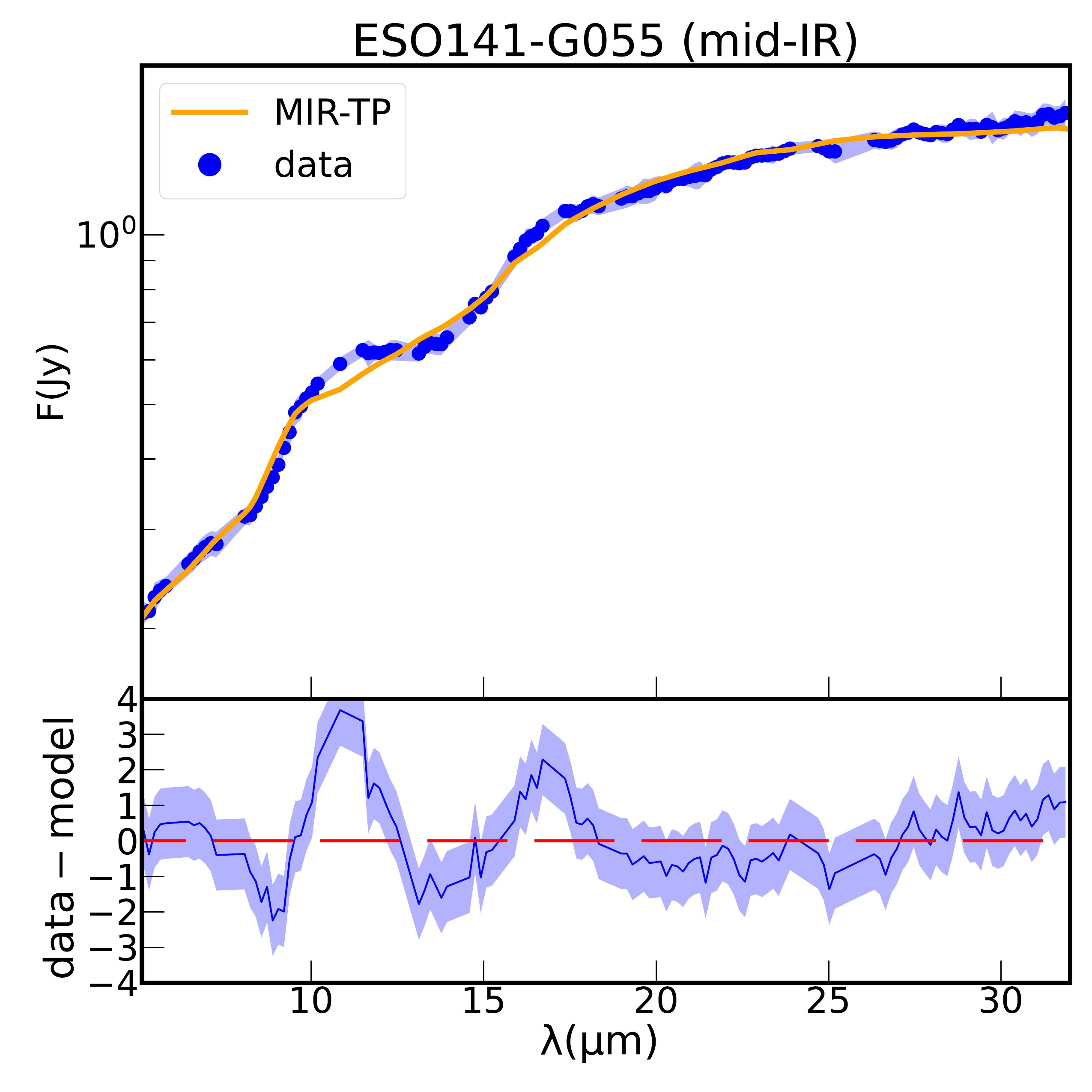}
    \includegraphics[width=0.48\columnwidth]{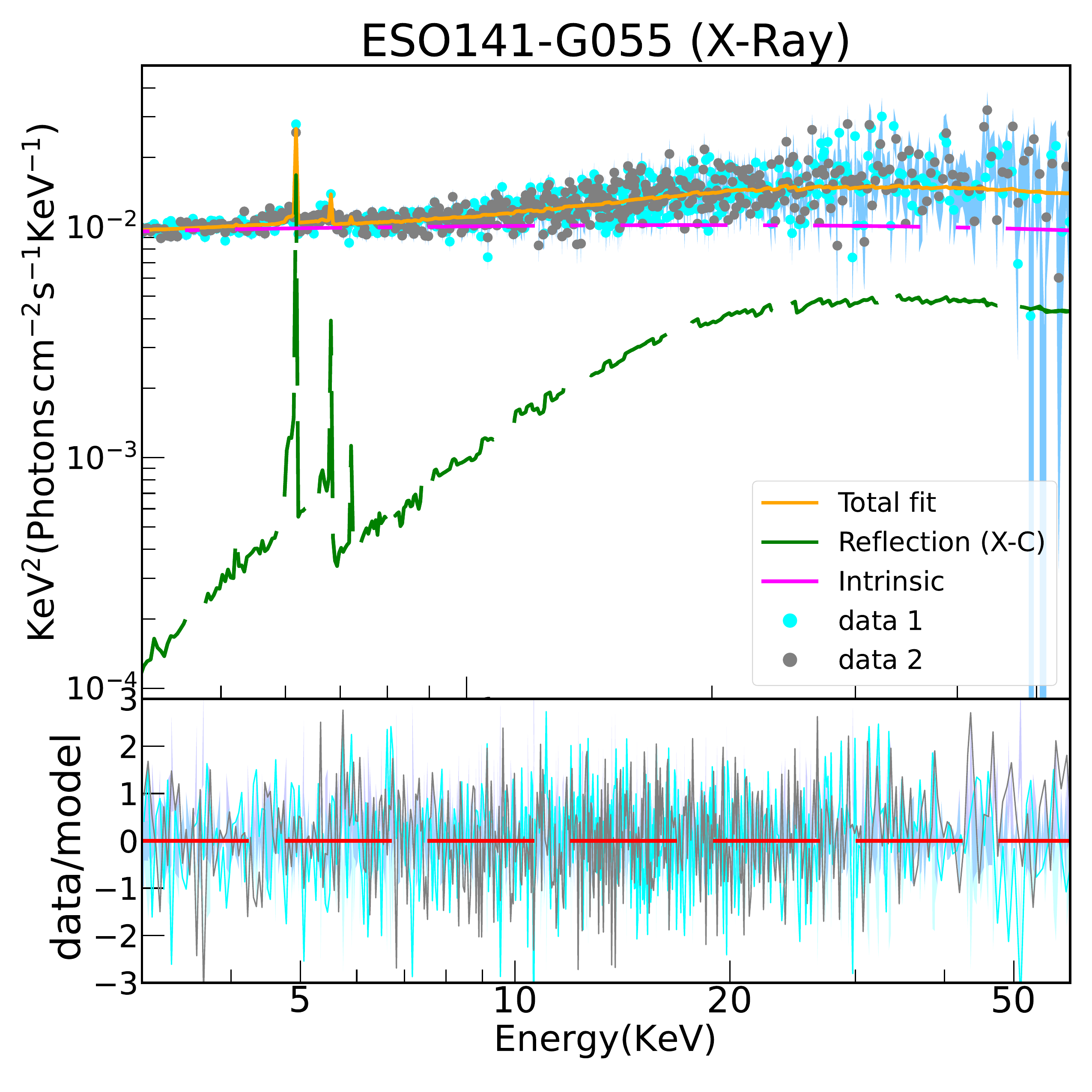}
    \includegraphics[width=0.48\columnwidth]{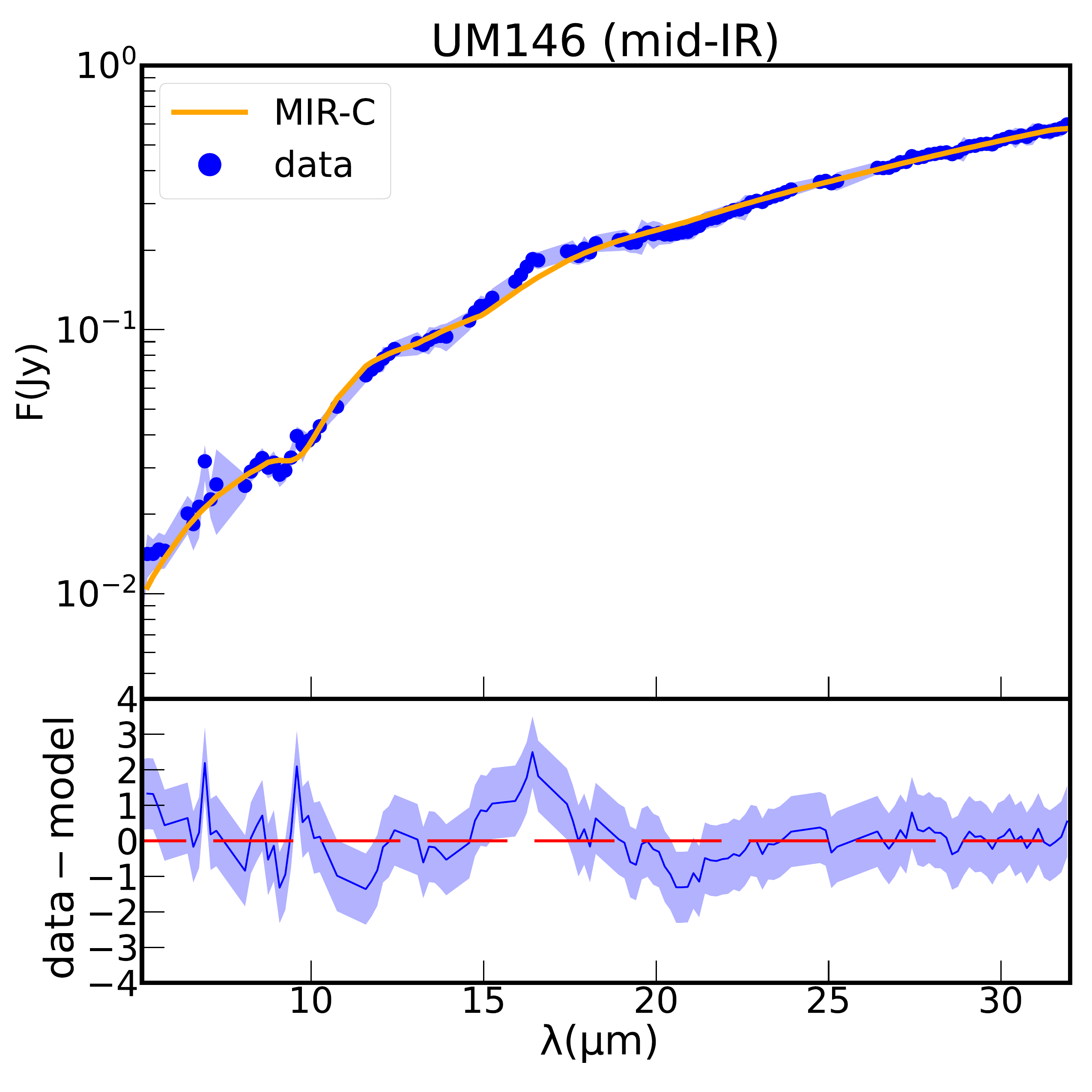}
    \includegraphics[width=0.48\columnwidth]{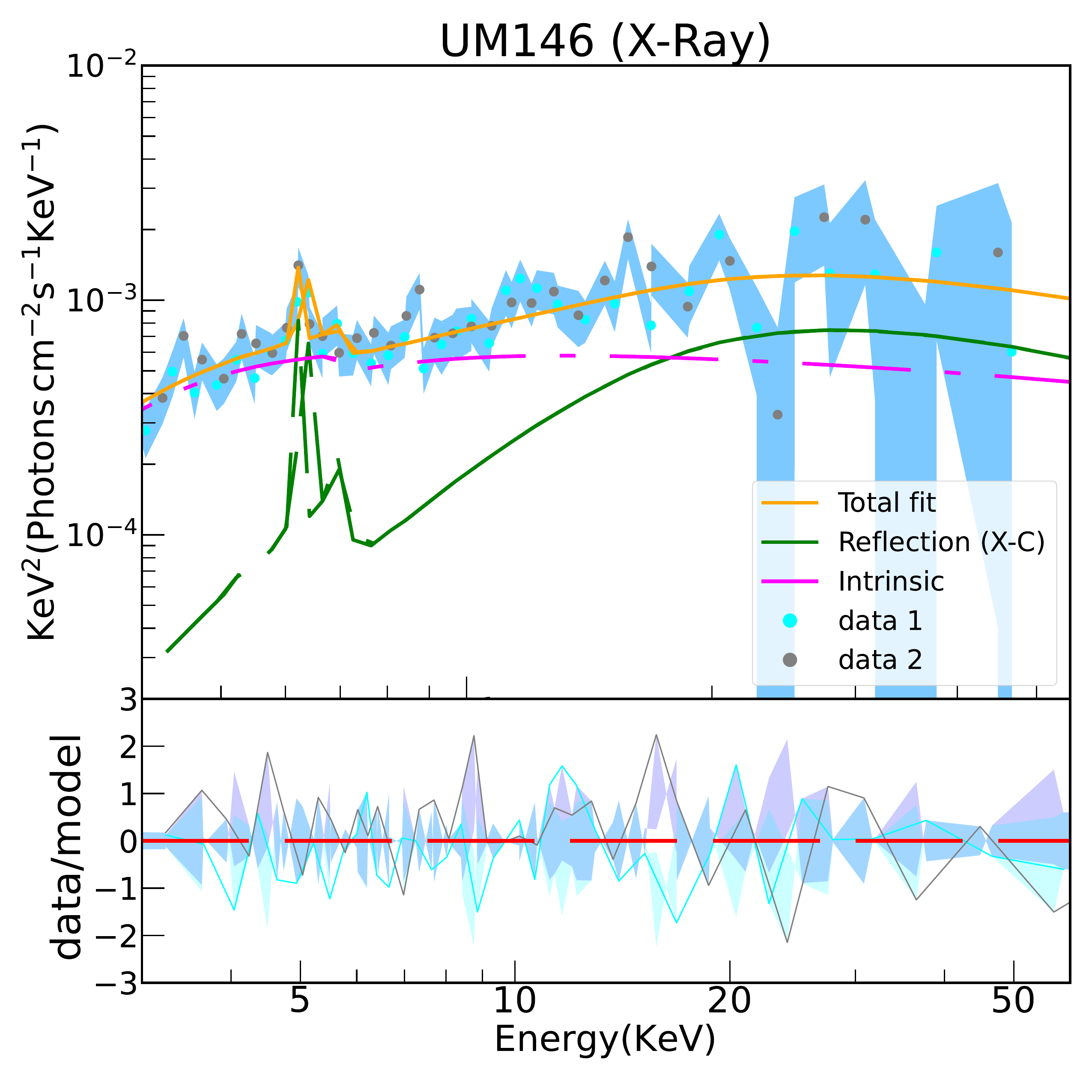}
    \includegraphics[width=0.48\columnwidth]{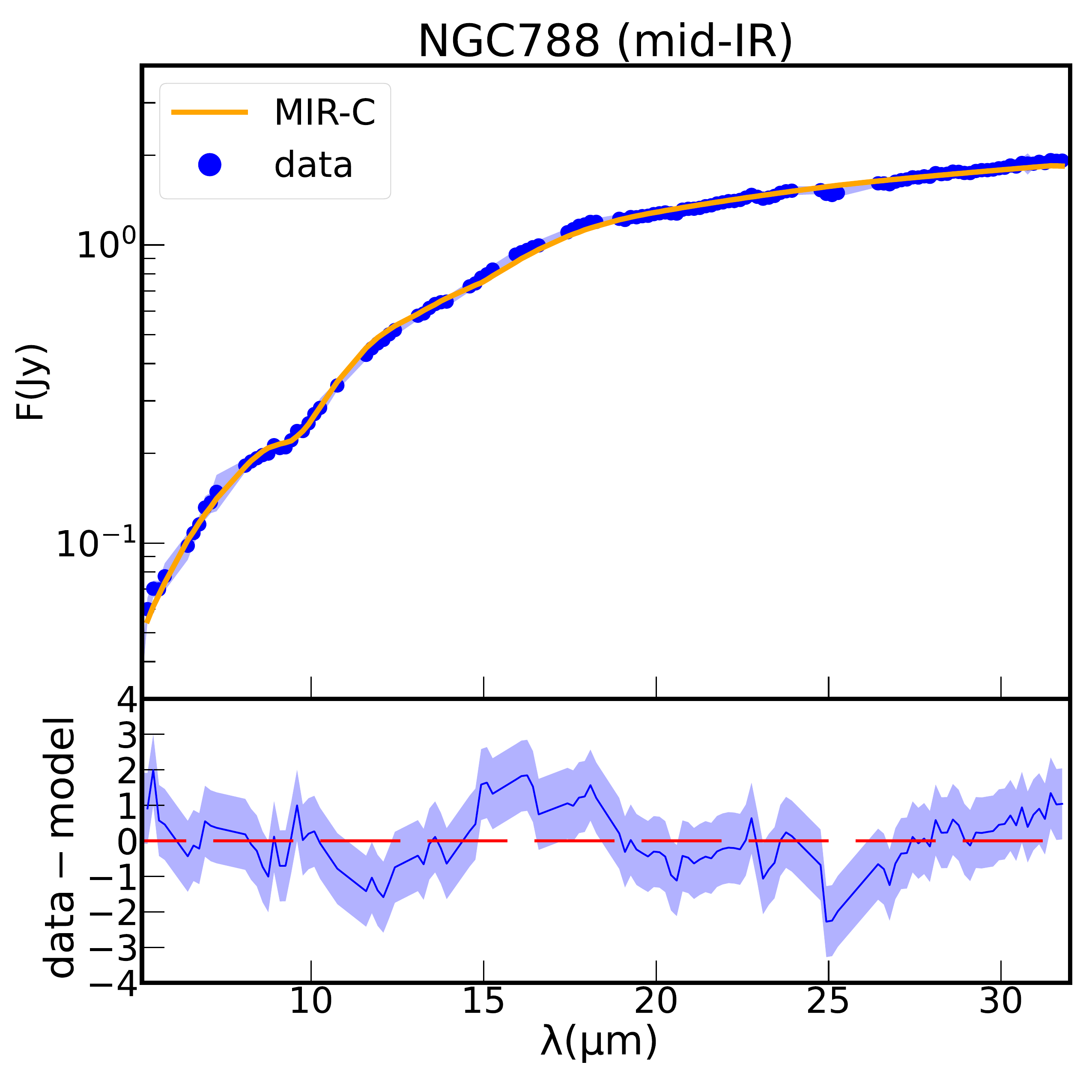}
    \includegraphics[width=0.48\columnwidth]{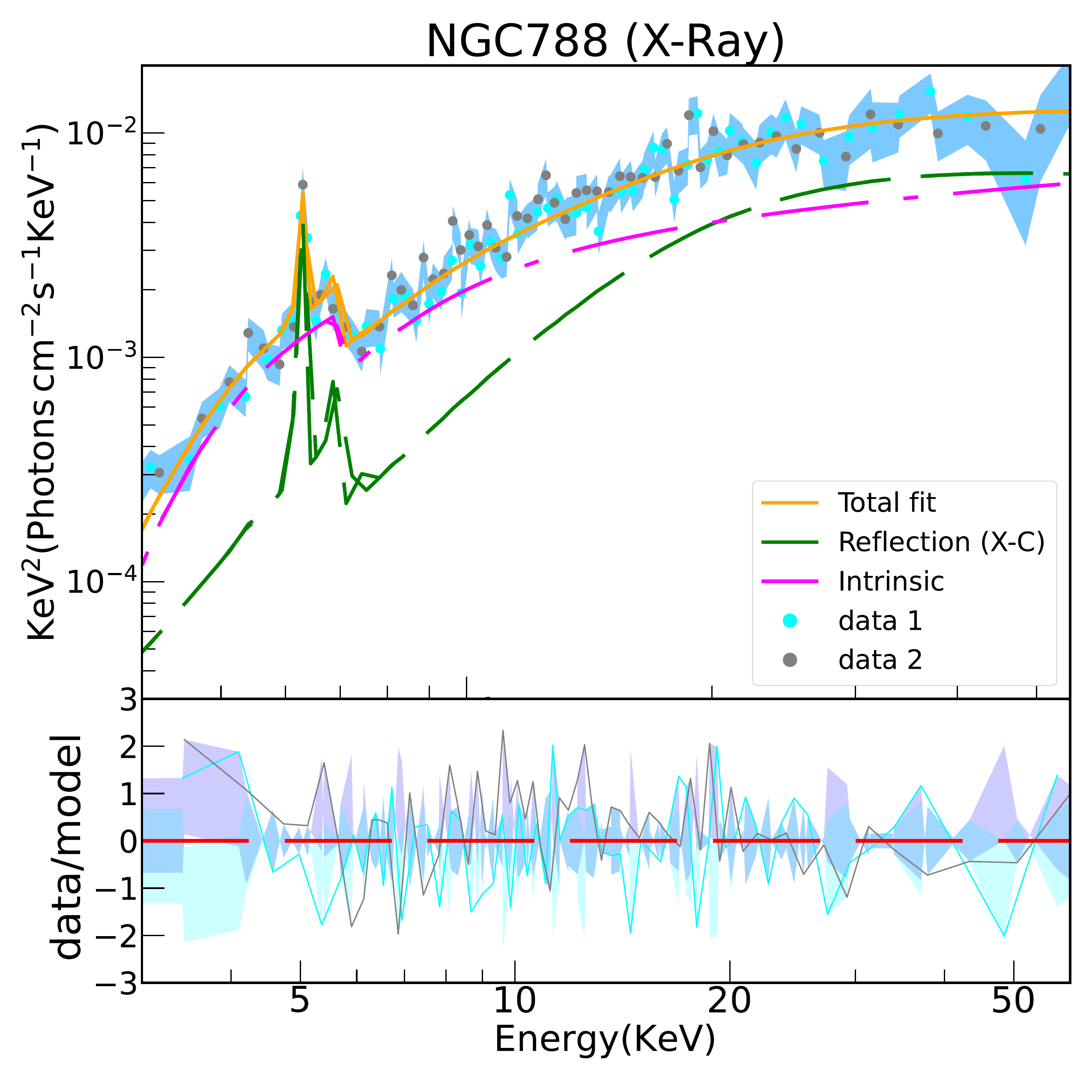}
    \caption{See the description at the beginning of this section.}
    \label{fig:Fits_2}
\end{figure}

\begin{figure}[!b]
    \includegraphics[width=0.48\columnwidth]{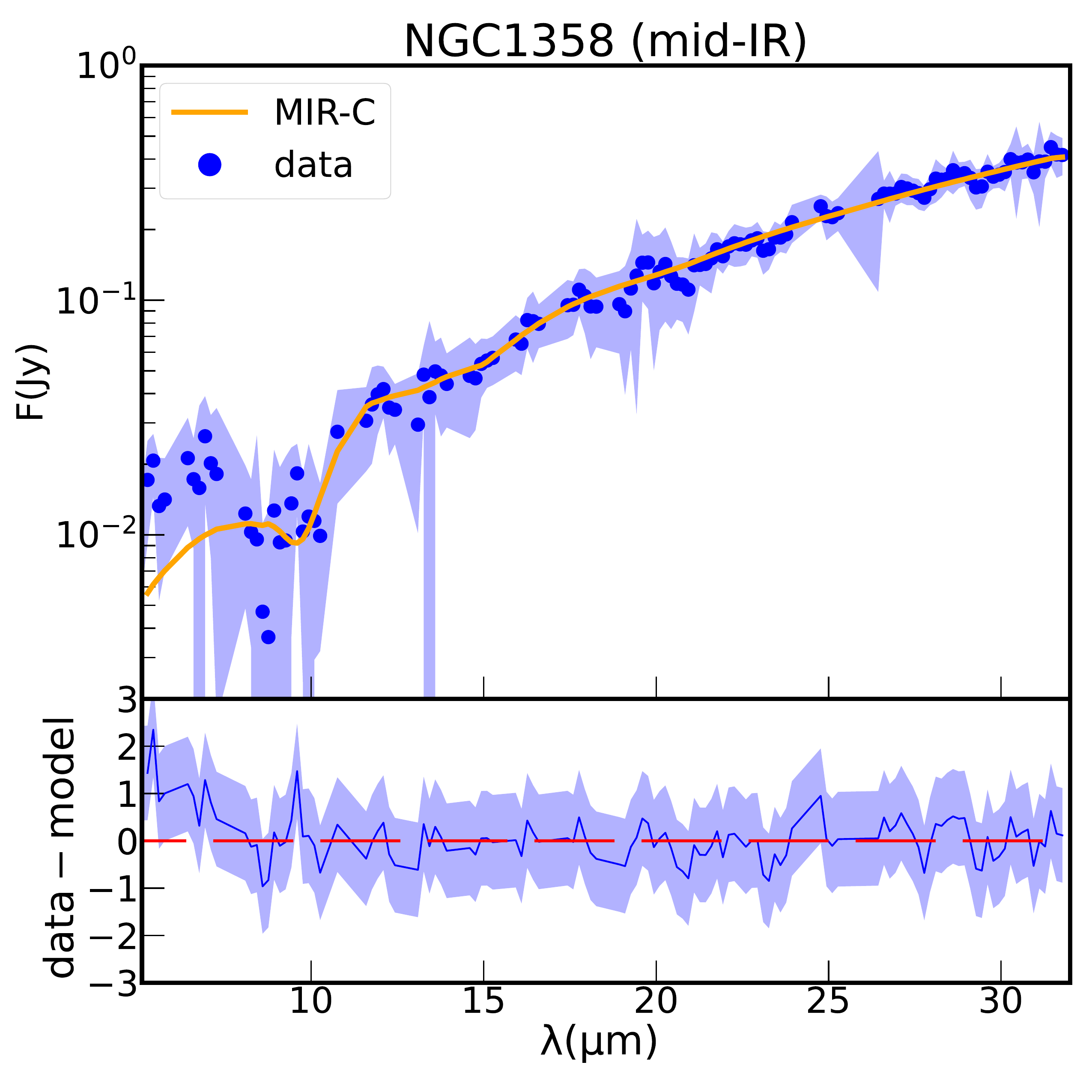}
    \includegraphics[width=0.48\columnwidth]{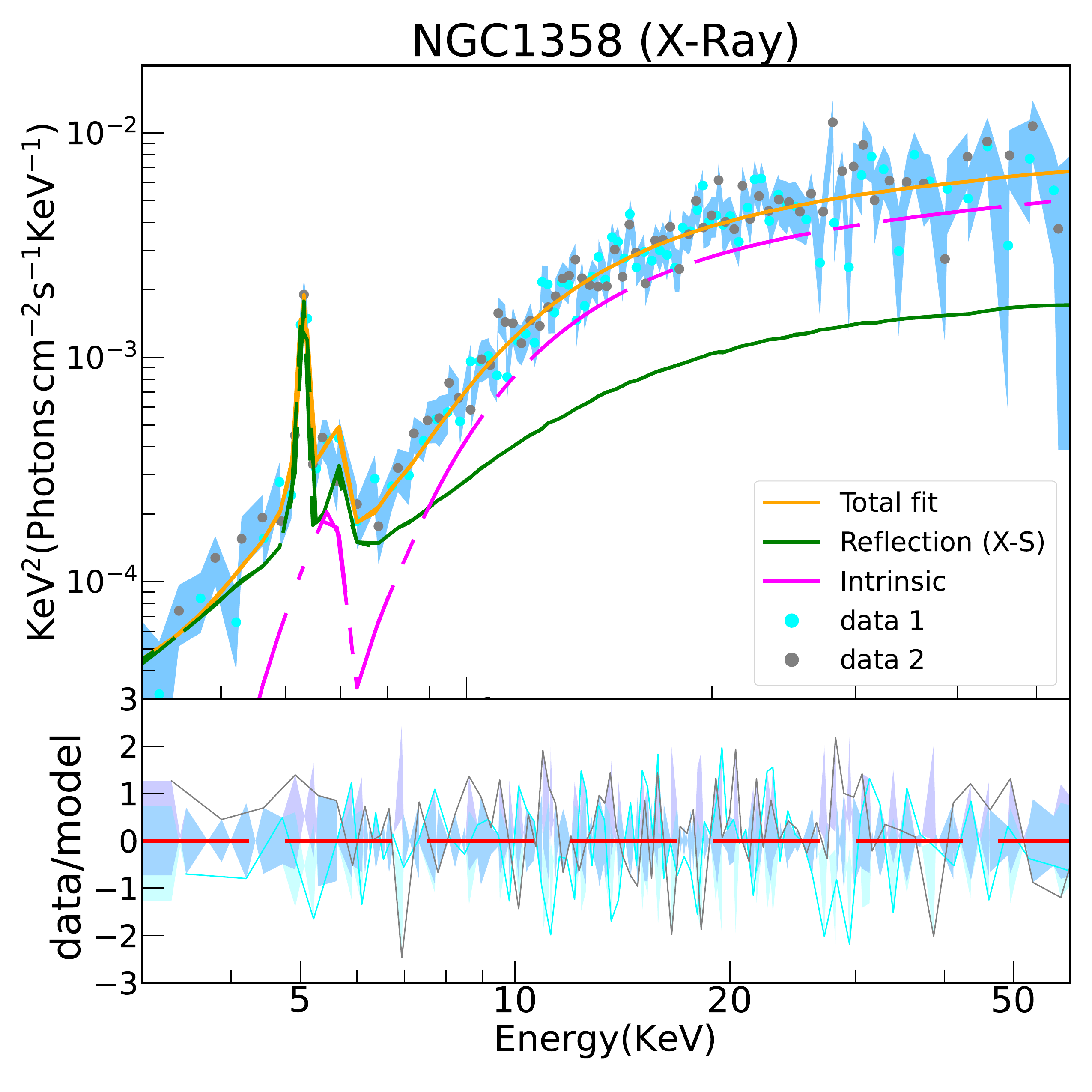}
    \includegraphics[width=0.48\columnwidth]{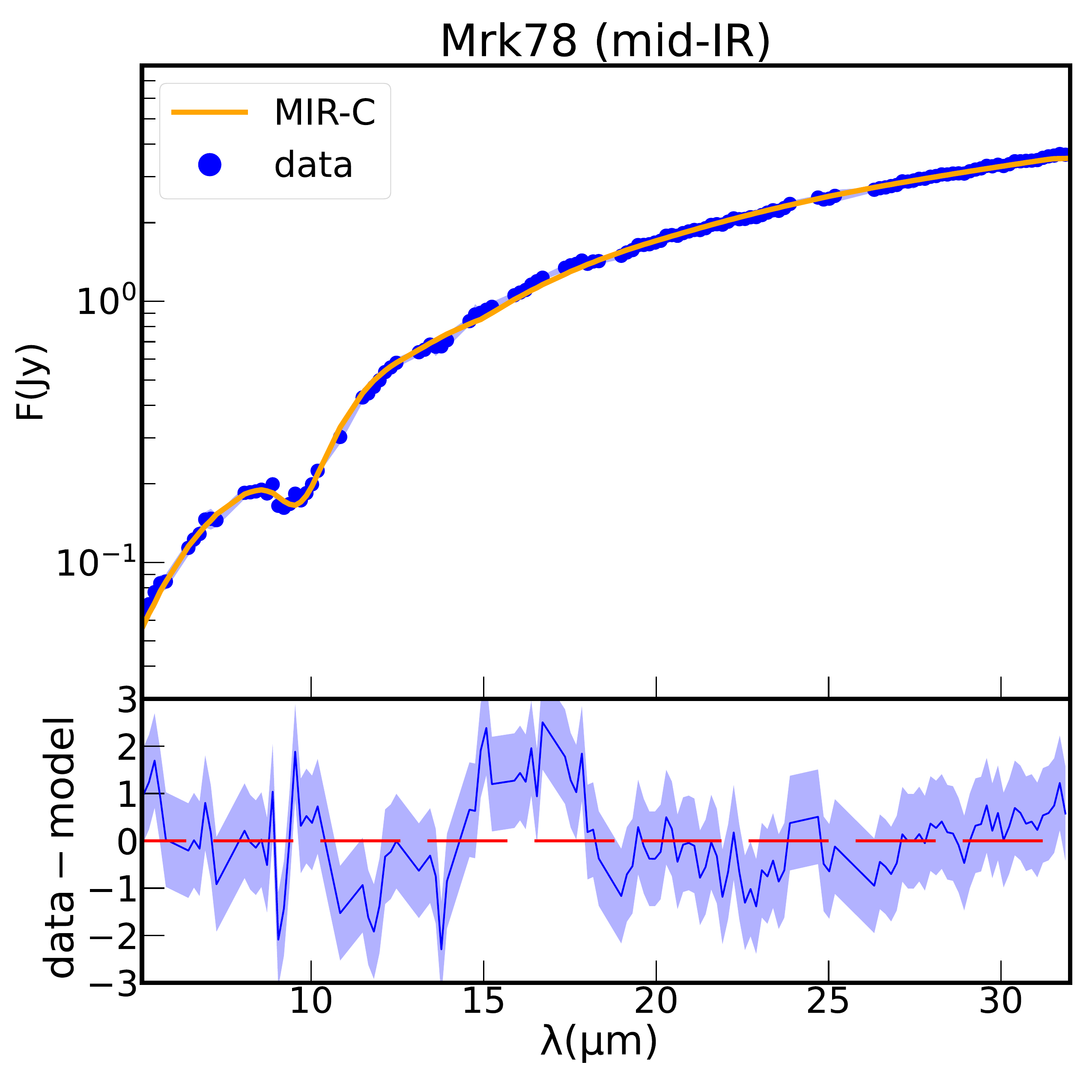}
    \includegraphics[width=0.48\columnwidth]{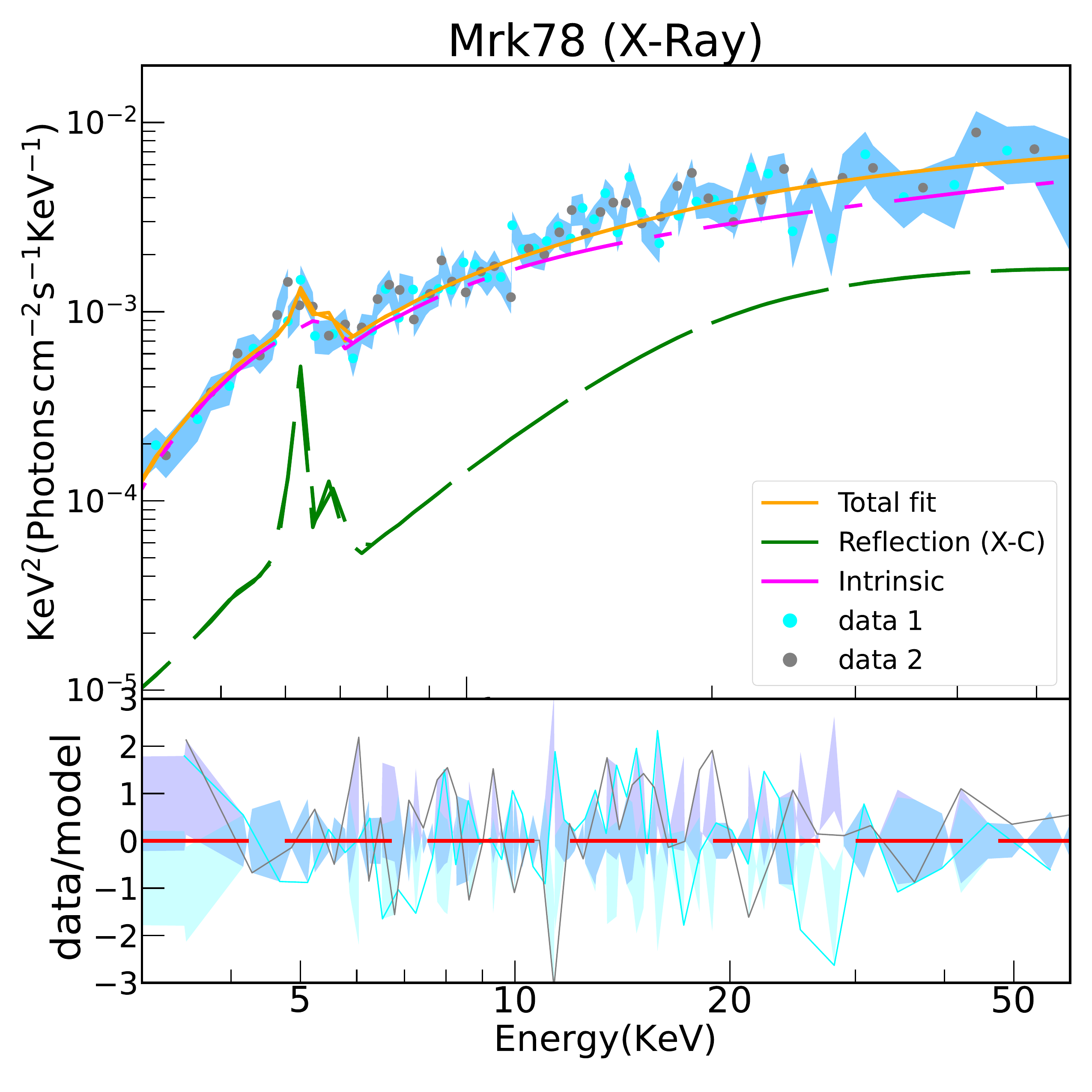}
    \includegraphics[width=0.48\columnwidth]{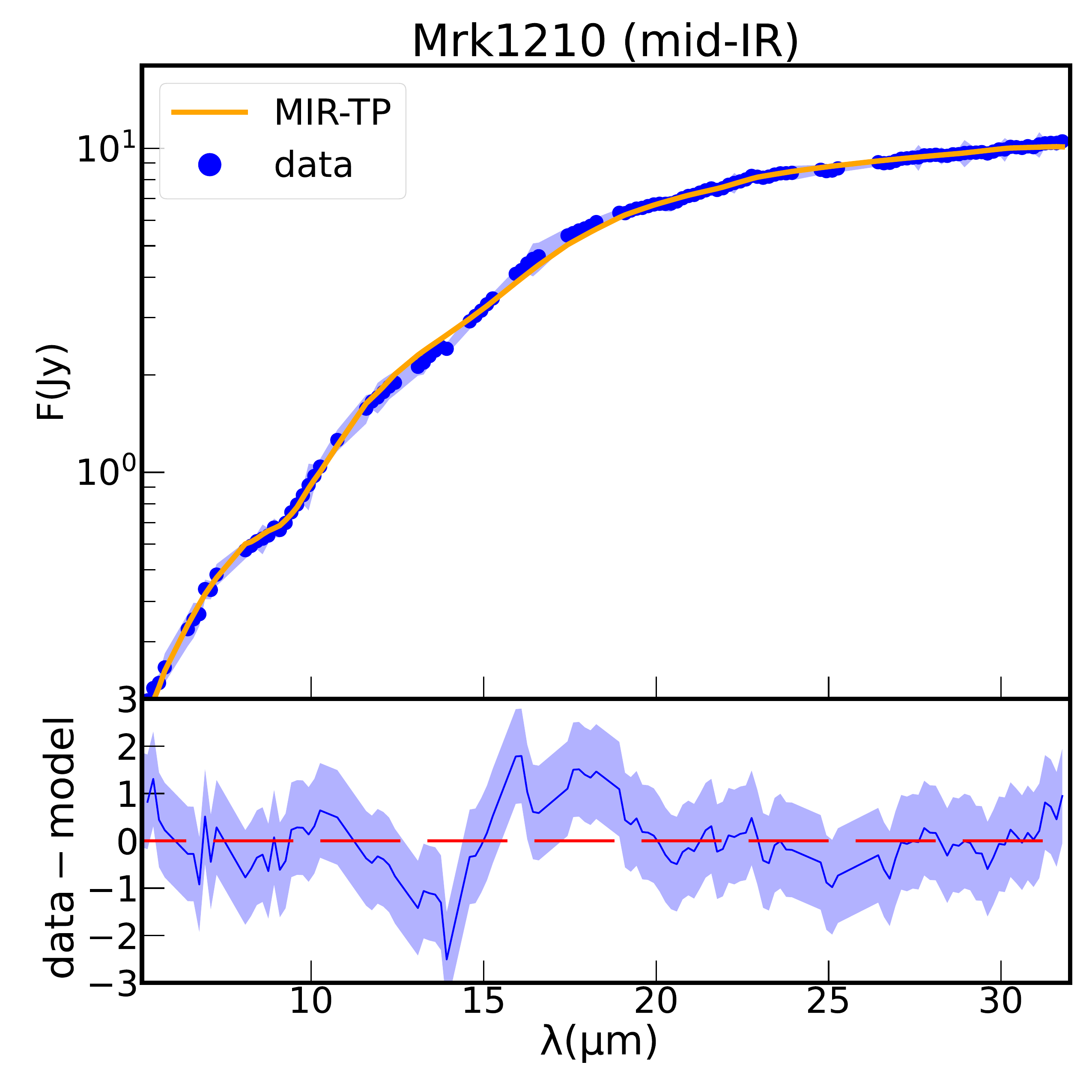}
    \includegraphics[width=0.48\columnwidth]{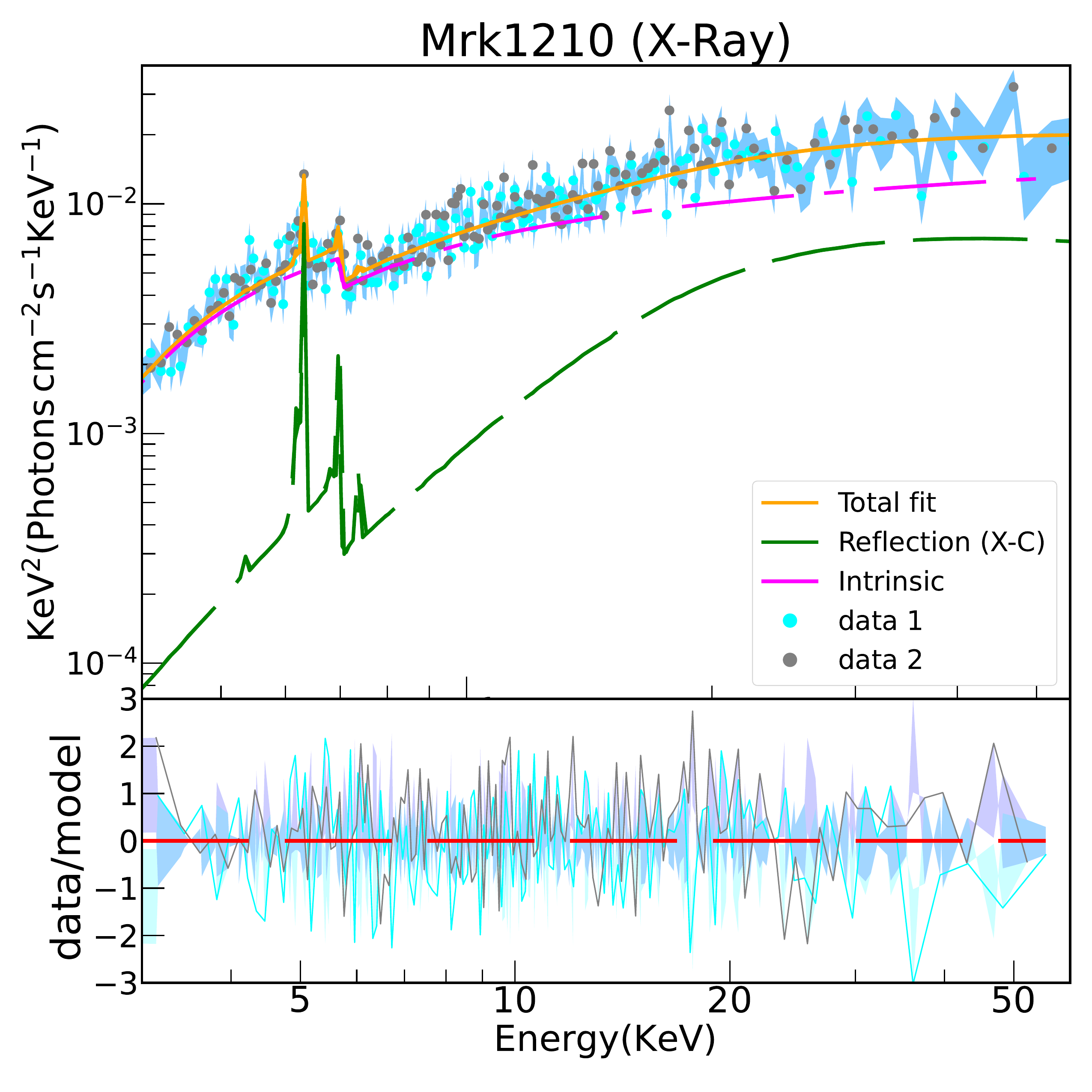}
    \includegraphics[width=0.48\columnwidth]{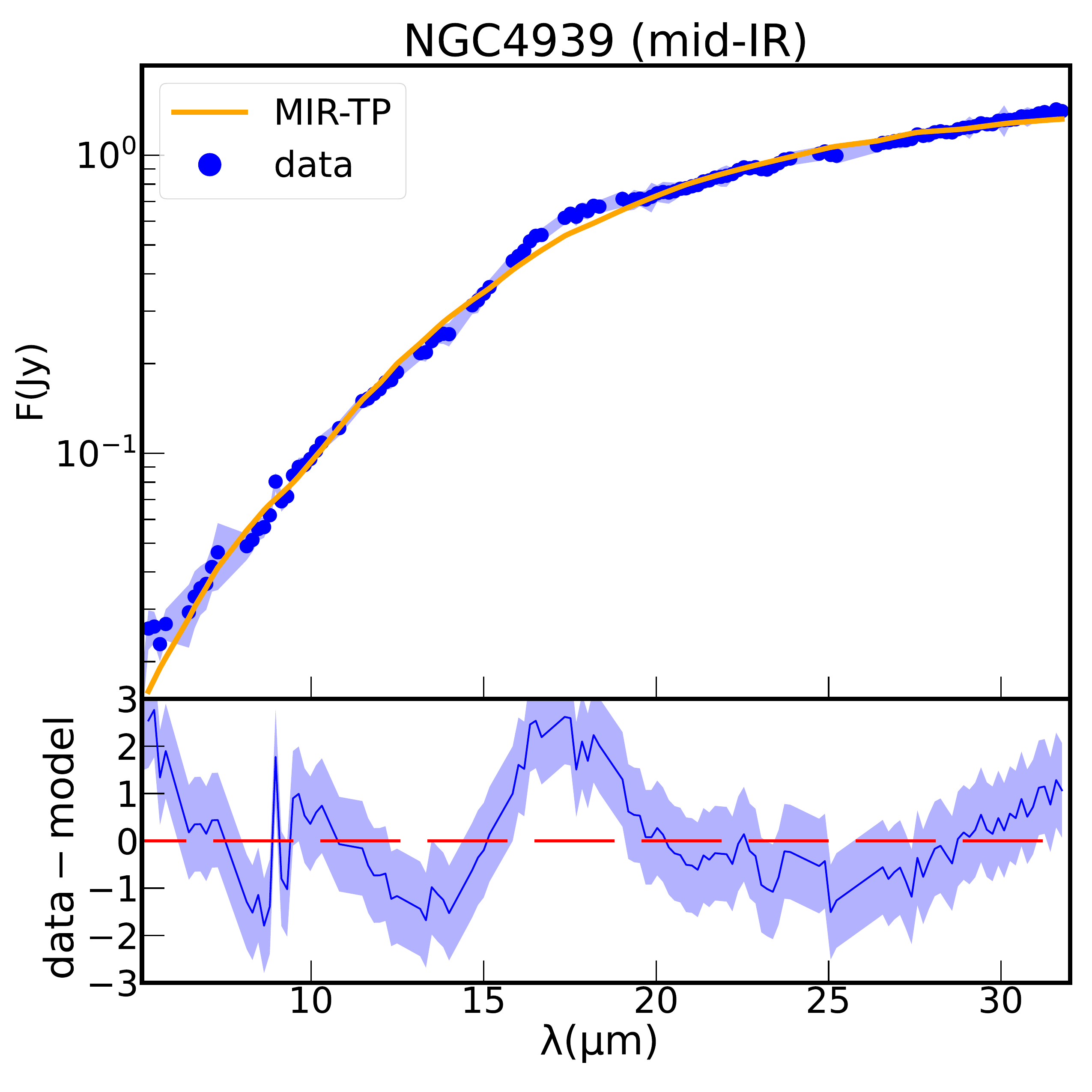}
    \includegraphics[width=0.48\columnwidth]{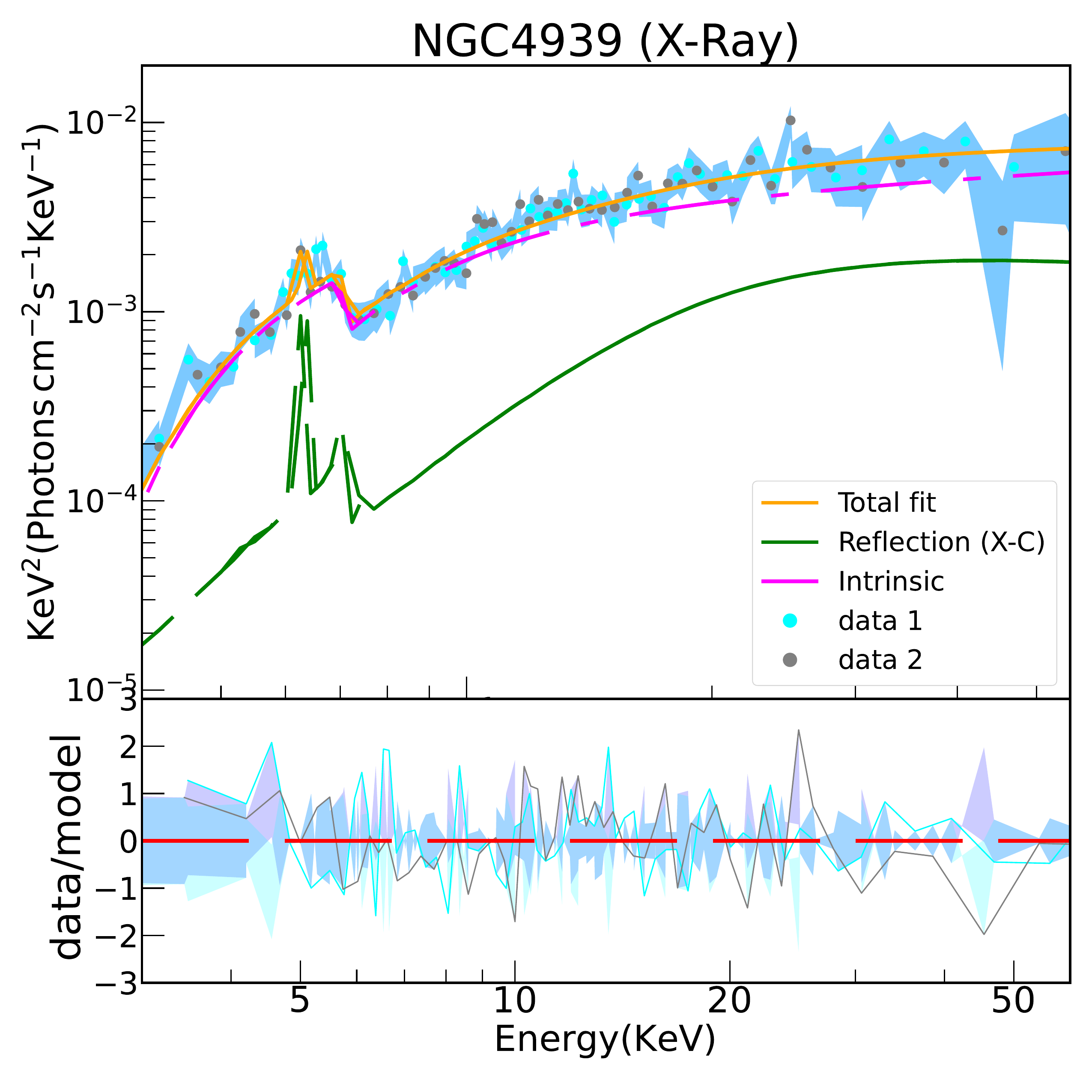}
    \includegraphics[width=0.48\columnwidth]{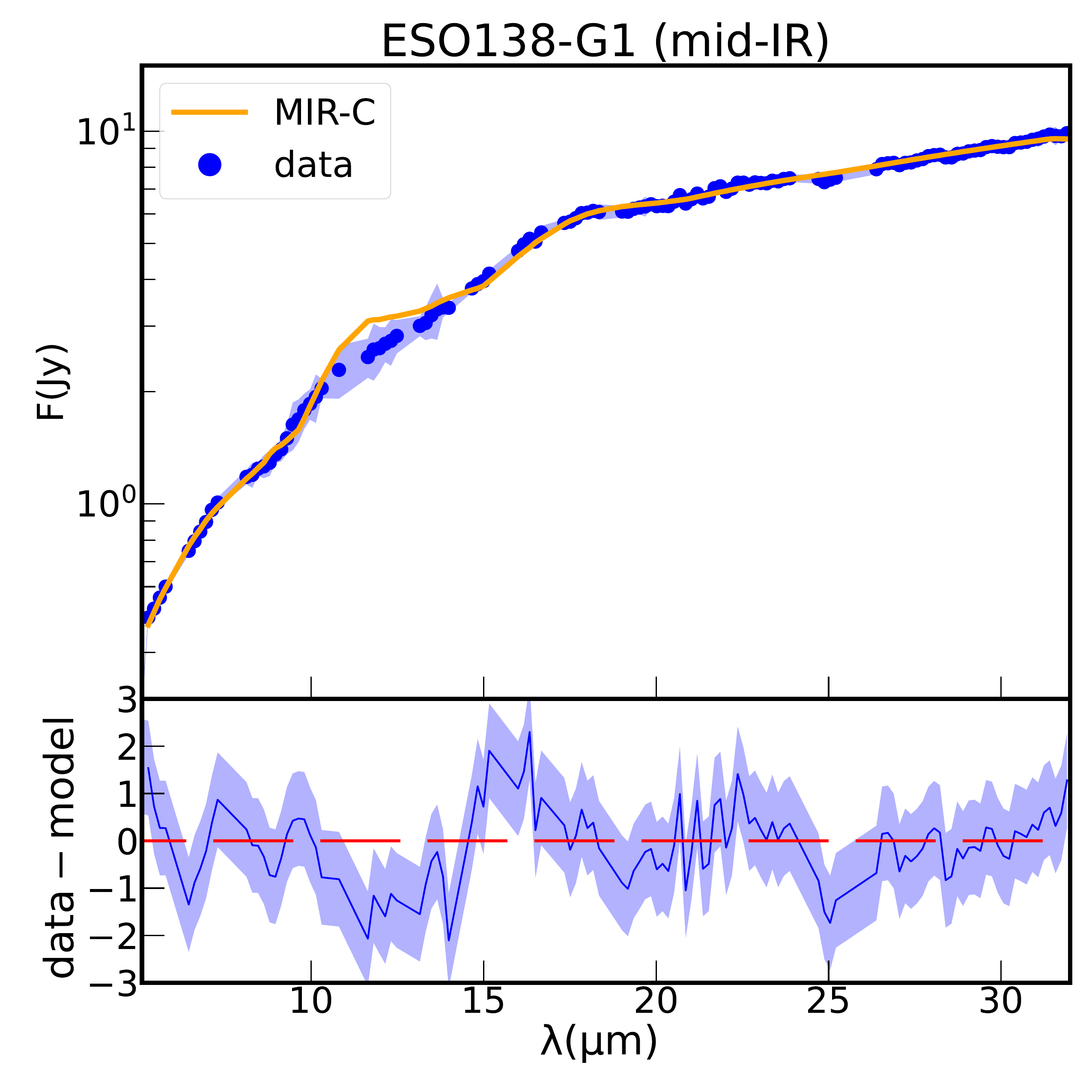}
    \includegraphics[width=0.48\columnwidth]{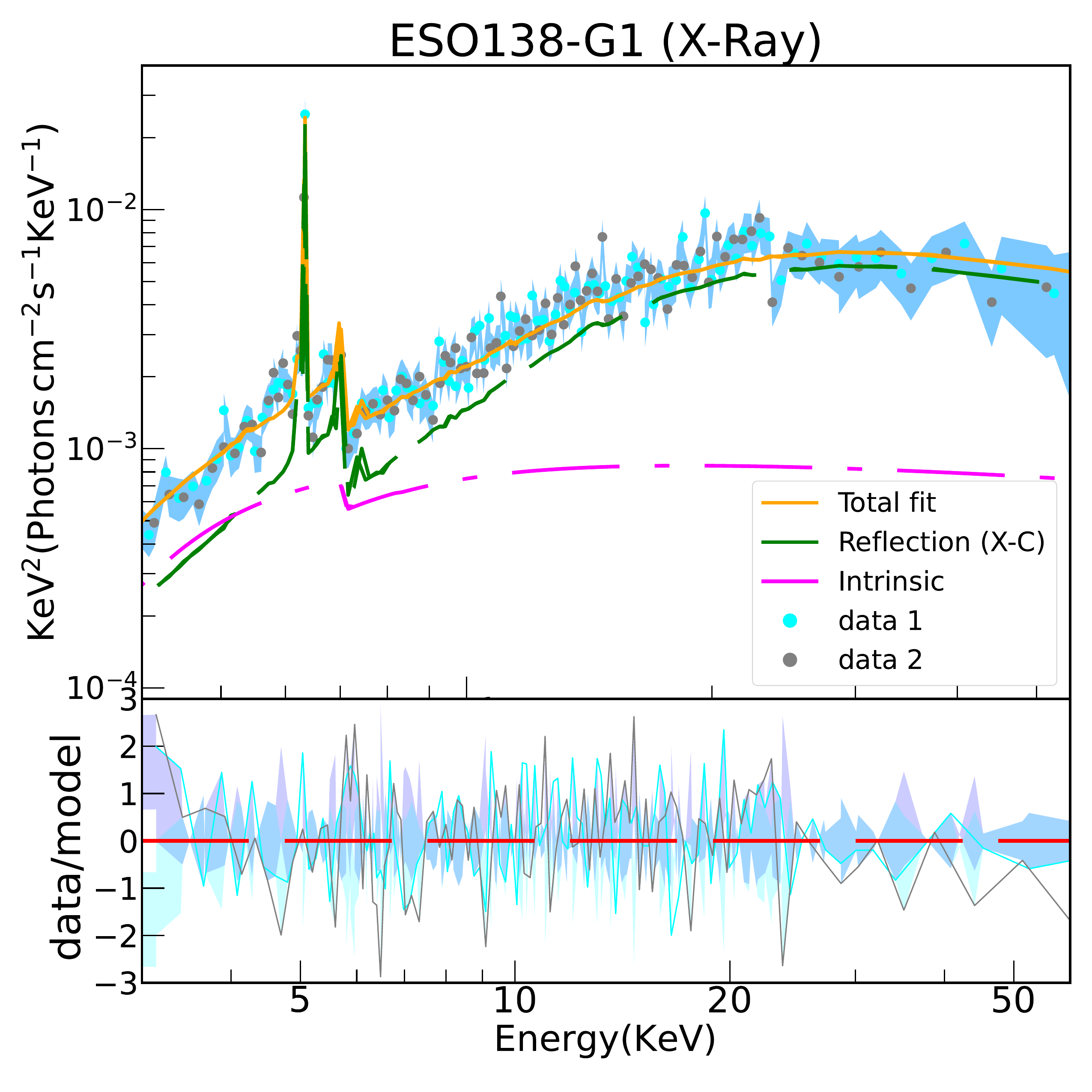}
    \caption{See the description at the beginning of this section.}
    \label{fig:Fits_3}
\end{figure}

\begin{figure}
 \includegraphics[width=0.48\columnwidth]{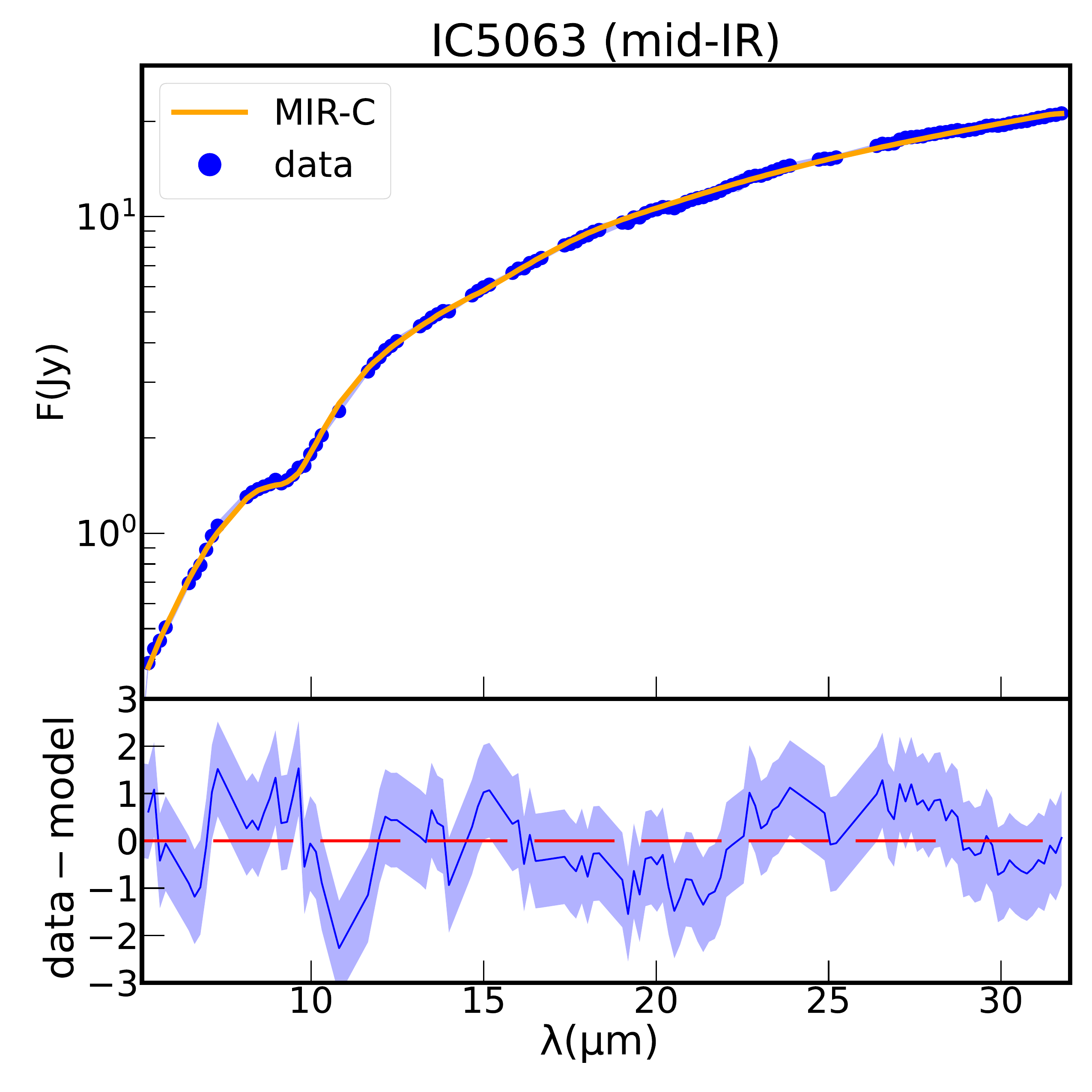}
\includegraphics[width=0.48\columnwidth]{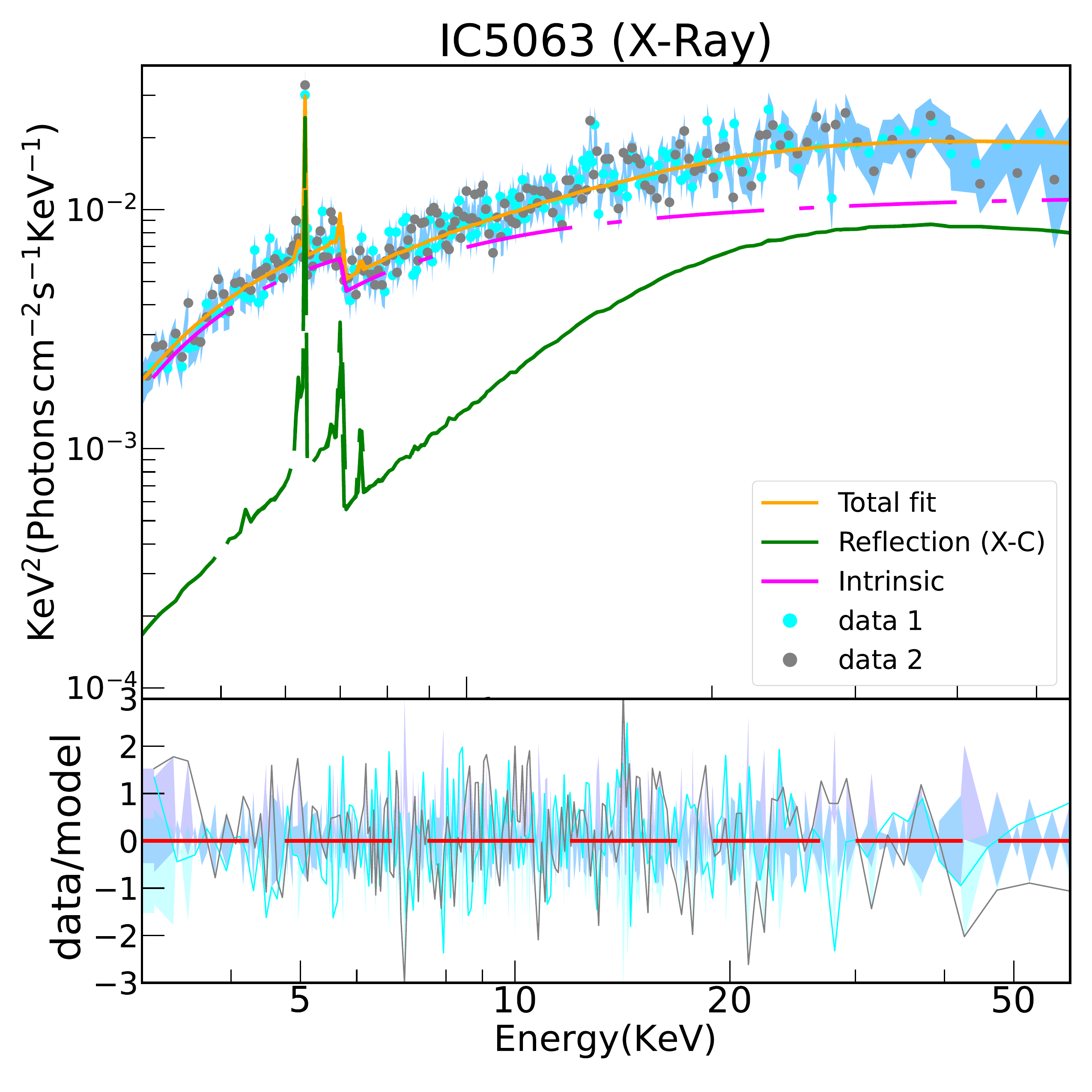}
\includegraphics[width=0.48\columnwidth]{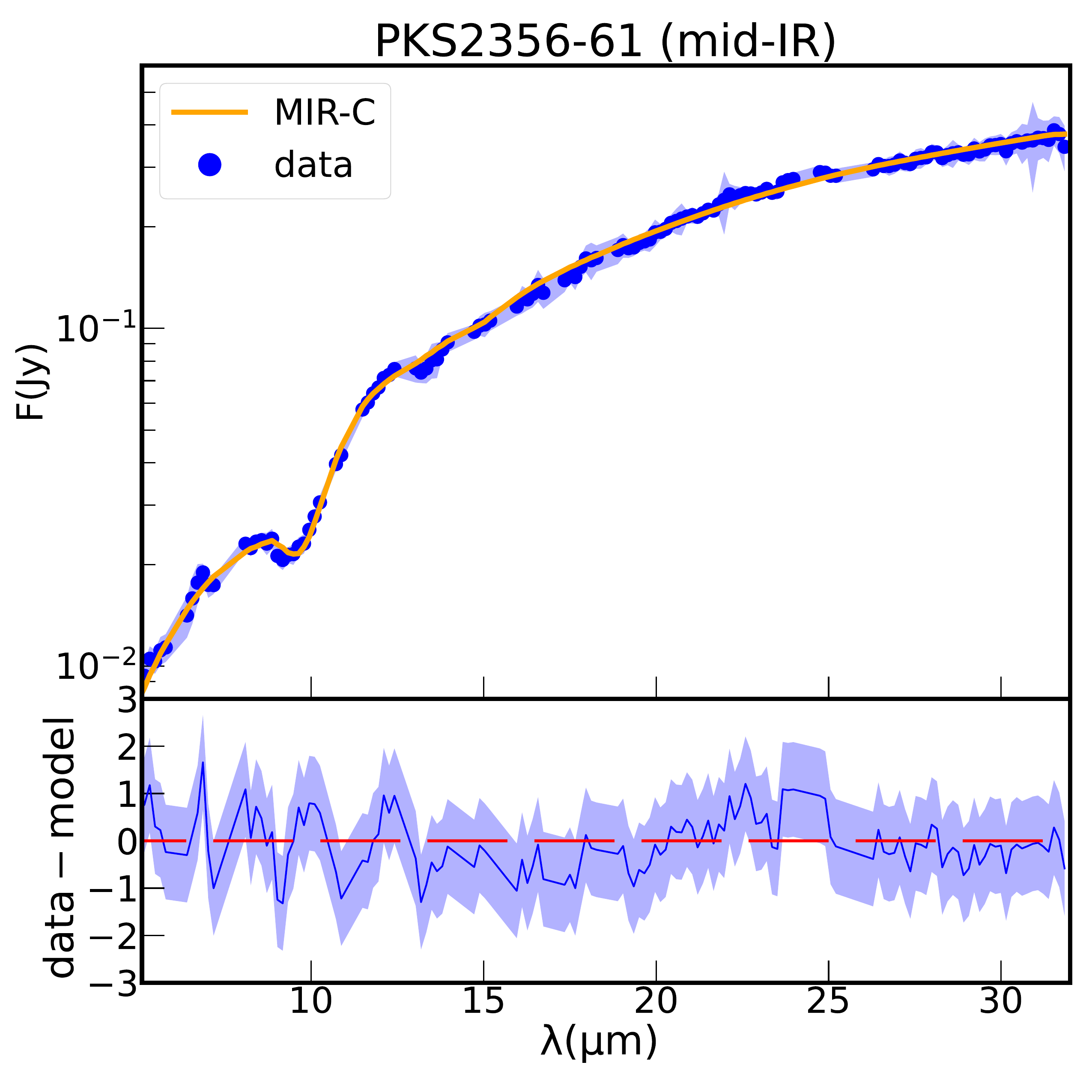}
    \includegraphics[width=0.48\columnwidth]{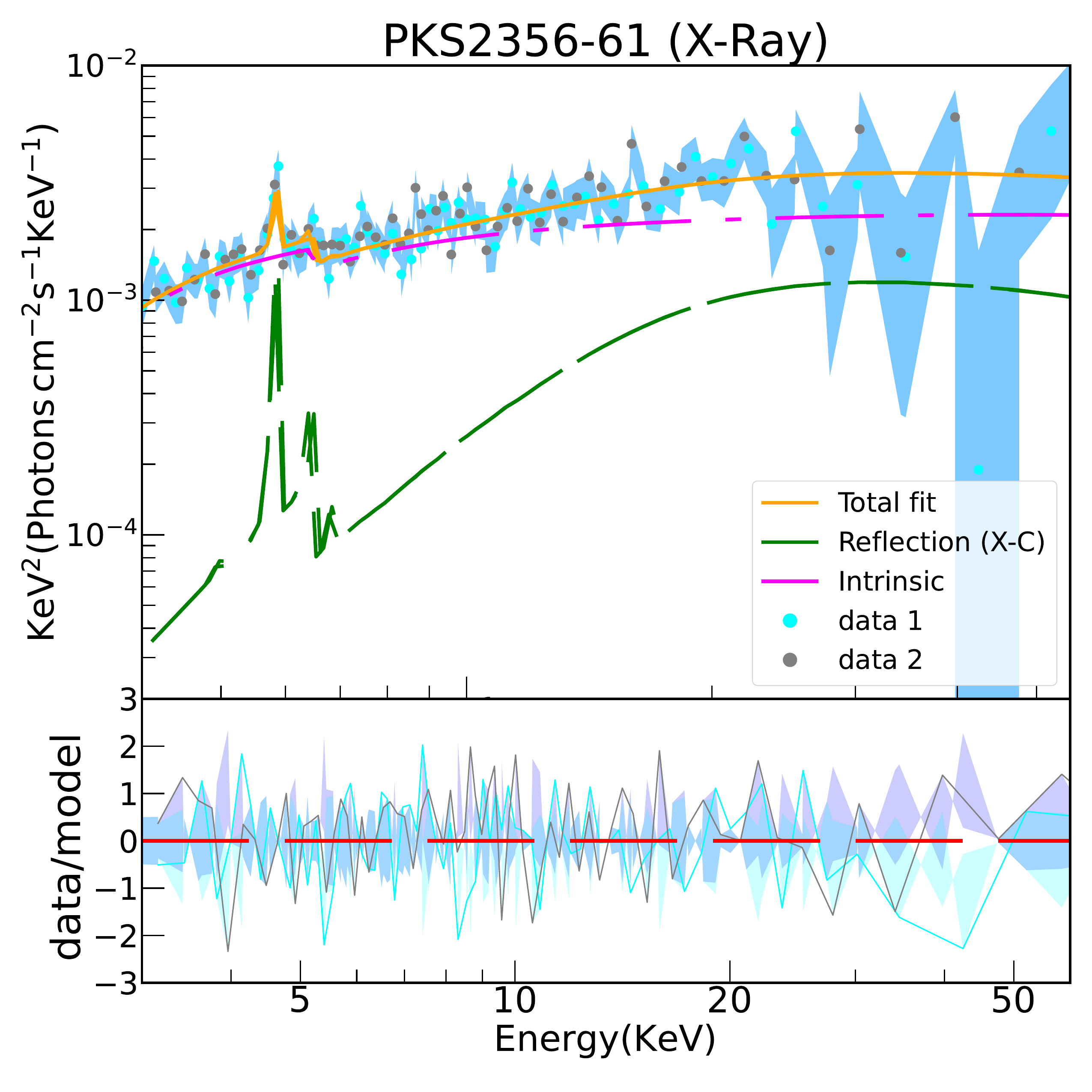}
\caption{See the description at the beginning of this section.}
    \label{fig:Fits_4}
\end{figure}

\clearpage

\onecolumn

\section{Comparison of confidence contours for free parameters before and after assuming that the dust and gas emission have the same origin structure.}

\label{sec:app2}


\begin{figure*}[ht!]
\centering
\includegraphics[width=1.\columnwidth]{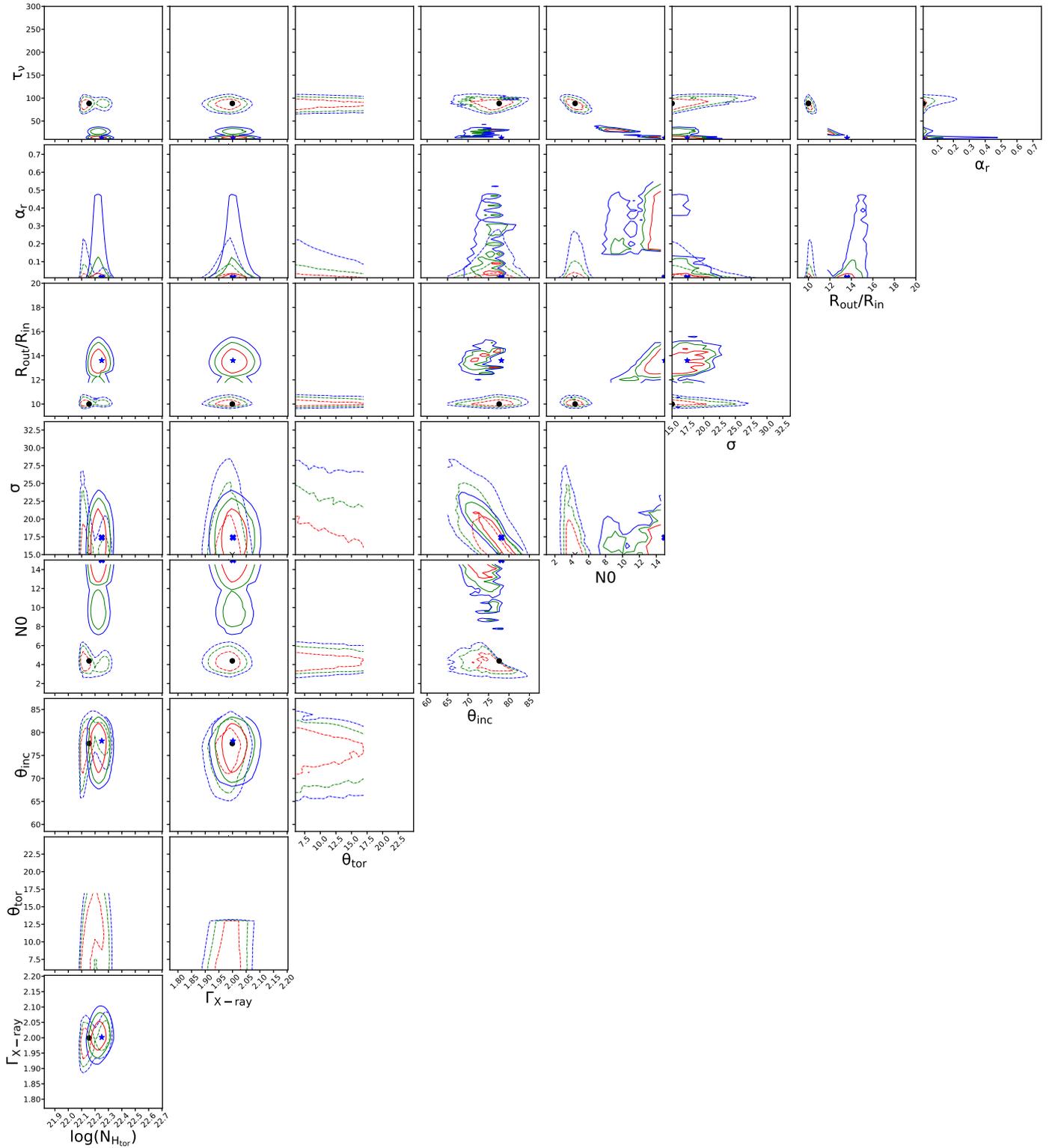}
\caption{Two-dimensional $\rm{\Delta \chi^2}$ contours for the resulting free parameters when we used the X-C/MIR-C baseline model before (dotted lines) and after the link (solid lines) the half-opening angle and torus angular width to fit Mrk590 spectra; Same description as in Fig.\,\ref{fig:Contours_ESO138-G01}}
    \label{fig:Contours_Mrk590}
\end{figure*}

\begin{figure*}[ht!]
\centering
\includegraphics[width=1.\columnwidth]{Figures/Contours_fits/PG0804+761_Clumpy_GoMar23.pdf}
\caption{Same description as in Figs.\,\ref{fig:Contours_ESO138-G01} and \ref{fig:Contours_Mrk590}, but for the case of PG0804+761 using the X-C/MIR-TP baseline model.}
    \label{fig:Contours_PG0804+761}
\end{figure*}

\begin{figure*}[ht!]
\centering
\includegraphics[width=1.\columnwidth]{Figures/Contours_fits/IRAS11119+3257_Smooth_GoMar23.pdf}
\caption{Same description as in Figs.\,\ref{fig:Contours_ESO138-G01} and \ref{fig:Contours_Mrk590}, but for the case of IRAS\,11119+3257 using the X-S/MIR-TP baseline model.}
    \label{fig:Contours_IRAS11119+3257}
\end{figure*}

\begin{figure*}[ht!]
\centering
\includegraphics[width=1.\columnwidth]{Figures/Contours_fits/Mrk231_Smooth_GoMar23.pdf}
\caption{Same description as in Figs.\,\ref{fig:Contours_ESO138-G01} and \ref{fig:Contours_Mrk590}, but for the case of Mrk\,231 using the X-S/MIR-TP baseline model.}
    \label{fig:Contours_Mrk231}
\end{figure*}

\begin{figure*}[ht!]
\centering
\includegraphics[width=1.\columnwidth]{Figures/Contours_fits/Mrk1383_Clumpy_GoMar23.pdf}
\caption{Same description as in Figs.\,\ref{fig:Contours_ESO138-G01} and \ref{fig:Contours_Mrk590}, but for the case of Mrk\,1383 using the X-C/MIR-TP baseline model.}
    \label{fig:Contours_Mrk1383}
\end{figure*}

\begin{figure*}[ht!]
\centering
\includegraphics[width=1.\columnwidth]{Figures/Contours_fits/Mrk1392_Clumpy_Clumpy.pdf}
\caption{Same description as in Fig.\,\ref{fig:Contours_ESO138-G01} and \ref{fig:Contours_Mrk590}, but for the case of Mrk\,1392 using the X-C/MIR-C baseline model.}
    \label{fig:Contours_Mrk1392}
\end{figure*}

\begin{figure*}[ht!]
\centering
\includegraphics[width=1.\columnwidth]{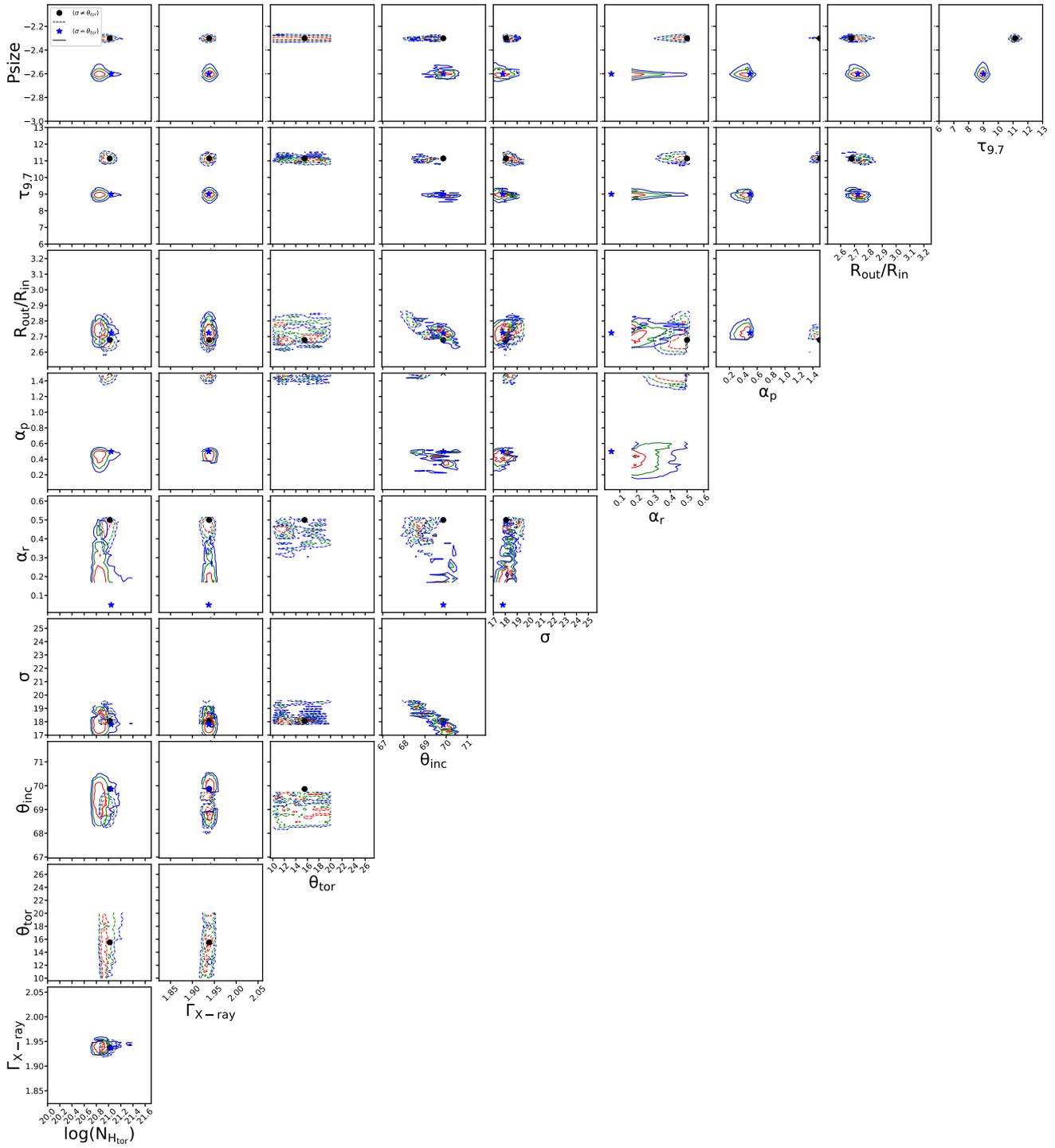}
\caption{Same description as in Figs.\,\ref{fig:Contours_ESO138-G01} and \ref{fig:Contours_Mrk590}, but for the case of ESO\,141-G055 using the X-C/MIR-TP baseline model.}
    \label{fig:Contours_ESO141-G055}
\end{figure*}

\begin{figure*}[ht!]
\centering
\includegraphics[width=1.\columnwidth]{Figures/Contours_fits/NGC7213_Smooth_Clumpy.pdf}
\caption{Same description as in Figs.\,\ref{fig:Contours_ESO138-G01} and \ref{fig:Contours_Mrk590}, but for the case of NGC\,7213 using the X-S/MIR-C baseline model.}
    \label{fig:Contours_NGC7213}
\end{figure*}


\begin{figure*}[ht!]
\centering
\includegraphics[width=1.\columnwidth]{Figures/Contours_fits/UM146_Clumpy_Clumpy.pdf}
\caption{Same description as in Figs.\,\ref{fig:Contours_ESO138-G01} and \ref{fig:Contours_Mrk590}, but for the case of UM\,146 using the X-C/MIR-C baseline model.}
    \label{fig:Contours_UM146}
\end{figure*}

\begin{figure*}[ht!]
\centering
\includegraphics[width=1.\columnwidth]{Figures/Contours_fits/NGC788_Clumpy_Clumpy.pdf}
\caption{Same description as in Figs.\,\ref{fig:Contours_ESO138-G01} and \ref{fig:Contours_Mrk590}, but for the case of NGC\,788 using the X-C/MIR-C baseline model.}
    \label{fig:Contours_NGC788}
\end{figure*}

\begin{figure*}[ht!]
\centering
\includegraphics[width=1.\columnwidth]{Figures/Contours_fits/NGC1358_Smooth_Clumpy.pdf}
\caption{Same description as in Figs.\,\ref{fig:Contours_ESO138-G01} and \ref{fig:Contours_Mrk590}, but for the case of NGC\,1358 using the X-S/MIR-C baseline model.}
    \label{fig:Contours_NGC1358}
\end{figure*}

\begin{figure*}[ht!]
\centering
\includegraphics[width=1.\columnwidth]{Figures/Contours_fits/Mrk78_Clumpy_Clumpy.pdf}
\caption{Same description as in Figs.\,\ref{fig:Contours_ESO138-G01} and \ref{fig:Contours_Mrk590}, but for the case of Mrk\,78 using the X-C/MIR-C baseline model.}
    \label{fig:Contours_Mrk78}
\end{figure*}

\begin{figure*}[ht!]
\centering
\includegraphics[width=1.\columnwidth]{Figures/Contours_fits/Mrk1210_Clumpy_GoMar23.pdf}
\caption{Same description as in Figs.\,\ref{fig:Contours_ESO138-G01} and \ref{fig:Contours_Mrk590}, but for the case of Mrk\,1210 using the X-C/MIR-TP baseline model.}
\label{fig:Contours_Mrk1210}
\end{figure*}

\begin{figure*}[ht!]
\centering
\includegraphics[width=1.\columnwidth]{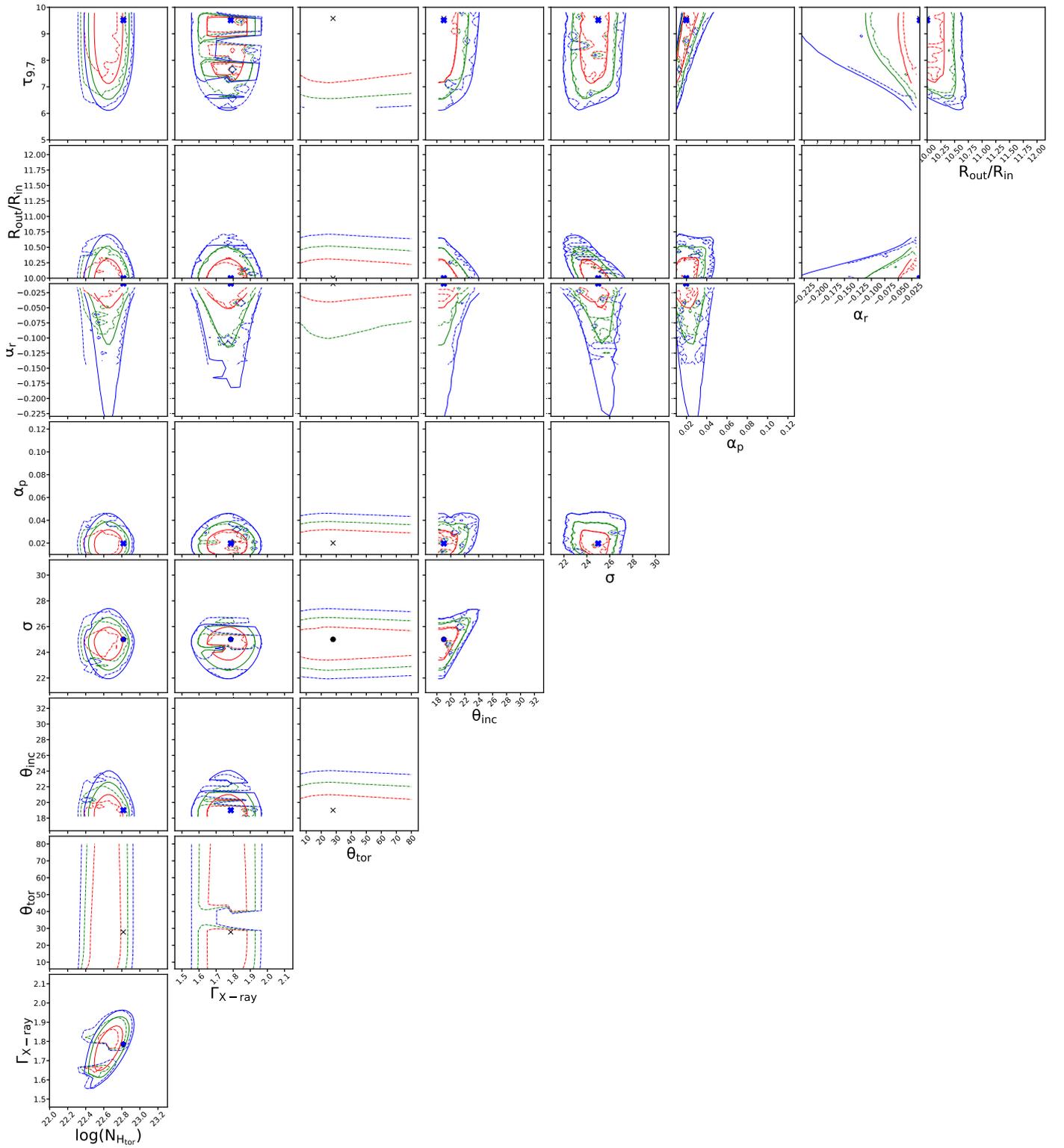}
\caption{Same description as in Figs.\,\ref{fig:Contours_ESO138-G01} and \ref{fig:Contours_Mrk590}, but for the case of 2MASX\,J10594361+6504063 using the X-C/MIR-S baseline model.}
    \label{fig:Contours_J10594361+6504063}
\end{figure*}

\begin{figure*}[ht!]
\centering
\includegraphics[width=1.\columnwidth]{Figures/Contours_fits/NGC4939_Clumpy_GoMar23.pdf}
\caption{Same description as in Figs.\,\ref{fig:Contours_ESO138-G01} and \ref{fig:Contours_Mrk590}, but for the case of NGC\,4939 using the X-C/MIR-TP baseline model.}
    \label{fig:Contours_NGC4939}
\end{figure*}

\begin{figure*}[ht!]
\centering
\includegraphics[width=1.\columnwidth]{Figures/Contours_fits/IC4518W_Clumpy_GoMar23.pdf}
\caption{Same description as in Figs.\,\ref{fig:Contours_ESO138-G01} and \ref{fig:Contours_Mrk590}, but for the case of IC\,4518W using the X-C/MIR-TP baseline model.}
    \label{fig:Contours_IC4518W}
\end{figure*}

\begin{figure*}[ht!]
\centering
\includegraphics[width=1.\columnwidth]{Figures/Contours_fits/NGC6300_Smooth_GoMar23.pdf}
\caption{Same description as in Figs.\,\ref{fig:Contours_ESO138-G01} and \ref{fig:Contours_Mrk590}, but for the case of NGC\,6300 using the X-S/MIR-TP baseline model.}
    \label{fig:Contours_NGC6300}
\end{figure*}

\begin{figure*}[ht!]
\centering
\includegraphics[width=1.\columnwidth]{Figures/Contours_fits/ESO103-G35_Clumpy_Clumpy.pdf}
\caption{Same description as in Figs.\,\ref{fig:Contours_ESO138-G01} and \ref{fig:Contours_Mrk590}, but for the case of ESO\,138-G01 using the X-C/MIR-C baseline model.}
\label{fig:Contours_ESO103-G35}
\end{figure*}

\begin{figure*}[ht!]
\centering
\includegraphics[width=1.\columnwidth]{Figures/Contours_fits/IC5063_Clumpy_Clumpy.pdf}
\caption{Same description as in Figs.\,\ref{fig:Contours_ESO138-G01} and \ref{fig:Contours_Mrk590}, but for the case of IC\,5063 using the X-C/MIR-C baseline model.}
\label{fig:Contours_IC5063}
\end{figure*}

\begin{figure*}[ht]
\centering
\includegraphics[width=1.\columnwidth]{Figures/Contours_fits/PKS2356-61_Clumpy_Clumpy.pdf}
\caption{Same description as in Figs.\,\ref{fig:Contours_ESO138-G01} and \ref{fig:Contours_Mrk590}, but for the case of PKS\,2356-61 using the X-C/MIR-C baseline model.}
\label{fig:Contours_PKS2356-61}
\end{figure*}

\end{appendix}

\label{LastPage}
\end{document}